\documentclass[times]{nmeauth}
\pdfoutput=1 
\usepackage{moreverb}
\usepackage[colorlinks,bookmarksopen,bookmarksnumbered,citecolor=red,urlcolor=red]{hyperref}
\newcommand\BibTeX{{\rmfamily B\kern-.05em \textsc{i\kern-.025em b}\kern-.08em
T\kern-.1667em\lower.7ex\hbox{E}\kern-.125emX}}

\usepackage{stmaryrd} 
\usepackage{xcolor,graphicx}                                    
\usepackage{algorithm}     
\usepackage{algorithmic}
\usepackage{amssymb,amsmath,amstext, amsfonts, mathrsfs, mathtools} 
\usepackage{arydshln} 
\usepackage{cite}
\usepackage{anyfontsize}

\algsetup{linenosize=\scriptsize}
\definecolor{algorithmcolor}{gray}{0.55} 

\newcommand{\algorithmicbreak}{\color{algorithmcolor}\textbf{break}\color{black}}

\renewcommand{\algorithmicwhile}{\color{algorithmcolor}\textbf{while}\color{black}}

\newcommand{\algorithmicbreakwhile}{\algorithmicbreak\ \algorithmicwhile}

\renewcommand{\div}{\nabla \cdot}     
\newcommand{\Div}{\div}               
\newcommand{\grad}{\nabla}            
\newcommand{\Grad}{\grad}             

        % ...

\usepackage{letltxmacro}
\renewcommand*{\mathellipsis}{%
  \mathinner{{\ldotp}{\ldotp}{\ldotp}}%
}
\makeatletter
\@ifdefinable{\org@ldots}{%
  \LetLtxMacro\org@ldots\ldots
  \DeclareRobustCommand*{\ldots}{%
    \ifmmode
      \expandafter\my@ldots
    \else
      \expandafter\textellipsis
    \fi
  }%
}
\newcommand*{\neghalfmskip}{%
  \nonscript\mskip-.5\muexpr\thinmuskip\relax%
}
\newcommand*{\my@ldots}{%
  \mathellipsis
  \@ifnextchar,\neghalfmskip{%
  \@ifnextchar:\neghalfmskip{%
  \@ifnextchar;\neghalfmskip{%
  \@ifnextchar.\neghalfmskip{%
  \@ifnextchar!\neghalfmskip{%
  \@ifnextchar?\neghalfmskip{%
    \rightdelim@
    \ifgtest@
      \mskip-.5\muexpr\thinmuskip\relax
    \fi
  }}}}}}%
}
\makeatother

\newcommand \dinf[1]   { \,\textrm d{#1}                        }   
\newcommand \deriv[2]  { \tfrac{\mathrm d{#1}}{\mathrm d{#2}}    }   
\newcommand \derivsecond[2]  { \frac{\mathrm d^2{#1}}{\mathrm d{#2}^2}    }   
\newcommand \partderiv[2]{\tfrac{\partial #1}{\partial #2}      }   

\DeclareMathOperator{\dGamma}{\dinf{\Int}}

\newcommand{\DSecondTime}[1]{\derivsecond{#1}{t}}                   
\newcommand{\DTime}[1]{\deriv{#1}{t}}                   
\newcommand{\dTime}[1]{\partderiv{#1}{t}}               
\newcommand{\dtimedot}[1]{\dot{#1}}                             
\newcommand{\dsecondtimedot}[1]{\ddot{#1}}                             

\newcommand{\restr}[2]{ \left. #1 \right|_{#2}}

\DeclareMathOperator{\const}{const}

\DeclareMathOperator{\supp}{supp}

\DeclareMathOperator{\RE}{Re}		
\DeclareMathOperator{\CFL}{CFL}		

\newcommand{\jump}[1]{\ensuremath{[\![#1]\!]} }
\newcommand{\wavg}[1]{\ensuremath{\left\{#1\right\}}}
\newcommand{\wavginv}[1]{\ensuremath{\langle #1 \rangle} }
\newcommand{\meanavg}[1]{\ensuremath{\left\{#1\right\}}_{\mathrm{m}}}

\DeclareMathOperator{\traceop}{tr}		
\newcommand{\trace}[1]{\traceop{(#1)}}

\newcommand{\vel}{\boldsymbol{u}}				

\newcommand{\x}{\boldsymbol{x}}
\newcommand{\n}{\boldsymbol{n}}
\newcommand{\bodyf}{\bff}

\newcommand{\stress}{\boldsymbol{\sigma}}
\newcommand{\visc}[1]{\boldsymbol{\epsilon}(#1)}
\newcommand{\I}{\boldsymbol{I}}

\newcommand{\scalL}[2]{(#1,#2 )}                       
\newcommand{\scalLDom}[3]{(#1,#2 )_{#3}}               
\newcommand{\scalbound}[3]{\langle #1,#2 \rangle_{#3}} 

\newcommand{\norm}[1]{\|#1\|}                                     

\newcommand{\onehalf}{\frac{1}{2}}

\newcommand{\eqdef}{:=}      

\newcommand{\sd}{\mathrm{s}} 
\newcommand{\fd}{\mathrm{f}} 
\newcommand{\fs}{\mathrm{fs}} 

\newcommand{\Dom}{\Omega}      
\newcommand{\Domh}{\Dom_h}     

\newcommand{\Domi}{\Dom^i}          
\newcommand{\Domj}{\Dom^j}          
\newcommand{\Int}{\Gamma}           
\newcommand{\Boundi}{\Int^i}        
\newcommand{\Intij}{\Int^{ij}}      

\newcommand{\Domf}{\Dom^\fd}          
\newcommand{\Doms}{\Dom^\sd}          
\newcommand{\DomsRef}{\Dom_0^\sd}          

\newcommand{\IntN}{\Int_{\mathrm{N}}}          
\newcommand{\IntD}{\Int_{\mathrm{D}}}          

\newcommand{\Intfs}{\Int^{\fs}}      

\newcommand{\IntDRef}{\Int_{\mathrm{D},0}}     
\newcommand{\IntNRef}{\Int_{\mathrm{N},0}}     

\newcommand{\bfhN}{\bfh_{\mathrm{N}}} 
\newcommand{\bfgD}{\bfg_{\mathrm{D}}} 

\newcommand{\bfsigma}{{\pmb\sigma}}

\newcommand{\bfepsilon}{{\pmb\epsilon}}

\newcommand{\bfc}{\boldsymbol{c}}
\newcommand{\bfu}{\boldsymbol{u}}
\newcommand{\bff}{\boldsymbol{f}}
\newcommand{\bfg}{\boldsymbol{g}}
\newcommand{\bfh}{\boldsymbol{h}}
\newcommand{\bfv}{\boldsymbol{v}}

\newcommand{\bfw}{\boldsymbol{w}}

\newcommand{\bfz}{\boldsymbol{z}}
\newcommand{\bfd}{\boldsymbol{d}}
\newcommand{\bfa}{\boldsymbol{a}}
\newcommand{\bfn}{\boldsymbol{n}}
\newcommand{\bfni}{\bfn^i}
\newcommand{\bfnj}{\bfn^j}
\newcommand{\bfnij}{\bfn^{ij}}

\newcommand{\bfr}{\boldsymbol{r}}

\newcommand{\bfP}{\boldsymbol{P}} 
\newcommand{\bfS}{\boldsymbol{S}} 
\newcommand{\bfF}{\boldsymbol{F}} 
\newcommand{\bfR}{\boldsymbol{R}} 
\newcommand{\bfU}{\boldsymbol{U}} 
\newcommand{\bfC}{\boldsymbol{C}} 
\newcommand{\bfE}{\boldsymbol{E}} 
\newcommand{\bfN}{\boldsymbol{N}} 
\newcommand{\bfG}{\boldsymbol{G}} 
\newcommand{\bfH}{\boldsymbol{H}} 

\newcommand{\bfM}{\boldsymbol{M}} 
\newcommand{\bfD}{\boldsymbol{D}} 
\newcommand{\bfA}{\boldsymbol{A}} 

\newcommand{\bfL}{\boldsymbol{L}} 

\newcommand{\bfx}{\boldsymbol{x}} 
\newcommand{\bfX}{\boldsymbol{X}} 
\newcommand{\bfxi}{\boldsymbol{\xi}} 
\newcommand{\bfChi}{\boldsymbol{\chi}} 
\newcommand{\bfPsi}{\boldsymbol{\Psi}} 
\newcommand{\bfPhi}{\boldsymbol{\Phi}} 
\newcommand{\bfvarphi}{\boldsymbol{\varphi}} 

\newcommand{\bfzero}{\boldsymbol{0}}

\newcommand{\R}{\mathbb{R}}

\newcommand{\RR}{\mathbb{R}}

\newcommand{\foralls}{\forall\,}

\newcommand{\mcG}{\mathcal{G}}
\newcommand{\mcF}{\mathcal{F}}
\newcommand{\mcT}{\mathcal{T}}
\newcommand{\mcV}{\mathcal{V}}
\newcommand{\mcW}{\mathcal{W}}
\newcommand{\mcX}{\mathcal{X}}
\newcommand{\mcQ}{\mathcal{Q}}
\newcommand{\mcE}{\mathcal{E}}
\newcommand{\mcN}{\mathcal{N}}
\newcommand{\mcS}{\mathcal{S}}
\newcommand{\mcC}{\mathcal{C}}

\newcommand{\mcA}{\mathcal{A}}

\newcommand{\mcR}{\mathcal{R}}
\newcommand{\mcD}{\mathcal{D}}
\newcommand{\mcB}{\mathcal{B}}

\newcommand{\mcL}{\mathcal{L}}

\newcommand{\apriori}{\emph{a~priori}}

\newcommand{\ndof}{\textrm{ndof}} 

\newcommand{\nablan}{\partial_{\bfn}}

\newcommand{\breakeq   }[3]{\ifthenelse{\equal{#1}{break}}{\nonumber\\&#2}{#3}}
\newcommand{\addifbreak}[2]{\ifthenelse{\equal{#1}{break}}{}{#2}}

\newcommand{\lat}[1]{#1} 
\newcommand{\ie}{\lat{i.e.\@}} 
 
\newcommand{\eg}{\lat{e.g.\@}}
\newcommand{\viz}{\lat{viz.\@}}
\newcommand{\st}{\lat{s.t.\@}} 
\newcommand{\etal}{\emph{et al.}}

\newcommand{\name}[1]{{\textsc{#1}}} 

\newcommand{\Secref}[1]{Section\,\ref{#1}}   
\newcommand{\Figref}[1]{Figure\,\ref{#1}}    
\newcommand{\Eqref}[1]{\eqref{#1}}           
\newcommand{\Algoref}[1]{Algorithm\,\ref{#1}}

\newcounter{defcount}
\newcounter{approachcount}
\newcounter{remcount}

\newtheorem{approach}[approachcount]{\textbf{Approach}}  

\newtheorem{definition}[defcount]{Definition}  
\newtheorem{remark}[remcount]{Remark}
\definecolor{darkblue}{rgb}{0.27 0.52 0.60}

\begin{document}
\runningheads{B.~Schott et al.}{Monolithic cut finite element based approaches for FSI}
\title{Monolithic cut finite element based approaches for\\ fluid-structure interaction}
\author{B.~Schott\corrauth, C.~Ager and W.A.~Wall}
\address{Institute for Computational Mechanics, Technical University of Munich,\linebreak Boltzmannstra{\ss}e~15, 85747~Garching,~Germany}
\corraddr{B.~Schott, Institute for Computational Mechanics, Technical University of Munich, Boltzmannstra{\ss}e 15, D-85747 Garching, Germany. E-mail:~schott@lnm.mw.tum.de}
\begin{abstract}
Cut finite element method (\name{CutFEM}) based approaches towards challenging fluid-structure interaction (\name{FSI}) are proposed.
The different considered methods combine the advantages of competing novel Eulerian (fixed-grid) and established Arbitrary-Lagrangian-Eulerian (ALE)
(moving mesh) finite element formulations for the fluid.
The objective is to highlight the benefit of using cut finite element techniques
for moving domain problems and to demonstrate their high potential with regards to
simplified mesh generation, treatment of large structural motions in surrounding flows, capturing boundary layers, their ability to deal with topological changes in the fluid phase
and their general straightforward extensibility to other coupled multiphysics problems.
In addition to a pure fixed-grid \name{FSI} method, also advanced fluid domain decomposition techniques
are considered rendering in highly flexible discretization methods for the \name{FSI} problem.
All stabilized formulations include Nitsche-based weak coupling of the phases supported by the ghost penalty technique for the flow field.
For the resulting systems, monolithic solution strategies are presented.
Various 2D and 3D \name{FSI}-cases of different complexity validate the methods and demonstrate their
capabilities and limitations in different situations.
\end{abstract}
\maketitle

\section{Introduction}

Fluid-structure interaction (\name{FSI}) problems are of significant interest in various fields of engineering and applied sciences.
However, current and future problem configurations of interest take available approaches often to their limit.
For this purpose, the capabilities of different geometrically unfitted cut finite element method (\name{CutFEM}) based discretization concepts
will be discussed. Moreover, novel Nitsche-based approaches and important algorithmic aspects will be proposed.

We start with providing important prerequisites for achieving reliable results
and for having a competitive discretization method for \name{FSI}.
An appropriate mesh resolution in the boundary layer is mandatory in order to capture the wall normal gradients around the
wet structure surface accurately. An insufficient mesh quality at the fluid-structure interface likely
results in an overall corrupted solution of the coupled problem.
The ability to adequately deal with this is an essential advantage of the established Arbitrary-Lagrangian-Eulerian (ALE)-based FSI approach,
a technique which goes back to early works by, \eg, \cite{Hirt1974,Donea1977,Belytschko1978,Belytschko1980,Hughes1981,DoneaGiulianiHalleux1982}.
It however requires that the fluid mesh always follows and deforms with the structure.
Large structural motions can heavily distort the fluid mesh, such that costly and sometimes numerically problematic remeshing and mesh-updating needs to be regularly performed.
Such limitations have been addressed in several publications, see, \eg,
\cite{LeTallec2001,WallGerstenbergerGamnitzerEtAl2006}.
Different attempts to relax the strong constraints on interface motion in \name{FSI} have been made.
A dual mortar interface coupling method was integrated into \name{ALE}-based \name{FSI} frameworks
in \cite{Kloppel2011} to cope with non-conforming meshes at the fluid-structure interface and so to enable mesh sliding and hence for example large rotational motions.
Non-matching discretizations in the framework of isogeometric \name{FSI} have been considered in
\cite{Bazilevs2012} and successfully applied to simulate the \name{FSI} of wind turbines.
Recently, in \cite{Farhat2014} an \name{ALE} formulation of embedded boundary methods was proposed
for tracking boundary layers in turbulent \name{FSI} problems. The key idea of this approach is that
non-interface-fitted embedded meshes are rigidly translated and/or rotated to track the rigid body component of the dynamic solid motion.

A class of \name{FSI} approaches which enables to overcome the difficulty of mesh distortion inherent to moving mesh techniques
are so-called \emph{fixed-grid methods}.
While the structural part is still described in a Lagrangean formalism, the fluid flow is approximated on a fixed non-interface-fitted non-moving
computational grid. This allows the structure to arbitrarily move within the fluid mesh.
Fixed-grid schemes differ in the numerical treatment of this overlap region, in the enforcement of the coupling constraints
at the \name{FSI} interface, as well as in the solution techniques applied to the overall fluid-solid system.

\emph{Immersed Boundary} (\name{IB}) methods date back to \cite{Peskin1972}.
In this class of methods, fluid and solid phases are linked by equivalent volumetric force terms within the fluid domain.
Such incorporated interaction equations involve smoothed approximations to the Dirac-delta function
and appropriate transfer operators for data between the two meshes, see, \eg, \cite{Zhang2004,Wang2004,Mittal2005} and references therein for various variants. 
For most of the \name{IB} approaches, the coupling of fluid and solid phases takes place at the fictitious
fluid domain which does not exhibit physical meaning. Smeared approximations
and often induced artificial incompressibility to the structural field
are sources of inaccuracy and are desired to be avoided, as discussed in \cite{Wall2008}.
In the latter publication, different domain decomposition ideas for fluid-structure interaction based on fixed fluid grids
have been introduced. 
Similar to the \name{IB} methods are the Lagrange Multiplier based fictitious domain methods based on the works \cite{GlowinskiPanPeriaux1994,GlowinskiPanHeslaJoseph1999},
which have been applied to consider thin elastic and incompressible thick elastic structures \cite{Baaijens2001,LoonAndersonVosse2006,Yu2005}.

Over the recent decade, the so-called extended and cut finite element methods (\name{XFEM}, \name{CutFEM})
became quite popular numerical discretization schemes.
Cutting-off overlapped regions from the fluid meshes, which are covered by other solid or fluid phases, allows for an
accurate description of the physically meaningful fluid domain up to the moving boundary. It also avoids having artificial overlaps or 
the need for iterating between meshes, as is the case for Chimera methods \cite{Wang2004, Houzeaux2003, Steger1983},
and thus eases setting up a global system for the coupled \name{FSI} problem.
Successful application of geometrically unfitted extended/cut finite elements methods, see \cite{BurmanClausHansboEtAl2014} for an overview, have been provided
for crack problems \cite{Belytschko1999, Moes1999,Sukumar2000},
for single-phase flow \cite{SchottWall2014, MassingLarsonLoggEtAl2014}, incompressible two-phase flow \cite{Rasthofer2011, SchottRasthoferGravemeierWall2015, HansboLarsonZahedi2014, GrossReusken2007}
and surface PDEs \cite{Burman2016}.
With regards to \name{FSI},
different \name{CutFEM}/\name{XFEM}-based interface-coupling strategies have been developed over the recent years.
In \cite{Gerstenberger2008}, the use of Lagrange multipliers for unfitted overlapping discretizations has been investigated.
A combination of fixed-grid and \name{ALE} techniques applied to \name{FSI} has been proposed in \cite{BaigesCodina2009},
a technique which is called \emph{fixed-mesh ALE approach}.
In \cite{Gerstenberger2009}, a stress-based Lagrange-multiplier method was established for the weak constraint enforcement of coupling conditions.
The latter methodology has been expanded to \name{XFEM}-based fluid-structure-contact interaction problem settings in \cite{Mayer2010},
which enables simulating contact of submersed bodies.
A stabilized Lagrange-multiplier-based fictitious domain approach
applicable to \name{XFEM}-based Stokes flow approximations was developed in \cite{CourtFournieLozinski2014},
which was extended to \name{FSI} with flows governed by the incompressible Navier-Stokes equations
in \cite{CourtFournie2015}.
A Nitsche-based \name{CutFEM} for Stokes flow interacting with linear elastic structures has been analyzed in \cite{Burman2014a}.
A comparison of various coupling strategies of an incompressible fluid with immersed thin-walled structures has been recently provided
in \cite{Alauzet2016}.
Moreover, different formulations applicable to fixed-grid methods
based on utilizing an additional embedded fluid patch, which fits to the fluid-structure interface,
but overlaps with a fixed background fluid grid in an unfitted fashion, have been developed in \cite{Shahmiri2011, SchottShahmiriKruseWall2015,MassingLarsonLoggEtAl2014}.

The objective of this article is to provide a generalized unfitted framework for \name{FSI} solvers
that allows to relax strong limitations of most computational \name{FSI} approaches existing so far
and might build the basis for future developments on more advanced fluid-structure interaction applications.
While the structure is always approximated in a Lagrangean setting with an interface-fitted moving mesh,
for the flow field different fitted mesh and geometrically unfitted approaches, including Nitsche-based overlapping domain decomposition techniques, are provided and juxtaposed.
Major differences to aforementioned \name{FSI} approaches
consist in the multitude of potential Nitsche-based discretization concepts for \name{FSI}
and detailed descriptions of algorithmic aspects with regards to a desired monolithic solution of unfitted \name{FSI} approximations.
Thereby, non-linear structural behavior is considered and the flow is described by the full non-linear incompressible Navier-Stokes equations,
in contrast to previous publications which mainly considered Stokes flow, linear structural models
or restrictions that only allow thin-walled structures.
Challenging two- and three-dimensional \name{FSI} problem configurations will demonstrate
the capabilities of the proposed stabilized formulations and the algorithmic procedures.

This paper is outlined as follows:
In \Secref{sec:discretization_concepts}, different established fitted and novel unfitted mesh CutFEM based discretization concepts are proposed
and their limitations and capabilities for moving domain multifield problem settings are discussed.
\Secref{sec:problem_formulation_FSI} reviews the fluid-structure interaction problem formulation for
which a Nitsche-coupled numerical scheme will be provided in \Secref{sec:spatial_discretization}.
Algorithmic peculiarities of monolithically solved unfitted FSI approaches will be treated in \Secref{sec:fsi:monolithic}.
Numerical examples in \Secref{sec:numerical_examples} will demonstrate the robustness of the proposed FSI schemes
and highlight their capabilities for certain challenging problem configurations, as finally summarized in \Secref{sec:summary}.

\section{Geometrically Fitted and Unfitted Finite Element Discretization Concepts}
\label{sec:discretization_concepts}

To begin with, we introduce general notation and conventions for multifield problems,
which serve as a basis for the presentation of different computational approaches for
fluid-structure interaction problems.
Afterwards, terminology and notation frequently used in the context of mesh approximation techniques are introduced,
which mainly address some terms describing the mesh location with respect to the physical domain and
the positioning of different meshes relative to each other.
Finally, the limitations and capabilities of different geometrically fitted and unfitted discretization concepts
for demanding fluid-structure interaction problem configurations are
compared to each other.

\subsection{Domains, Boundaries and Conventions}

\begin{figure}
  \begin{center}
  \includegraphics[width=0.48\textwidth]{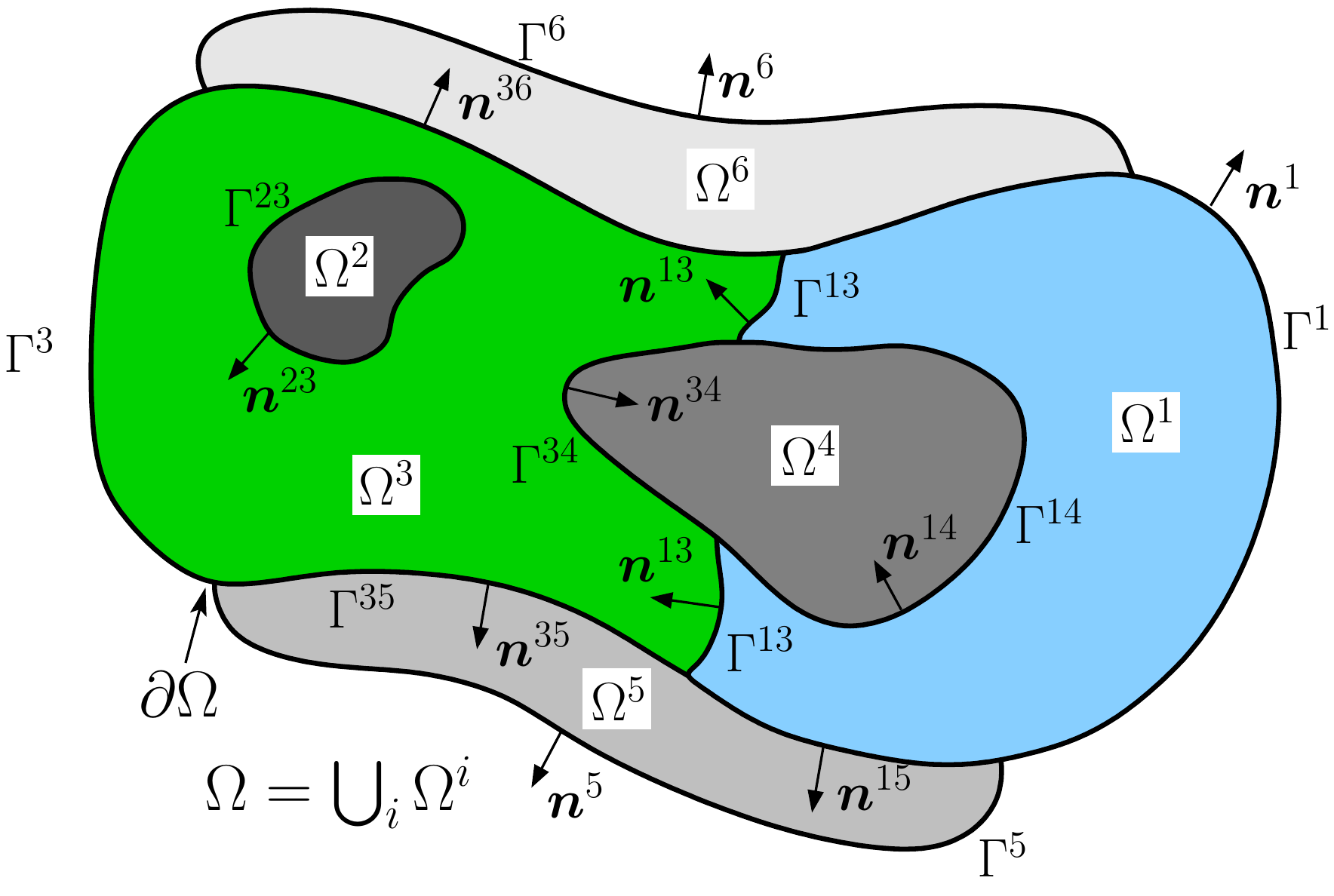} 
  \end{center}
\vspace{-12pt}
  \caption{
Multifield problem setting consisting of fluid and solid subdomains~$\Domi$, \mbox{$1\leqslant i \leqslant N_{\textrm{dom}}$}, exterior boundaries~$\Gamma^i$
and interfaces \mbox{$\Int^{ij} (i<j)$} with respective unit normal vectors $\bfn^i$ and $\bfn^{ij}$.}
  \label{fig:chap_2_multifield_setting_notation}
\end{figure}

Let \mbox{$\Dom(t) \eqdef \substack{\bigcup}_{1\leqslant i \leqslant N_{\textrm{dom}}} \Domi(t)$} be the overall physical domain of a multifield problem
consisting of a partition into \mbox{$N_{\textrm{dom}} \geqslant 1$} disjoint, possibly time-dependent subdomains
\mbox{$\Domi(t)\subset \R^d,~1\leqslant i \leqslant N_{\textrm{dom}}$}, with \mbox{$d=2,3$} and
\mbox{$\Intij(t) \eqdef \overline{\Domi(t)}\cap\overline{\Domj(t)}~(i<j)$} the separating interfaces between partitions, with the most prominent being the interface between fluid and solid phases, respectively.
Note that, if unmistakable, occasionally the temporal variable~$t$ is omitted to shorten the presentation.
For the subdomain boundaries $\partial \Domi$, which are assumed as \mbox{$(d-1)$}-manifolds,
partitions \mbox{$\partial \Domi = \Boundi \cup (\bigcup_{j>i}{\Intij})$}
with exterior boundaries \mbox{$\Boundi \subset \partial\Dom$} are considered.
Respective Dirichlet and Neumann boundary parts are denoted with \mbox{$\IntD^i,\IntN^i$}
and satisfy \mbox{$\Boundi = \IntD^i\cup\IntN^i$}.
Outward-pointing unit normal vectors defined on the manifolds are denoted with~$\bfni$ for boundaries and
with \mbox{$\bfnij \eqdef \bfni = -\bfnj~(i<j)$} for interfaces, respectively.
The notation is depicted in \Figref{fig:chap_2_multifield_setting_notation}.

For possibly discontinuous scalar quantities \mbox{$f:\Dom\rightarrow\R$},
like for instance material parameters, weighted average operators and a jump operator are defined as
\begin{alignat}{2}
  \wavg{f(\bfx)}    &=  \lim_{t\rightarrow0^+}(w^i f(\bfx-t\bfnij) + w^j f(\bfx+t\bfnij))  \quad&&\foralls \bfx\in\Intij, \label{eq:jump-average-definition_1} \\
  \wavginv{f(\bfx)} &=  \lim_{t\rightarrow0^+}(w^j f(\bfx-t\bfnij) + w^i f(\bfx+t\bfnij))  \quad&&\foralls \bfx\in\Intij, \label{eq:jump-average-definition_2} \\
  \jump{f(\bfx)}    &=  \lim_{t\rightarrow0^+}(f(\bfx-t\bfnij)-f(\bfx+t\bfnij)) \quad&&\foralls \bfx\in\Intij\label{eq:jump-average-definition_3}
\end{alignat}
with positive weights \mbox{$w^i,w^j\in[0,1]$} satisfying \mbox{$w^i=1-w^j$}.
The weights will be specified later depending on the computational approach.
Irrespective of their choice, the following relation holds
\begin{equation}
\label{eq:jump-average-relation}
 \jump{fg} = \jump{f}\wavg{g} + \wavginv{f}\jump{g}
\end{equation}
for functions~\mbox{$f,g$} with a straightforward extension to vector-valued quantities, like velocities or tractions.

\subsection{Notation on Function Spaces, Norms and Inner Products}

Deriving a weak formulation for fluid-structure interaction requires the definition of appropriate functional spaces.
For any time \mbox{$t\in(T_0,T]$} and \mbox{$U \in \{\Dom(t), \Int(t) \}$},
let $H^m(U)$ and $[H^m(U)]^d$ be the standard Sobolev space of order \mbox{$m \in \RR$}
and their $\RR^d$-valued equivalents.
Their corresponding inner products are denoted by
\mbox{$(\cdot,\cdot)_{m,\Dom}$} for the domain and by \mbox{$\langle\cdot,\cdot\rangle_{m,\Int}$} for the boundary.
It is occasionally written \mbox{$(\cdot,\cdot)_{U}$}, \mbox{$\langle \cdot,\cdot \rangle_{U}$} and $\|\cdot \|_{U}$
for the inner products and norms associated with $L^2(U)$, with $U$ being a Lebesque-measurable subset of $\RR^d$.
For \mbox{$m > 1/2$}, the
notation $[H_{\bfg}^m(\Dom)]^d$ is used to designate the set of all functions in $[H^m(\Dom)]^d$ whose
$\RR^d$-valued boundary traces are equal to~$\bfg$.

\subsection{Notation and Terminology on Computational Domains and Meshes}

\label{sec:dis_cut_fem:motivation_concepts:overview_domain_decomposition:notation_terminology}

\begin{figure}
  \begin{center}
    \includegraphics[width=0.47\textwidth]{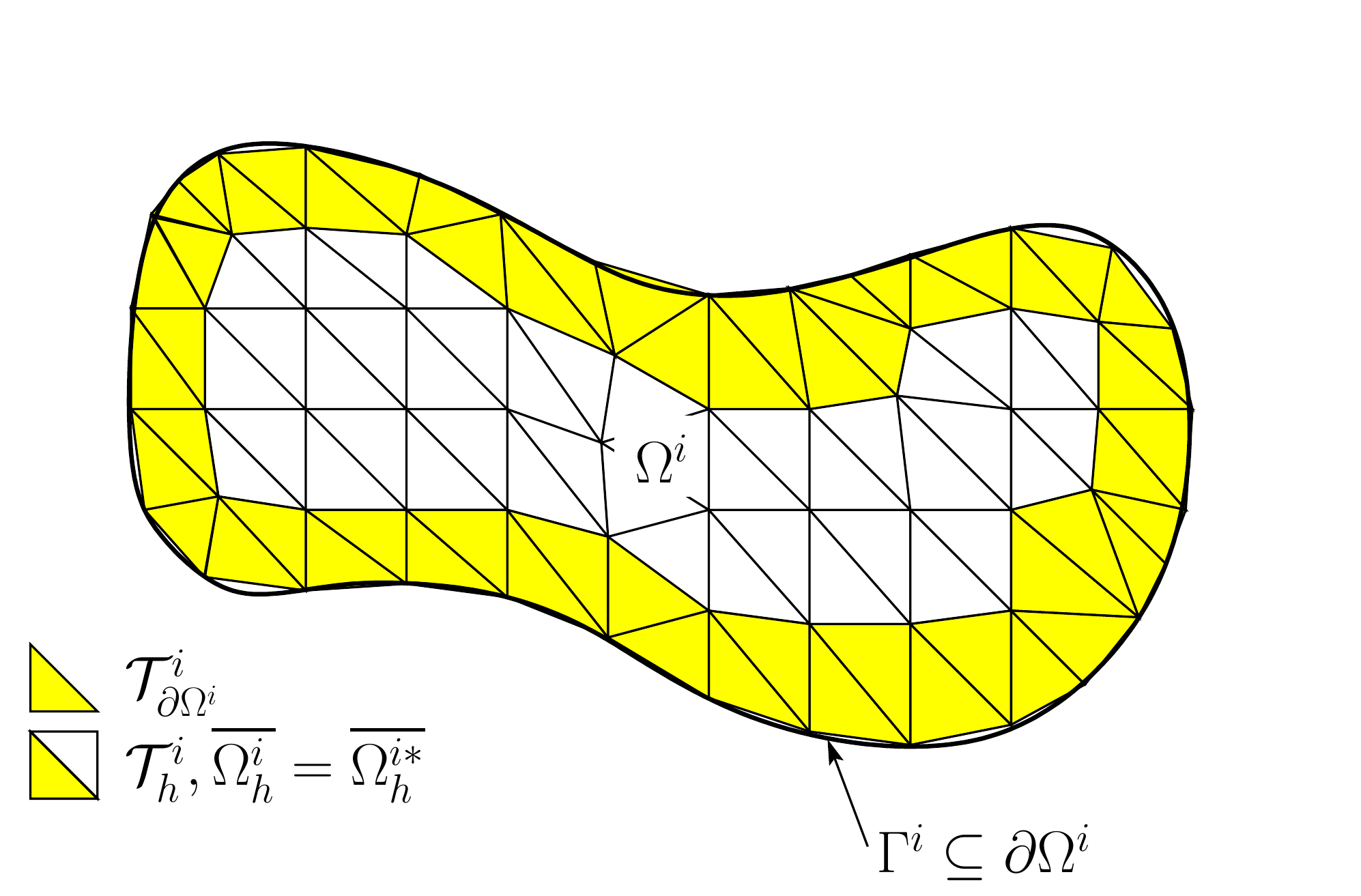} \qquad
    \includegraphics[width=0.47\textwidth]{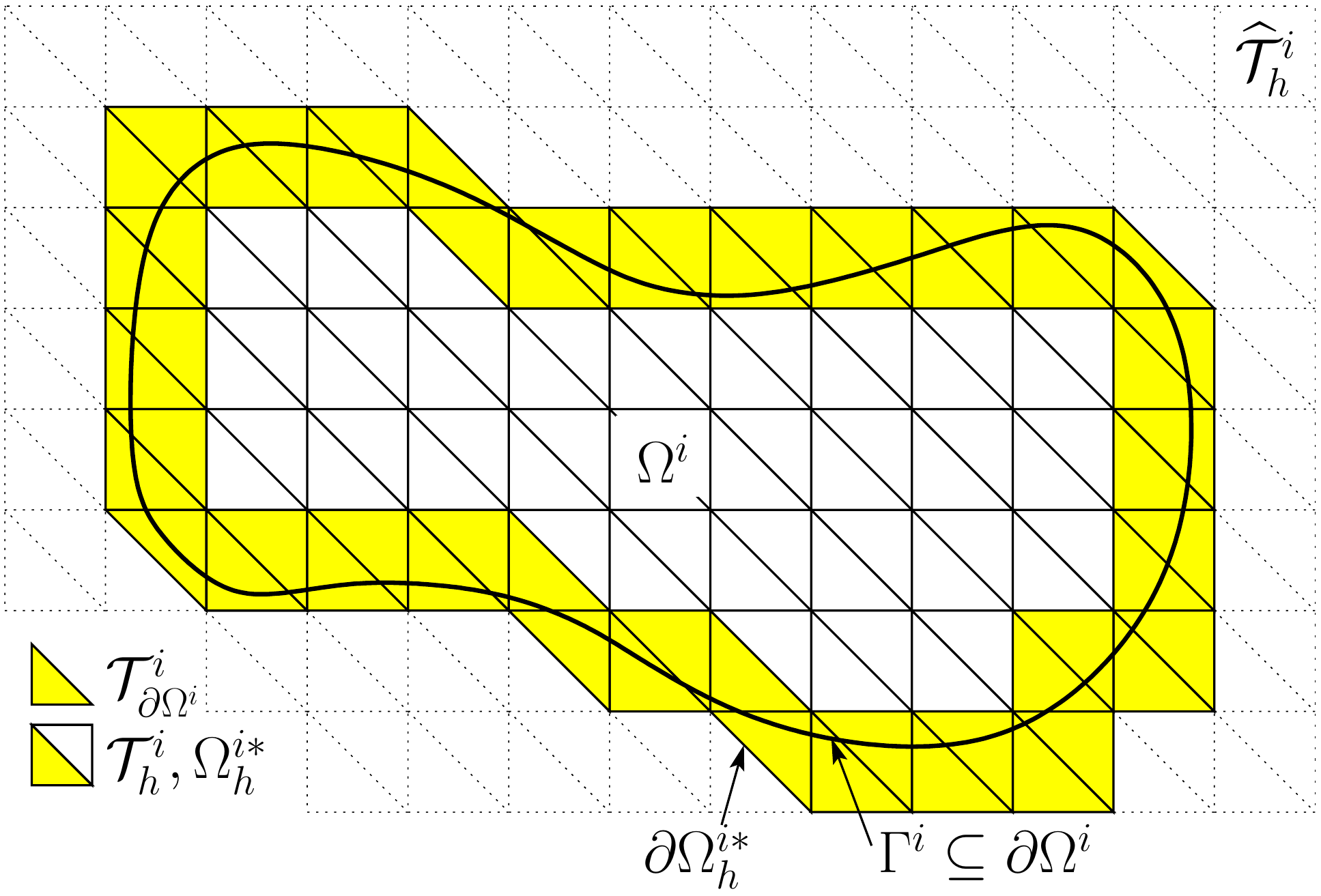}
  \end{center}
\vspace{-12pt}
  \caption{Boundary-fitted versus non-boundary-fitted meshes:
(Left) A computational grid~$\widehat{\mcT}_h^i$ is (geometrically) fitted to the boundary~$\partial\Domi$ of a physical domain~$\Domi$
and its active part~$\mcT_h^i$ fits to the domain such that \mbox{$\overline{\Domh^i} = \overline{\Domh^{i\ast}}$}.
(Right) The physical subdomain~$\Domi$ is defined as the
interior of a given boundary~$\partial\Domi$ embedded into an unfitted
background mesh~$\widehat{\mcT}_h^i$, the fictitious
domain~$\Dom_h^{i\ast}$ is the union of the minimal active subset~$\mcT_h^i
\subset \widehat{\mcT}_h^i$ covering~$\Domi$.
}
\label{fig:computational-domain}
\end{figure}

Let $\widehat{\mcT}_h^i$ be a background mesh with mesh size parameter \mbox{$h^i>0$} which covers an open and bounded physical subdomain~$\Domi$.
The \emph{active} part of a computational mesh is defined as
\begin{align}
  \label{eq:define-active-mesh}
  \mcT_h^i = \{ T \in \widehat{\mcT}_h^i : T \cap \Domi \neq \emptyset \}
\end{align}
and consists of all elements in $\widehat{\mcT}_h^i$ which intersect~$\Domi$.
Denoting the union of all elements \mbox{$T \in \mcT_h^i$} by $\Domh^{i\ast}$,
then $\mcT_h^i$ is called a \emph{(geometrically) fitted} or more precisely a \emph{boundary-fitted} mesh if
\mbox{$\overline{\Domh^i} = \overline{\Domh^{i\ast}}$}
and an \emph{unfitted} or \emph{non-boundary-fitted} mesh if
\mbox{$\overline{\Domh^i} \subsetneq \overline{\Domh^{i\ast}}$}.
To each mesh, the subset of elements that intersect the boundary~$\partial\Domi$
\begin{equation}
  \label{notation:p2:eq:define-cutting-cell-mesh}
  \mcT_{\partial\Dom}^i = \{T \in \mcT_h^i: T \cap \partial\Domi \neq \emptyset \}
\end{equation}
is associated, which plays a particular role for unfitted meshes.
Notation according to fitted and unfitted computational meshes is depicted in~\Figref{fig:computational-domain}.
For a given mesh $\mcT_h^i$, the space
\begin{equation}
\label{eq:isoparametric_continuous_function_space}
 \mcX_h^i(t) \eqdef \{ x_h \in C^0(\overline{\Dom_h^i}(t)): \restr{x_h}{T}={v}_{\hat{T}}\circ S_T^{-1}(t) \text{ with }{v}_{\hat{T}}\in \mathbb{V}^k(\hat{T})\, \foralls T \in \mcT_h^i \}
\end{equation}
denotes the finite element approximation space consisting of continuous piecewise
polynomials of order \mbox{$k\geqslant 1$} on tetrahedral, hexahedral or wedge-shaped elements,
where \mbox{$S_T(t)$} denote affine or isoparametric mappings from the respective element parameter spaces~$\bfxi\in\hat{T}$
to the spatial coordinates of~$T\subsetneq\Dom_h^i(t)$ at time~$t$.

For inter-element faces in~$\mcT_h^i$, let~$\mcF^i$ be the set of \emph{interior faces}~$F$ which are
shared by exactly two elements, denoted by $T^+_F$ and $T^-_F$.
Furthermore, introduce the notation~$\mcF_{\partial\Domi}^i$ for the set of all interior faces
belonging to elements intersected by the boundary~$\partial\Domi$,
\begin{equation}
  \label{eq:ghost-penalty-facets}
  \mcF^i_{\partial\Dom} = \{ F \in \mcF^i ~|~
  T^+_F \cap \partial\Domi \neq  \emptyset
  \vee
  T^-_F \cap \partial\Domi \neq  \emptyset
  \},
\end{equation}
which is empty for boundary-fitted approximations.

\begin{remark}
If a boundary~$\partial\Domi$ intersects elements \mbox{$T\in \mcT_{\partial\Domi}^i$} of a non-boundary-fitted mesh~$\mcT_h^i$, it
subdivides them into several arbitrary formed polyhedra.
Note that for consistency reasons the weak form integration associated to a subdomain takes place only
in the parts covered by the physical domain.
Mesh parts lying outside of the domain~$\Domi$ are not considered for bulk integrals, nevertheless,
play an important role in \name{CutFEM}s as will be elaborated in \Secref{ssec:stabilized_fitted_unfitted_FEM_fluids}.
For possible integration techniques, see, \eg, \cite{SudhakarWall2013,SudhakarMoitinhoWall2014, Sudhakar2017}.
\end{remark}

\begin{remark}
Note that in \name{CutFEM}s, geometric entities like active element or facet sets and associated function spaces are usually time-dependent
and may be regularly updated for instationary moving domain problems. Algorithmic issues regarding this topic will be addressed in detail
in \Secref{sec:fsi:monolithic}.
\end{remark}

\subsection{Limitations and Capabilities of Fitted and Unfitted Discretization Techniques}

\paragraph*{Single-phase boundary value problems -- fitted or unfitted meshes?}

Classical boundary-fitted methods, as visualized in the left part of \Figref{fig:computational-domain}, are the state-of-the-art in finite element based
approximations of single-phase settings. Among many other advantages, their most important features can be summarized as follows:
Fitting mesh techniques allow to easily obtain higher-order geometric approximations using isoparametric concepts,
enable the straightforward incorporation of essential boundary conditions
into the discrete function space and come along with well-established stability and best-approximation properties
for a variety of partial differential equations modeling continuum mechanics.
Such schemes are undisputed the method of choice for the approximation of classical solid mechanics.
Nevertheless, for complex three-dimensional domains, generating high quality computational grids that
conform to the domain boundary can often be time-consuming and difficult.
This becomes particularly cumbersome and costly for geometries
given, for instance, by \name{CAD} data from industrial applications or by image data from, \eg, geological, biological or medical sources.

In contrast, utilizing geometrically unfitted finite element methods that compute the solution to
the problem on the active part~$\mcT_h$ of mostly fixed grids, which are intersected by the boundary, can drastically
simplify meshing of the computational domain.
Such approximations have been visualized in the right of \Figref{fig:computational-domain}.
In contrast to boundary-fitted meshes,
special measures are required to impose the different types of essential and natural boundary conditions in a consistent weak sense
due to the unfittedness.
In particular ensuring robustness, stability and accuracy of the resulting numerical scheme
becomes more challenging when intersecting elements.
As one of the major issues,
small parts of intersected elements are often not sufficiently controlled depending on the interface location,
an aspect which will be elaborated in more detail for fluids in~\Secref{ssec:stabilized_fitted_unfitted_FEM_fluids}.
\begin{figure}
  \centering
\includegraphics[width=0.31\textwidth]{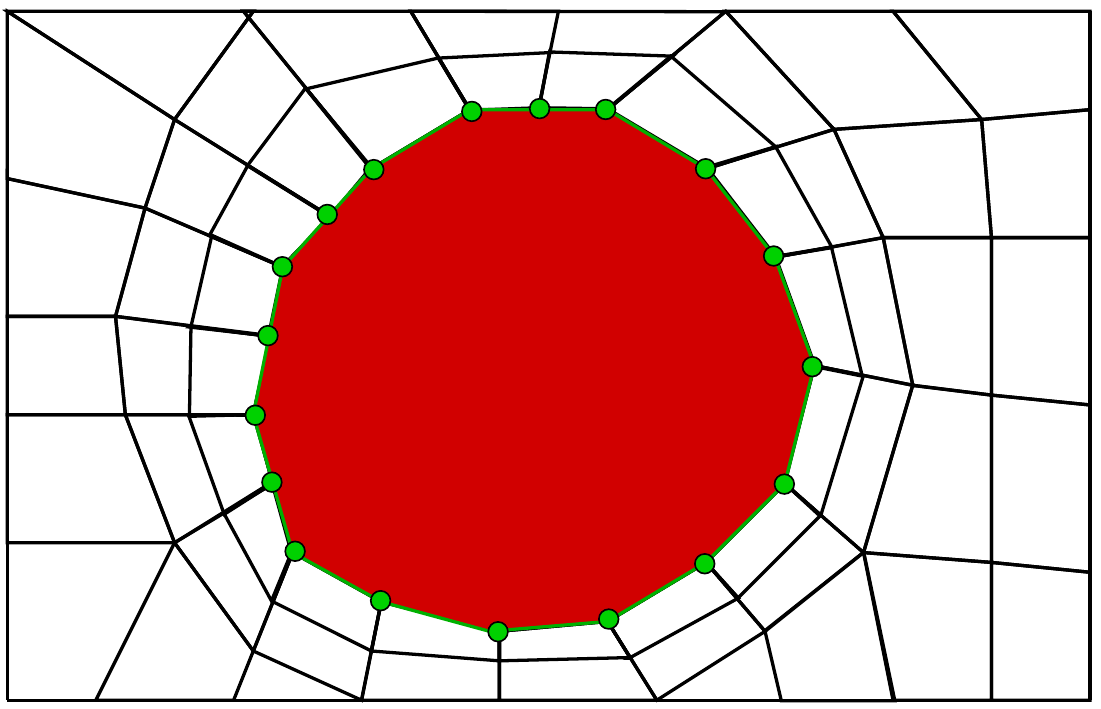}
\quad
\includegraphics[width=0.31\textwidth]{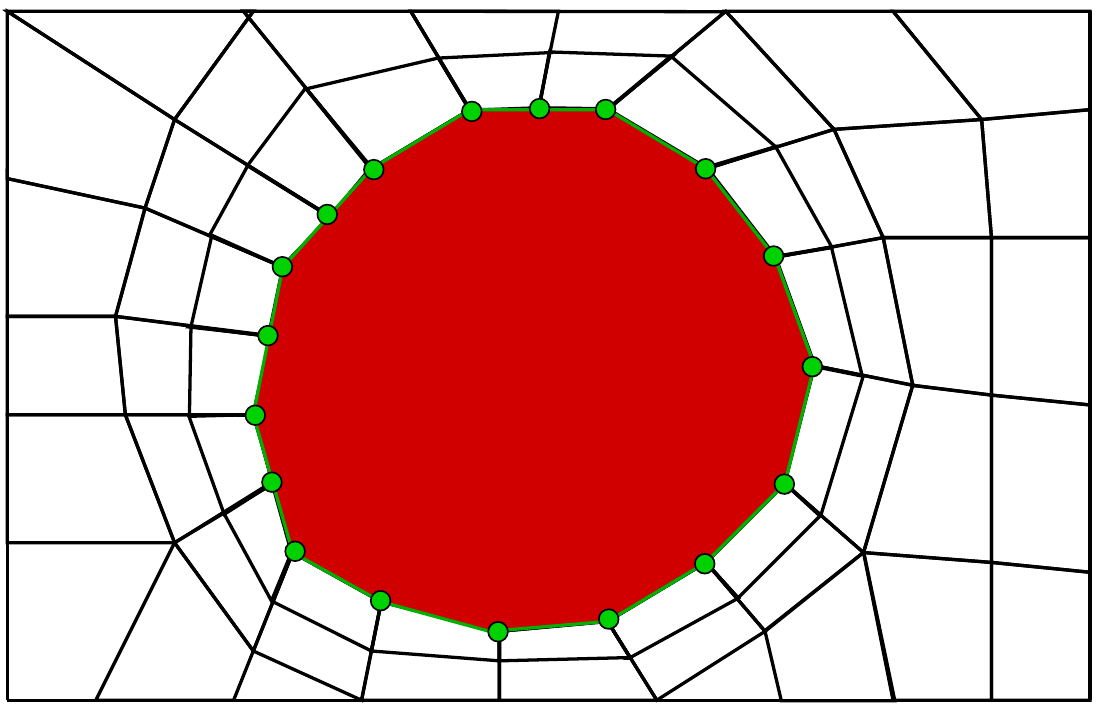}
\quad
\includegraphics[width=0.31\textwidth]{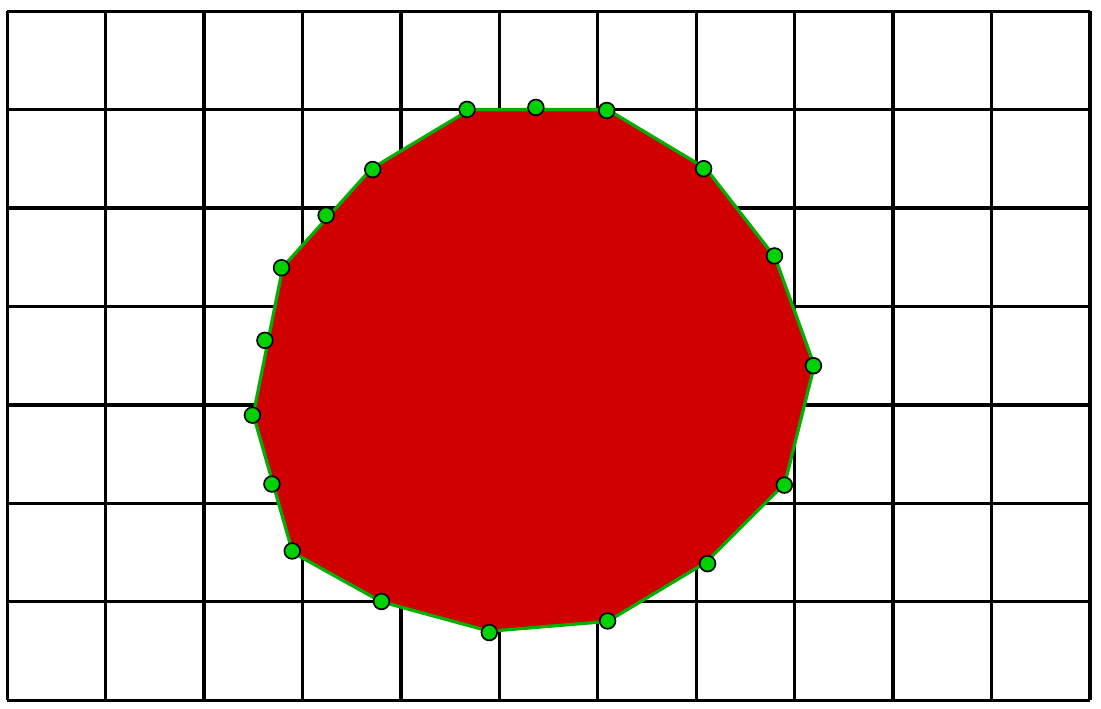}
\\
\includegraphics[width=0.31\textwidth]{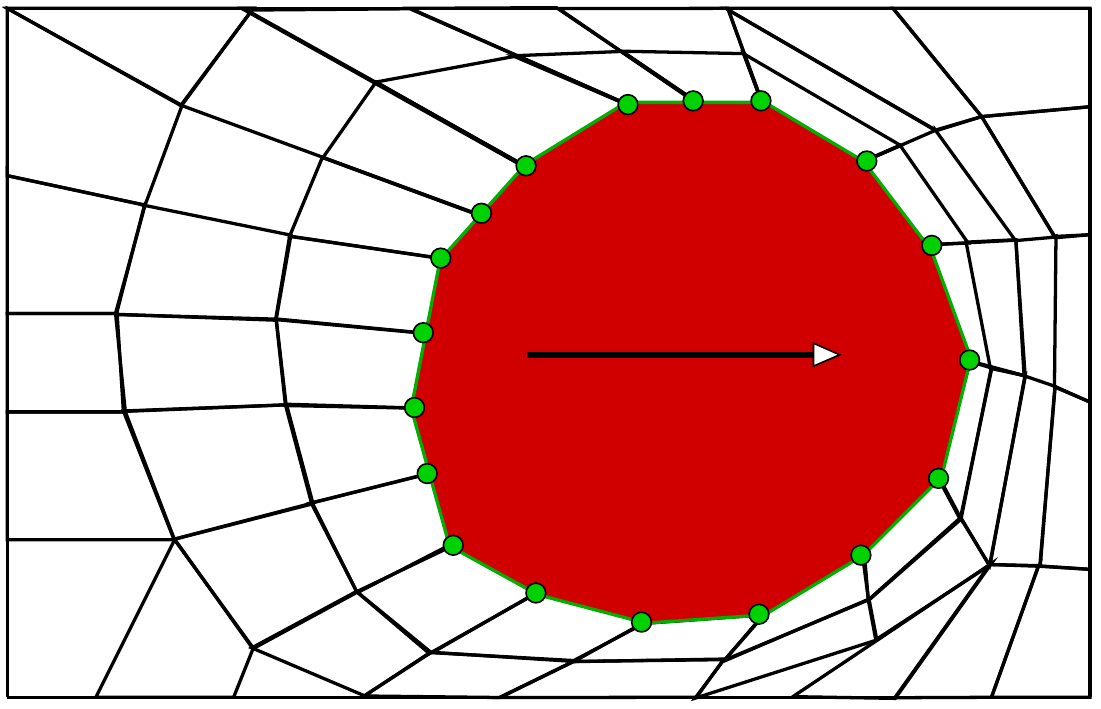}
\quad
\includegraphics[width=0.31\textwidth]{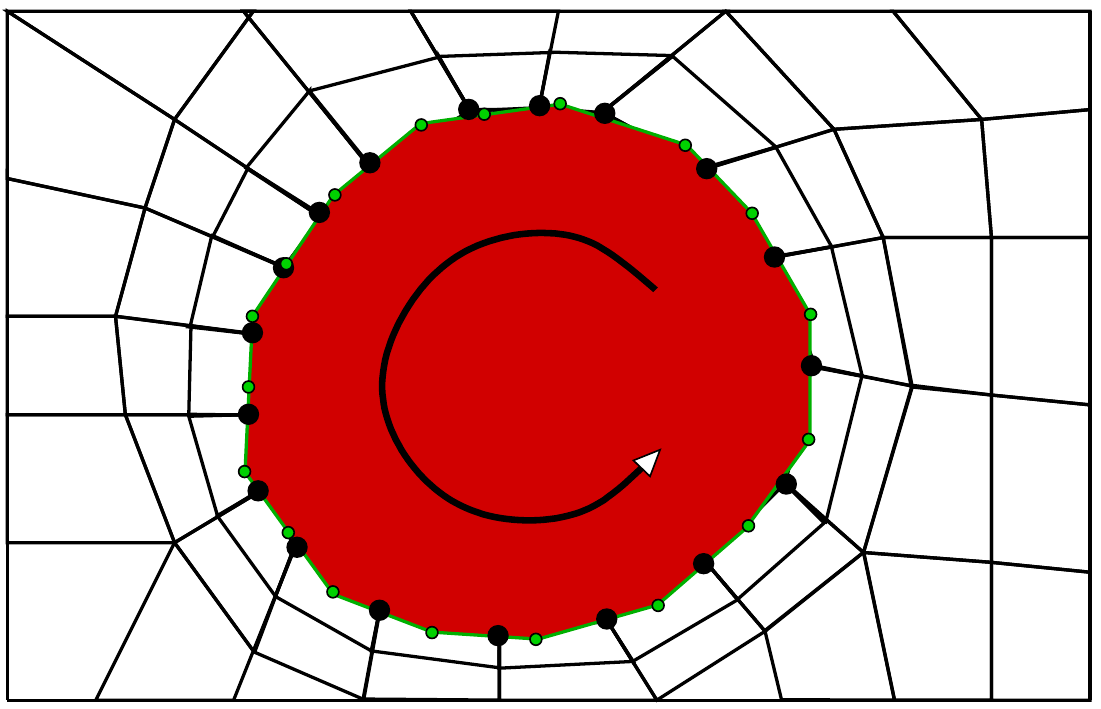}
\quad
\includegraphics[width=0.31\textwidth]{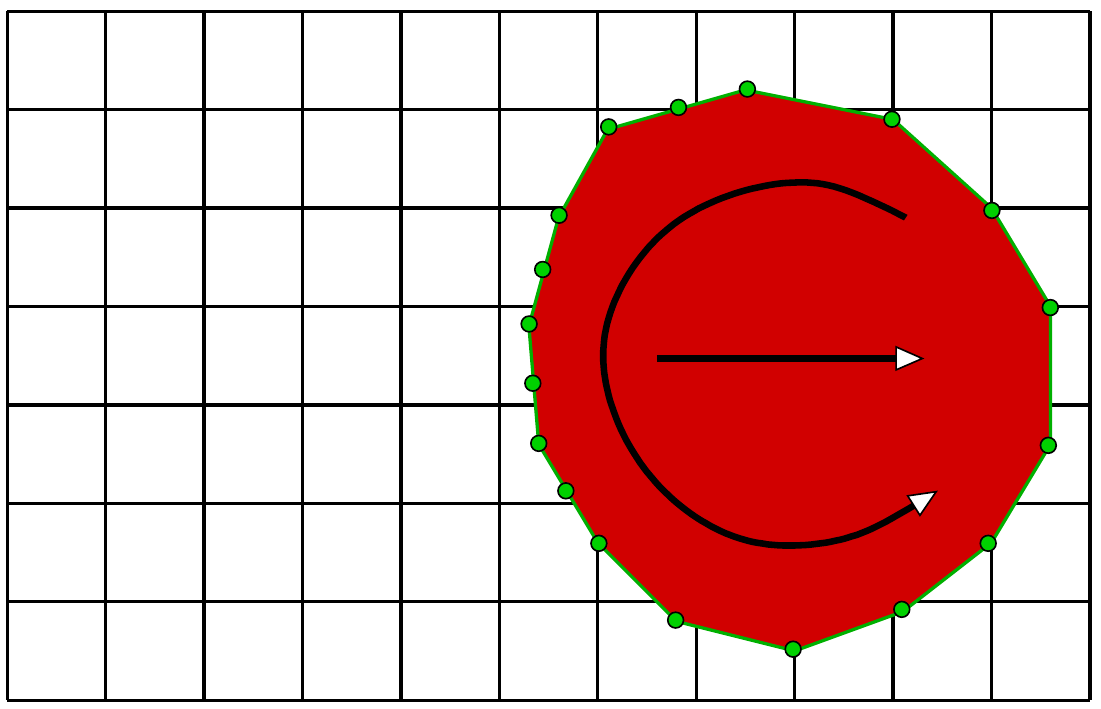}
  \caption{Multiphysics settings with moving domains:
(Left)~\emph{Node-matching moving meshes} are subjected to strict limitations regarding interface motion and deformation, otherwise the meshes can drastically deteriorate.
(Middle)~\emph{Interface-fitted non-node-matching meshes} enable mesh motion in interface-tangential directions, but
can fail for large mesh movement in interface-normal directions.
(Right)~\emph{Unfitted non-matching meshes} allow for arbitrary motion of the meshes relative to each other.
Green markers indicate boundary nodes of the inner structural mesh, black markers denote interface nodes of the outer fluid mesh.}
  \label{fig:computational-mesh-fitted-matching-nonmatching-moving-domains}
  \vspace{-12pt}
\end{figure}

The capabilities and limitations of interface-fitted approximation techniques are almost obvious when considering transient
coupled multiphysics applications, like for example fluid-structure interaction.

\paragraph*{Moving multidomain problems -- classical node-matching meshes.}

Let us assume a split of the entire domain~$\Dom$ into two disjoint subdomains.
The use of a single \emph{node-matching} finite element function space for the entire domain~$\Dom$
provides one of the most easiest and feasible finite element discretization methods.
Its simplicity with regard to stability and continuity requirements, which can be either incorporated directly to the function space
or accounted for in a weak sense, makes this technique attractive.
However, it comes along with very restrictive limitations for transient problems where moving domains are involved,
as sketched in~\Figref{fig:computational-mesh-fitted-matching-nonmatching-moving-domains}~(left).
These limitations become particularly painful for demanding fluid-structure interaction problem configurations,
where structural bodies undergo large deformations, rotations or even contact in a surrounding flow.
As the finite elements need to follow the interface in its evolution, the meshes can rapidly deteriorate
and time consuming remeshing and projection steps have to be performed regularly.

\paragraph*{Moving multidomain problems -- sliding interface-fitted meshes.}

Relaxation of the strong node-matching constraint, as visualized in~\Figref{fig:computational-mesh-fitted-matching-nonmatching-moving-domains}~(middle),
allows for independent mesh movements at least in interface-tangential directions.
While the subdomains are still approximated in an interface-fitted way, the meshes are not node-matching anymore and are thus allowed to slide.
Such techniques have been provided in the context of FSI by, \eg, \cite{Kloppel2011}.
They enable large structural rotations and thus widens the range of potential multifield problem settings by a multiple.
It needs to be pointed out that this non-match of the involved meshes put further demands on the weak constraint enforcement
and require additional geometric operations at the common interface \cite{Popp2010}.
Another interesting variant of sliding interface-fitted meshes is the so called shear-slip mesh update method introduced in \cite{Behr1999} that reconnects nodes in the element layer next to the interface.
Nevertheless, such sliding mesh techniques still show severe limitations for large motions in interface-normal direction,
a phenomenon which is almost omnipresent in demanding FSI problem settings.

\paragraph*{Moving multidomain problems -- advanced use of unfitted meshes.}

To overcome the shortcomings of interface-fitted meshes with regards to domain motions and deformations,
in the following the benefit of using unfitted mesh techniques is discussed.

As introduced in \Secref{sec:dis_cut_fem:motivation_concepts:overview_domain_decomposition:notation_terminology}
and indicated in~\Figref{fig:computational-mesh-fitted-matching-nonmatching-moving-domains}~(right),
a powerful strategy to reduce restrictions on the interface location and its temporal movement
consists in choosing computational meshes independent of the geometries to be approximated.
The fundamental idea consists in constructing finite element based computational grids,
which in the end are used to define the discrete function spaces and to represent the approximative solutions,
independent of the location of the separating interface.
Interface evolution and potential mesh motions can then be treated fully decoupled.
Considering a two-field problem, as a result of this desired independence, for time-dependent subdomains the active mesh parts \mbox{$\mcT^1_h, \mcT^2_h$}~\Eqref{eq:define-active-mesh}
will change over time.
As introduced in~\Eqref{notation:p2:eq:define-cutting-cell-mesh}, respective sets of elements intersected by the interface are given by
\mbox{$\mcT^1_{\Int} = \{T \in \mcT^1_h: T \cap \Int \neq \emptyset \}$} and \mbox{$\mcT^2_{\Int} = \{T \in \mcT^2_h: T \cap \Int \neq \emptyset \}$}.
Related fictitious subdomains \mbox{$\Dom_h^{1\ast}, \Dom_h^{2\ast}$} are defined accordingly.

It needs to be highlighted that treating the evolution of the interface independent from the potential motions of the computational grids
greatly widens the range of feasible interface-coupled multiphysics problems
and thus opens a variety of novel discretization concepts.
Some highly beneficial geometrically unfitted mesh approaches towards FSI will be discussed in more detail subsequently.

\subsection{Powerful Fitted and Unfitted Discretization Concepts for Fluid-Structure Interaction}
\label{ssec:powerful_fitted_and_unfitted_discretization_concepts}
For the numerical simulation of mutual interactions between fluids and solids,
different competitive approaches will be discussed in the following.
While for structures a Lagrangean description approximated with a boundary-fitted mesh is the most reasonable choice,
for flows, different mesh techniques are competing.
A temporally changing fluid domain can be approximated by means of
\begin{itemize}
\setlength{\itemindent}{1.5cm}
\item[\textbf{Approach~\ref{approach:1_ALE-FSI}}]\textbf{\textsc{fitted-ALE:}} a fitted, deforming mesh that follows the fluid-solid (FS) interface,
\item[\textbf{Approach~\ref{approach:2_XFSI}}]\textbf{\textsc{unfitted-Euler:}} an unfitted fluid mesh that remains fixed over time,
\item[\textbf{Approach~\ref{approach:3_ALE-XFSI}}]\textbf{\textsc{unfitted-ALE:}} an unfitted fluid mesh that is allowed to deform independent of the FS interface evolution,
\item[\textbf{Approach~\ref{approach:4_FXFSI}}]\textbf{\textsc{unfitted-Euler/embedded-unfitted-Euler:}} a fluid domain decomposition technique including several unfitted overlapping fluid meshes that remain fixed in time,
\item[\textbf{Approach~\ref{approach:5_XFFSI}}]\textbf{\textsc{unfitted-Euler/embedded-fitted-ALE}} or \textbf{\textsc{hybrid Eulerian-ALE:}} a fluid domain decomposition technique consisting of an FS-interface-fitted deforming fluid patch, which is embedded into an unfitted fluid background mesh.
\end{itemize}
Note that Approach~\ref{approach:3_ALE-XFSI} provides an extension to Approach~\ref{approach:2_XFSI} and allows the unfitted background mesh to move independently.
As this requires further essential algorithmic advancements, as will be elaborated in \Secref{sec:fsi:monolithic},
they are considered as two distinct approaches.
When choosing a domain decomposition technique for approximating the fluid domain, in principle, any combination and number of fitted and unfitted meshes
can be used, as long as they cover the entire fluid domain. In general, all involved meshes can be moving or fixed, deforming or non-deforming.
To discuss the challenges and to present principle algorithms for fluid domain decomposition in a clear fashion,
two beneficial approaches have been picked out of this class, designated as
Approach~\ref{approach:4_FXFSI} and Approach~\ref{approach:5_XFFSI}.
In the following, Approaches~\ref{approach:1_ALE-FSI}--\ref{approach:5_XFFSI} are discussed. 

\begin{approach}[\textbf{\textsc{fitted-ALE:}} fitted, moving fluid mesh technique]
\label{approach:1_ALE-FSI}
The structural mesh~$\mcT_h^\sd$ and the fluid-mesh~$\mcT_h^\fd$ are both fitted to the common FS interface, \ie,
\begin{align}
\label{eq:computational-mesh-fitted-matching-fsi-ALE-FSI}
\overline{\Doms_h} &= \overline{\Dom_h^{\sd\ast}}, \qquad
\overline{\Domf_h} = \overline{\Dom_h^{\fd\ast}}, \qquad
\Dom_h^{\sd\ast} \cap \Dom_h^{\fd\ast} = \emptyset,
\end{align}
as visualized in \Figref{fig:computational-mesh-fitted-matching-fsi-ALE}.

This classical fitted mesh technique, see \eg~\cite{Kuttler2008, Fernandez2011, Mayr2015}, which utilizes an ALE-based formalism for the flow description, enables accurate couplings between fluid and solid phase due to the node-match
of the involved meshes at the \name{FS} interface.
Weak constraint enforcement techniques for coupling the solution fields are thereby often superior over simpler strong impositions for
high-Reynolds-number flows, as discussed in, \eg, \cite{Bazilevs2007,Gamnitzer2012}.
Since the fluid mesh has to deform and follow the structural body over time to maintain the interface conformity, solids are limited to small motions and rotations,
otherwise the fluid mesh will rapidly distort.
Main difficulties arising from this technique consists in providing high quality meshes with sufficient mesh resolution in the vicinity of the interface as these small elements next to the moving interface are particularly vulnerable.
More advanced FSI problems, like those including contact of submersed solids, cannot be supported by this technique.
\begin{figure}[h!]
\centering
\includegraphics[width=0.47\textwidth]{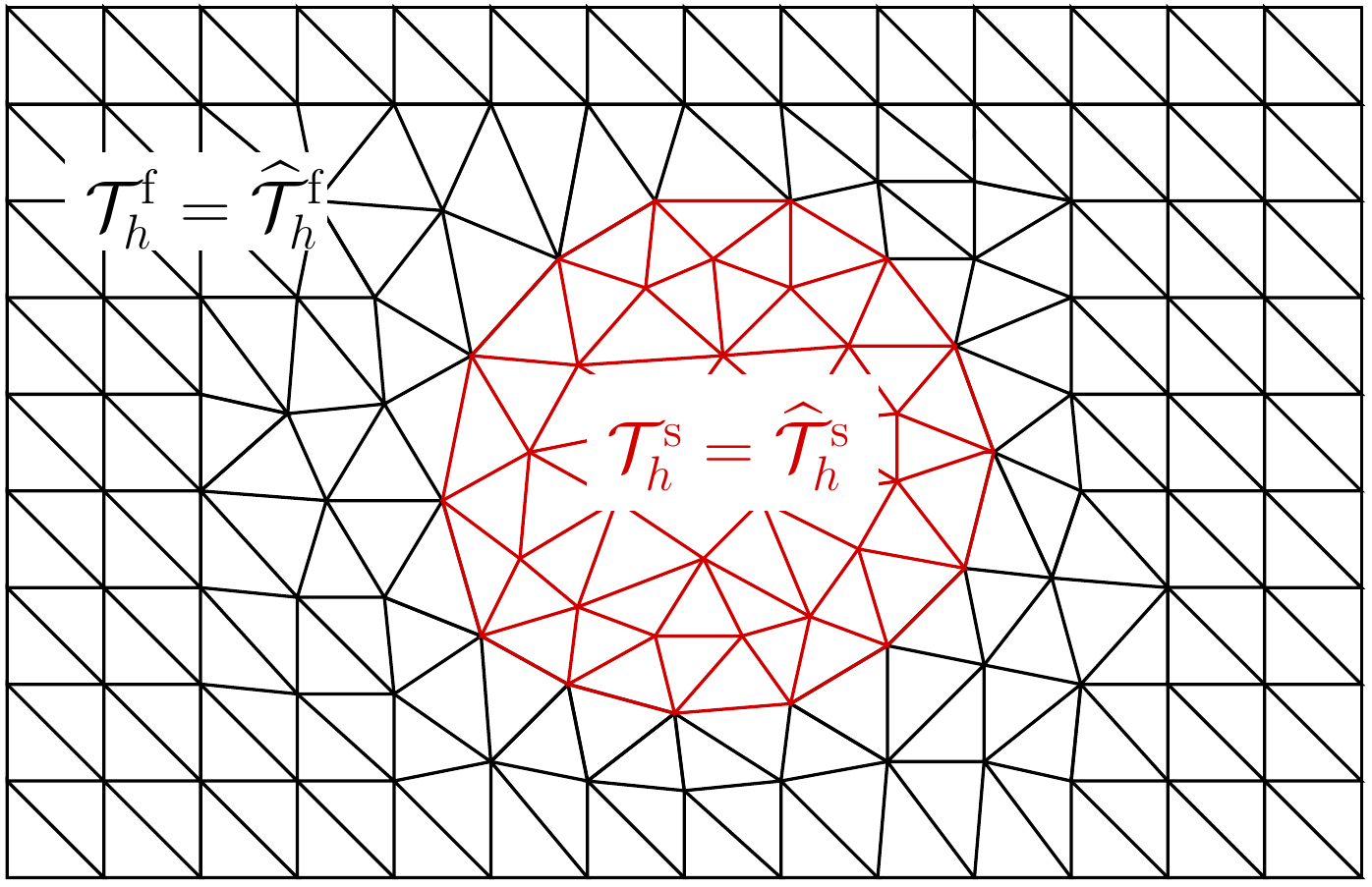}
\includegraphics[width=0.47\textwidth]{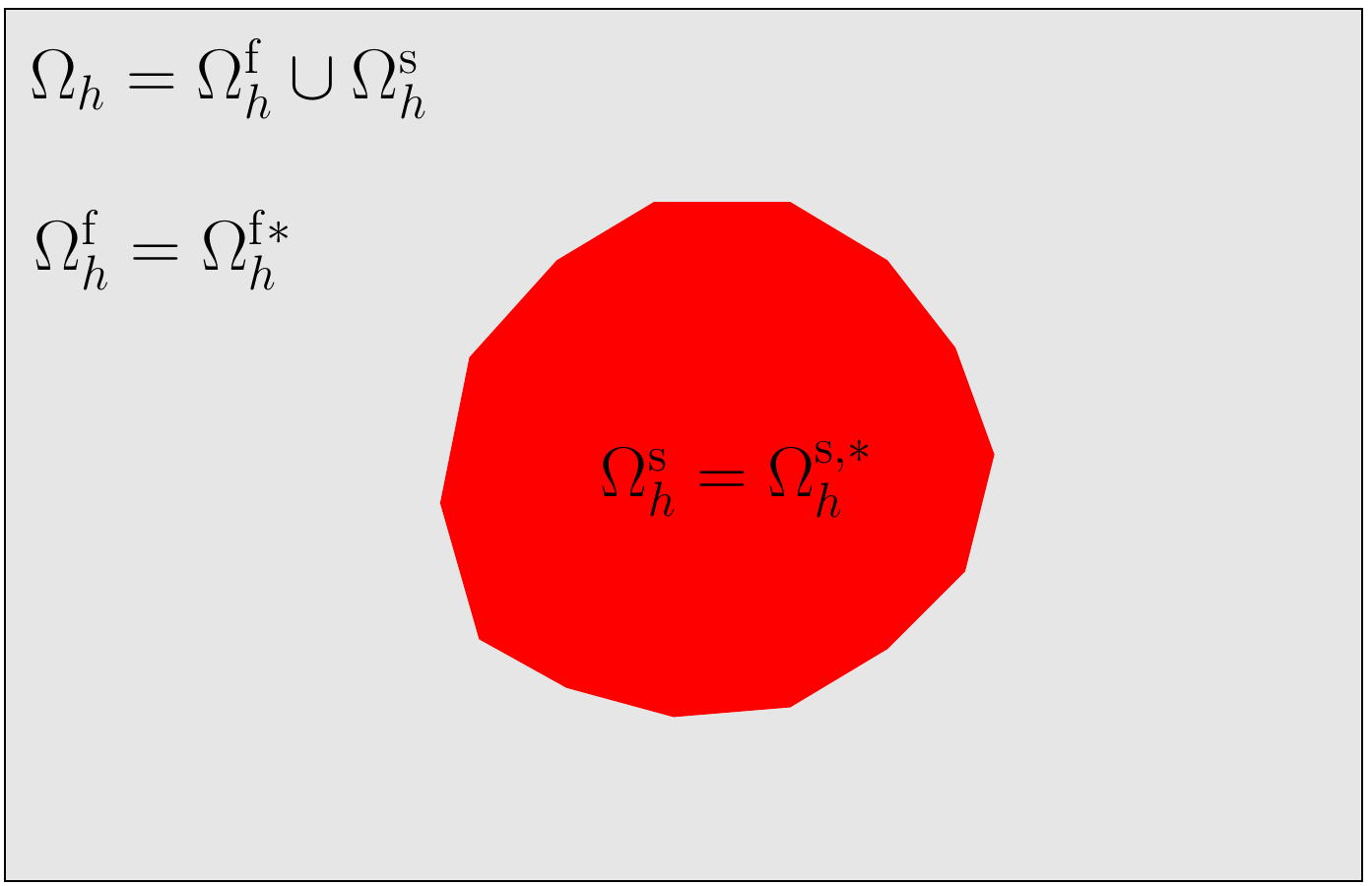}
\caption{Approach~\ref{approach:1_ALE-FSI}:
Computational meshes (left) and domain partition (right) for the FSI problem.}
\label{fig:computational-mesh-fitted-matching-fsi-ALE}
  \vspace{-12pt}
\end{figure}
\end{approach}

\begin{approach}[\textbf{\textsc{unfitted-Euler:}} unfitted, non-moving fluid mesh technique]
\label{approach:2_XFSI}
An FS-interface-fitted structural mesh~$\mcT_h^\sd$ for the solid overlaps with
a non-FS-interface-fitted background fluid mesh~$\mcT_h^\fd$, \ie,
\begin{align}
\label{eq:computational-mesh-unfitted-nonmatching-fsi-xfsi}
\overline{\Doms_h} &= \overline{\Dom_h^{\sd\ast}}, \qquad
\overline{\Domf_h} \subsetneq \overline{\Dom_h^{\fd\ast}}, \qquad
\Dom_h^{\sd\ast} \cap \Dom_h^{\fd\ast} \neq \emptyset,
\end{align}
as visualized in \Figref{fig:computational-mesh-unfitted-nonmatching-fsi-xfsi}.

Such an unfitted mesh technique has been considered in, \eg, \cite{Gerstenberger2008,Alauzet2016, FernandezLandajuela2016, Burman2014a}
and enables arbitrary solid motions within and interactions with the surrounding fluid.
In contrast to Approach~\ref{approach:1_ALE-FSI}, the fluid mesh is assumed to be fixed over time and thus is not subjected to any distortion
and it does not necessarily need to account for mesh motion by an ALE technique.
Furthermore, to simplify meshing next to outer boundaries,
fluid domain boundaries can be simply embedded in an unfitted way, as done, \eg, in the works \cite{SchottWall2014, MassingSchottWall2016_CMAME_Arxiv_submit}.
As a great benefit, this technique even allows for additional interactions between several structural bodies in a surrounding fluid.
Moreover, potential contact of submersed structures, as addressed by algorithms developed in \cite{Mayer2010, Ager2018},
can be supported in general, even though the presentation of the according more sophisticated algorithms go much beyond the scope of this work.

\begin{figure}[ht!]
\centering
\includegraphics[width=0.47\textwidth]{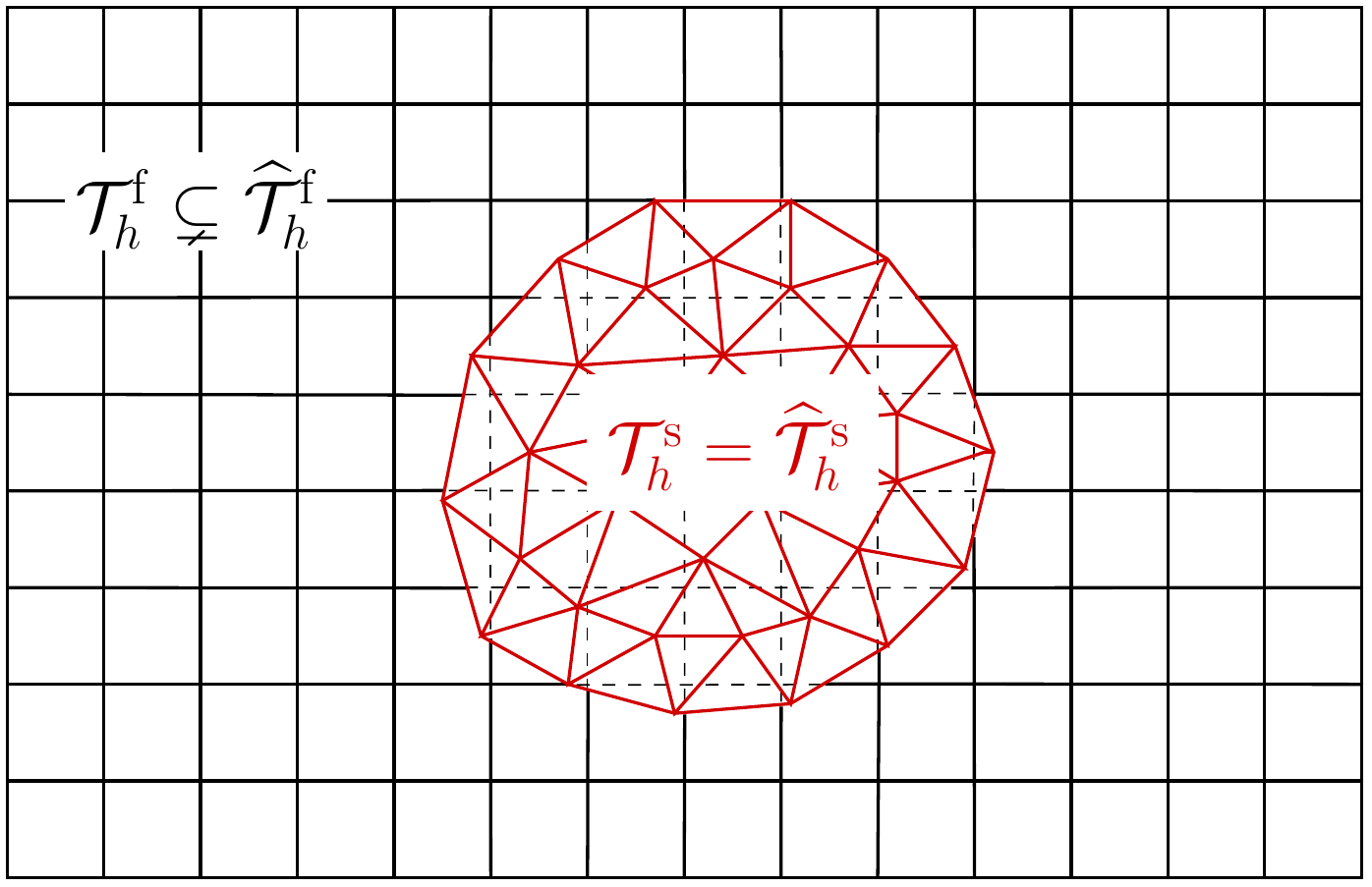}
\includegraphics[width=0.47\textwidth]{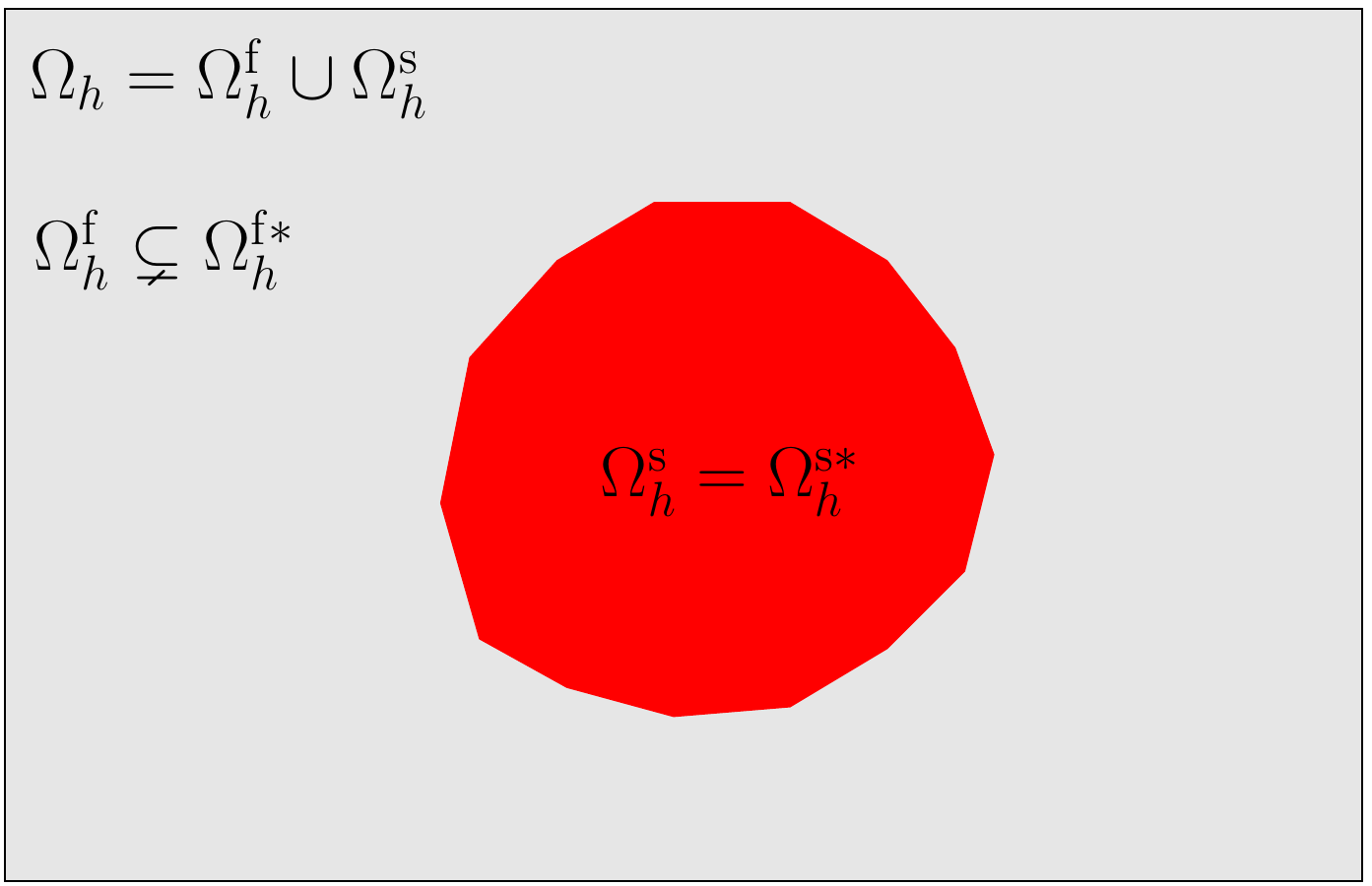}
\caption{Approach~\ref{approach:2_XFSI}:
Computational meshes (left) and domain partition (right) for the FSI problem.}
\label{fig:computational-mesh-unfitted-nonmatching-fsi-xfsi}
  \vspace{-12pt}
\end{figure}
\end{approach}

\begin{approach}[\textbf{\textsc{unfitted-ALE:}} unfitted, moving fluid mesh technique]
\label{approach:3_ALE-XFSI}
This approach provides a combination of the fitted, moving mesh Approach~\ref{approach:1_ALE-FSI}
and the unfitted technique described as Approach~\ref{approach:2_XFSI}.
To allow the unfitted background fluid mesh to move might be advantageous when several structural bodies interact with a fluid phase.
In particular for a solid body ($\Dom^{\sd_1}$) that undergoes only small deformations, the fluid mesh can be constructed being interface-fitted at $\Gamma^{\sd_1 \fd}$,
whereas for another body ($\Dom^{\sd_2}$), which might largely move, the fluid mesh needs to be chosen unfitted at $\Gamma^{\sd_2 \fd}$, resulting in
\begin{align}
\label{eq:computational-mesh-unfitted-nonmatching-fsi-alefsi}
\overline{\Dom_h^{\sd_1}} = \overline{\Dom_h^{\sd_1\ast}}, \qquad
\overline{\Dom_h^{\sd_2}} = \overline{\Dom_h^{\sd_2\ast}}, \qquad
\overline{\Domf_h} \subsetneq \overline{\Dom_h^{\fd\ast}}, \qquad
\Dom_h^{\sd_1\ast} \cap \Dom_h^{\fd\ast} = \emptyset, \qquad
\Dom_h^{\sd_2\ast} \cap \Dom_h^{\fd\ast} \neq \emptyset,
\end{align}
as visualized in \Figref{fig:computational-mesh-unfitted-nonmatching-fsi-ale-xfsi}.
Allowing for a more accurate node-matching, interface-fitted approximation at $\Gamma^{\sd_1 \fd}$ can make such a technique
superior to applying a pure unfitted scheme as proposed in Approach~\ref{approach:2_XFSI}.
This, however, is at the expense of providing a fitted fluid mesh at $\Gamma^{\sd_1 \fd}$
and to provide algorithms that allow to combine ALE moving mesh techniques with unfitted approximations in the fluid field.

\begin{figure}[h!]
\centering
\includegraphics[width=0.47\textwidth]{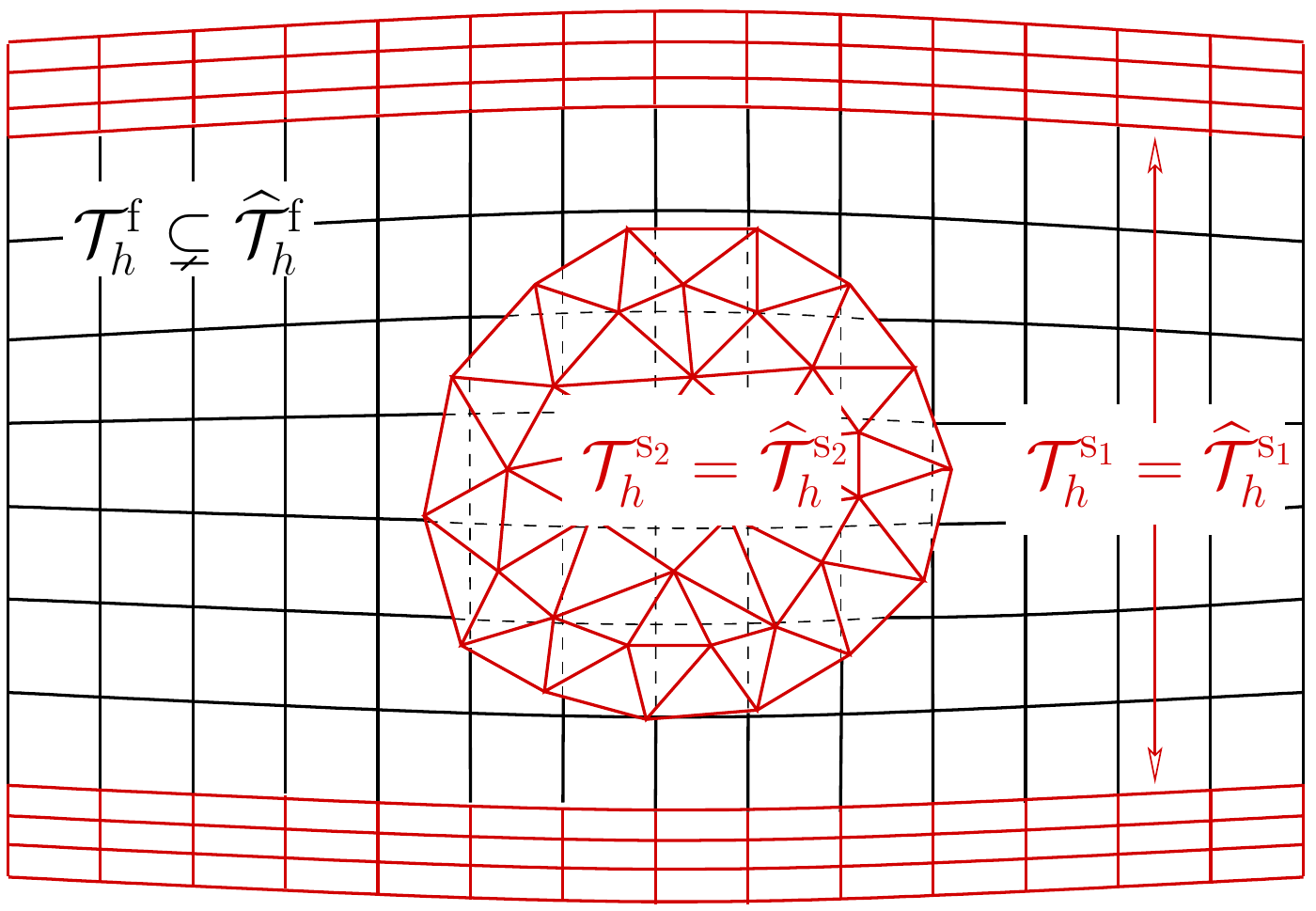}
\includegraphics[width=0.47\textwidth]{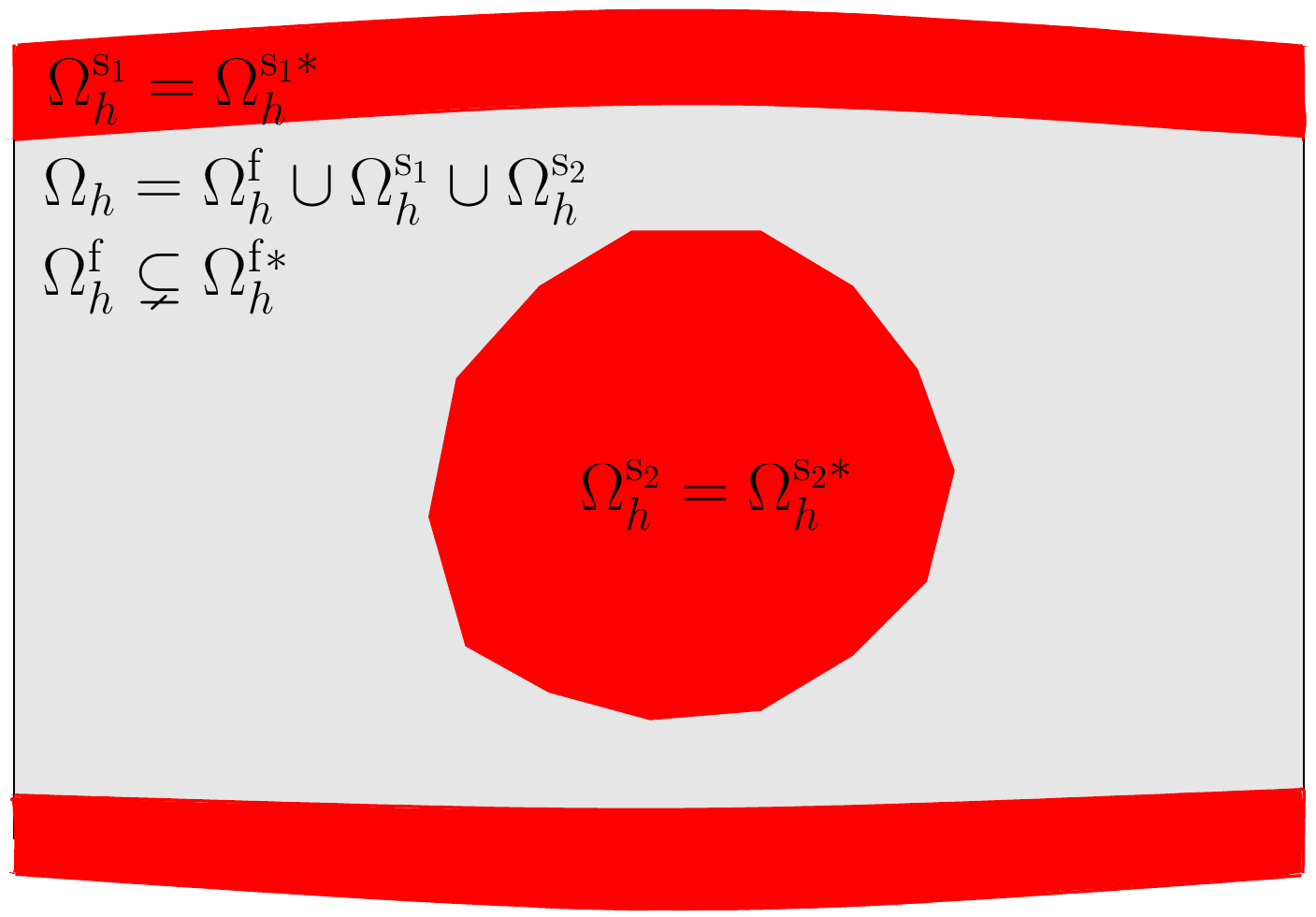}
\caption{Approach~\ref{approach:3_ALE-XFSI}:
Computational meshes (left) and domain partition (right) for the FSI problem.
}
  \vspace{-12pt}
  \label{fig:computational-mesh-unfitted-nonmatching-fsi-ale-xfsi}
\end{figure}
\end{approach}

\begin{approach}[\textbf{\textsc{unfitted-Euler/embedded-unfitted-Euler:}} unfitted, non-moving fluid mesh technique with unfitted, non-moving fluid patch]
\label{approach:4_FXFSI}
As in the unfitted Approach~\ref{approach:2_XFSI} uniform regular background fluid meshes are preferred
to fully exploit the advantages of simplified mesh generation or to allow for moving domains.
As described previously for regular meshes, either the mesh resolution can get very poor in the interface region when using coarse fluid meshes,
or the computational costs can enormously increase when using fine resolving meshes even in the far-field.
However, if \apriori~information about the potential movement of the structure is available,
all capabilities of an unfitted fluid-structure coupling can be conflated
with that of an unfitted overlapping fluid domain decomposition technique, as depicted in \Figref{fig:computational-mesh-unfitted-nonmatching-fsi-fxfsi}.
This still allows the structure to largely deform, unlike as for Approach~\ref{approach:1_ALE-FSI},
but is able to gain efficiency by means of using differently resolved fluid meshes.

For such a purpose, in a first construction step, the fitted solid mesh~$\mcT^\sd_h$ overlaps with a high resolution background mesh~$\mcT_h^{\fd_2}$.
This is usually defined within a region of expected structural motion.
In a second step, this fluid mesh is embedded into a coarser far-field background fluid mesh~$\mcT_h^{\fd_1}$ using unfitted
overlapping mesh techniques. The domain and mesh configurations can be summarized as
\begin{align}
\label{eq:computational-mesh-unfitted-nonmatching-fsi-fxfsi}
\overline{\Domh^{\fd_1}} \subsetneq \overline{\Dom_h^{\fd_1\ast}}, \qquad
\overline{\Domh^{\fd_2}} \subsetneq \overline{\Dom_h^{\fd_2\ast}}, \qquad
\overline{\Domh^\sd} = \overline{\Dom_h^{\sd\ast}}, \qquad
\Dom_h^{\fd_1\ast} \cap \Dom_h^{\fd_2 \ast} \neq \emptyset \qquad
\Dom_h^{\fd_2 \ast} \cap \Dom_h^{\sd\ast} \neq \emptyset.
\end{align}
Note that limitations newly introduced by incorporating a priori knowledge about the expected structural motion
can be overcome by enabling moving mesh techniques to allow the inner embedded fluid patch to follow the structure in its motion,
for instance in an approximate sense via tracking the center of gravity.

\begin{figure}[h!]
\centering
\includegraphics[width=0.47\textwidth]{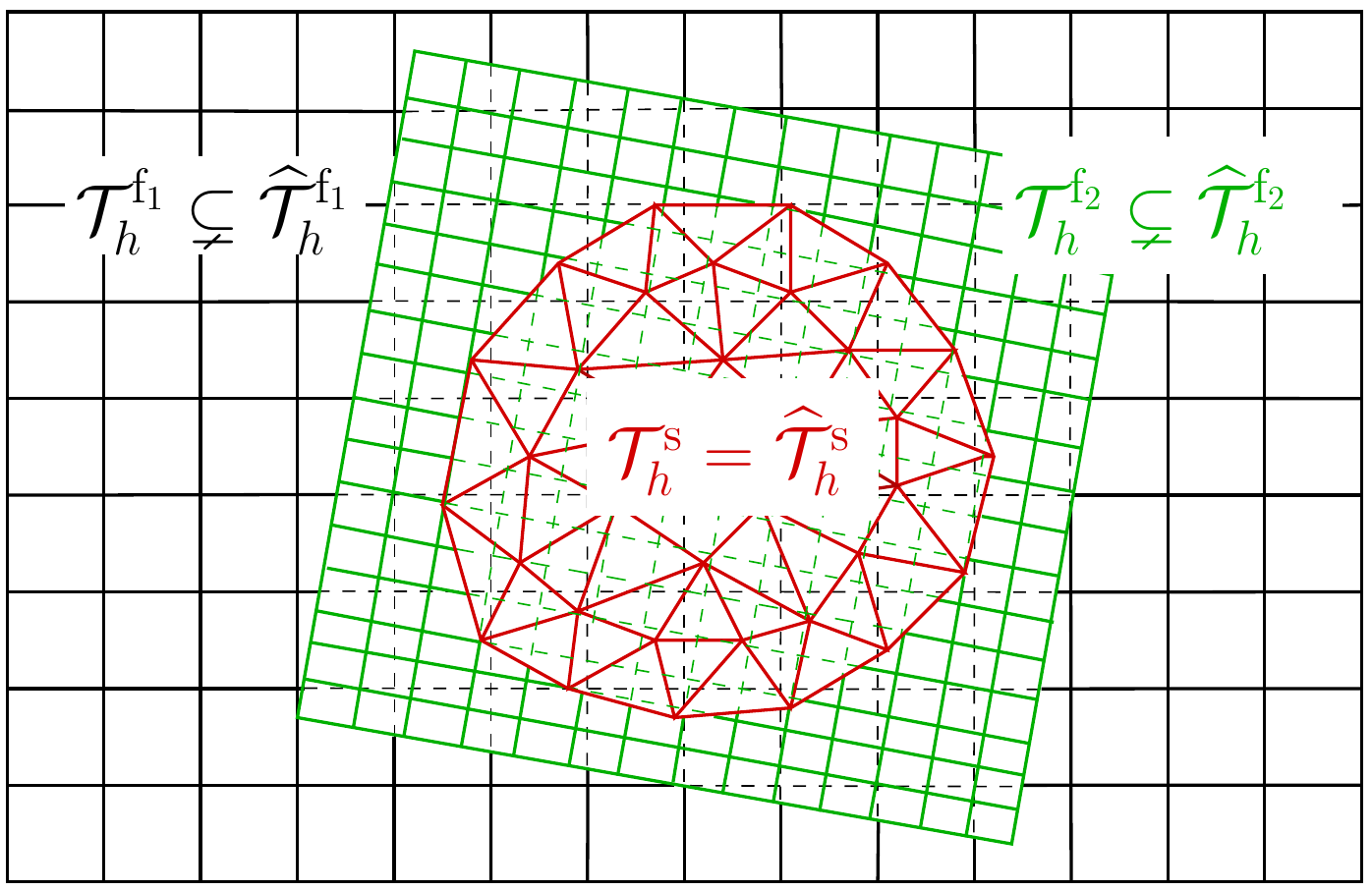}
\includegraphics[width=0.47\textwidth]{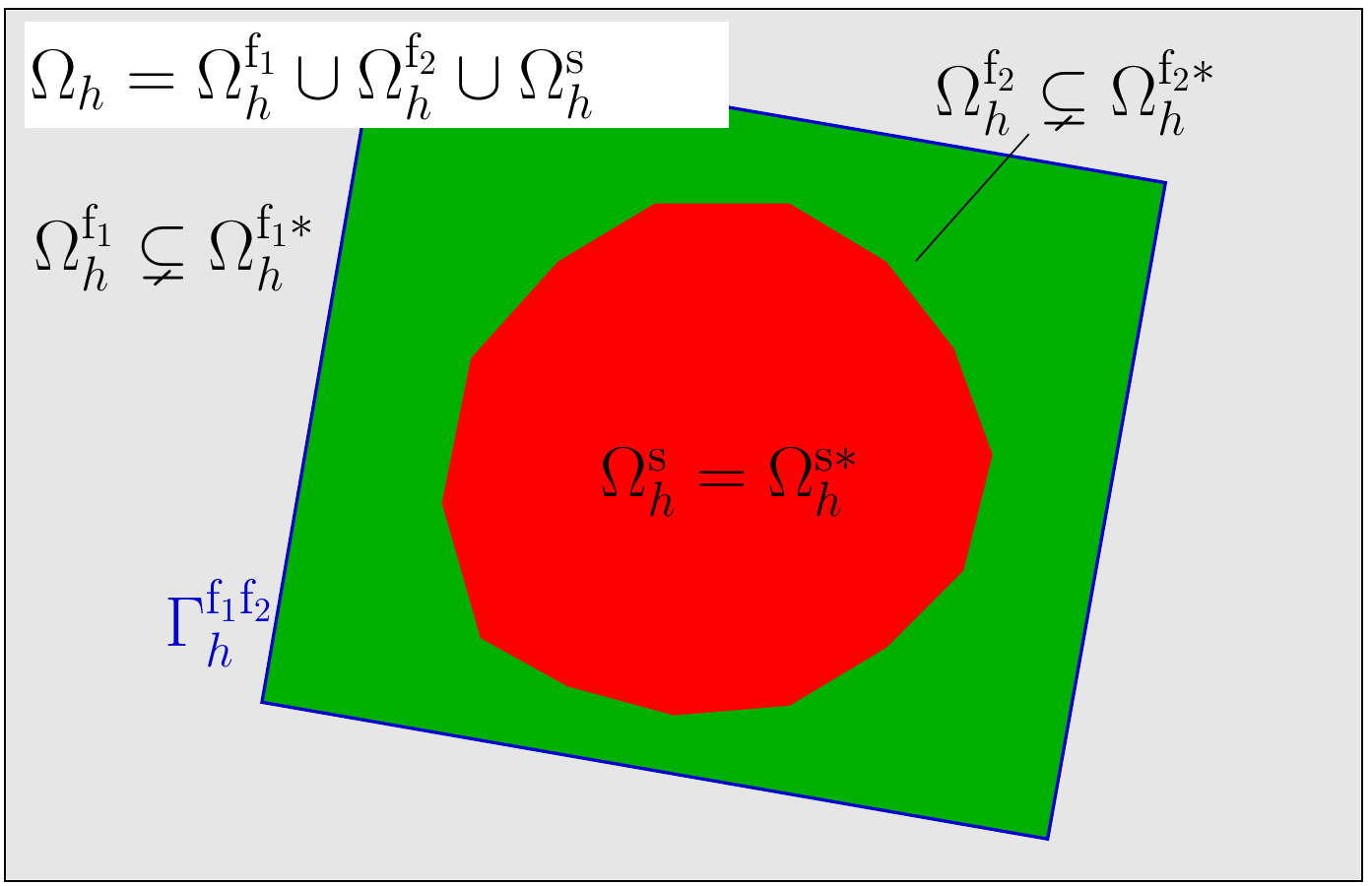}
\caption{Approach~\ref{approach:4_FXFSI}:
Computational meshes (left) and domain partition (right) for the FSI problem.}
\vspace{-12pt}
\label{fig:computational-mesh-unfitted-nonmatching-fsi-fxfsi}
\end{figure}
\end{approach}

\begin{approach}[\textbf{\textsc{unfitted-Euler/embedded-fitted-ALE}} or \textbf{\textsc{hybrid Eulerian-ALE:}} unfitted, non-moving fluid mesh technique with fitted, deforming fluid patch]
\label{approach:5_XFFSI}

Based on the idea to allow the embedded fluid patch to move and follow the structure over time
introduced in Approach~\ref{approach:4_FXFSI},
a powerful FSI algorithm can be set up when (self-)contact of structural bodies or topological changes of the fluid phase in the vicinity of the solid do not have to be supported.

Instead of an unfitted overlapping mesh approach at the fluid-solid interface~$\Int^{\fd_2 \sd}$,
one can also construct interface-fitted node-matching meshes~$\mcT^\sd_h$ and~$\mcT^{\fd_2}_h$.
The latter one can be simply provided by expanding the structural surface.
This fine resolved fluid patch~$\mcT^{\fd_2}_h$ will then be embedded into the background grid~$\mcT^{\fd_1}_h$ by means of an unfitted coupling.
The setting from Approach~\ref{approach:4_FXFSI} then changes to
\begin{align}
\label{eq:computational-mesh-unfitted-nonmatching-fsi-xffsi}
\overline{\Domh^{\fd_1}} \subsetneq \overline{\Dom_h^{\fd_1\ast}}, \qquad
\overline{\Domh^{\fd_2}} = \overline{\Dom_h^{\fd_2\ast}}, \qquad
\overline{\Domh^\sd} = \overline{\Dom_h^{\sd\ast}}, \qquad
\Dom_h^{\fd_1\ast} \cap \Dom_h^{\fd_2 \ast} \neq \emptyset \qquad
\Dom_h^{\fd_2 \ast} \cap \Dom_h^{\sd\ast}= \emptyset,
\end{align}
as visualized in~\Figref{fig:computational-mesh-unfitted-nonmatching-fsi-xffsi}.

Applying a moving mesh framework to the embedded fluid patch, the solid-fitted fluid mesh~$\mcT^{\fd_2}_h$
can follow the body in its movement.
This enables to perfectly capture flow effects at the fluid-structure interface
independent of the fluid patch location within the background mesh~$\mcT^{\fd_1}_h$.
This renders such approaches highly attractive, \eg, for
interactions of solids with turbulent incompressible flows,
in which capturing boundary layer effects in the vicinity of solids is a crucial task.
For a detailed presentation including numerical examples of Approach~\ref{approach:5_XFFSI} \textbf{\textsc{hybrid Eulerian-ALE}} within a monolithic framework the reader is referred to \cite{Schott2018b}. 
\begin{figure}[h!]
\centering
\includegraphics[width=0.47\textwidth]{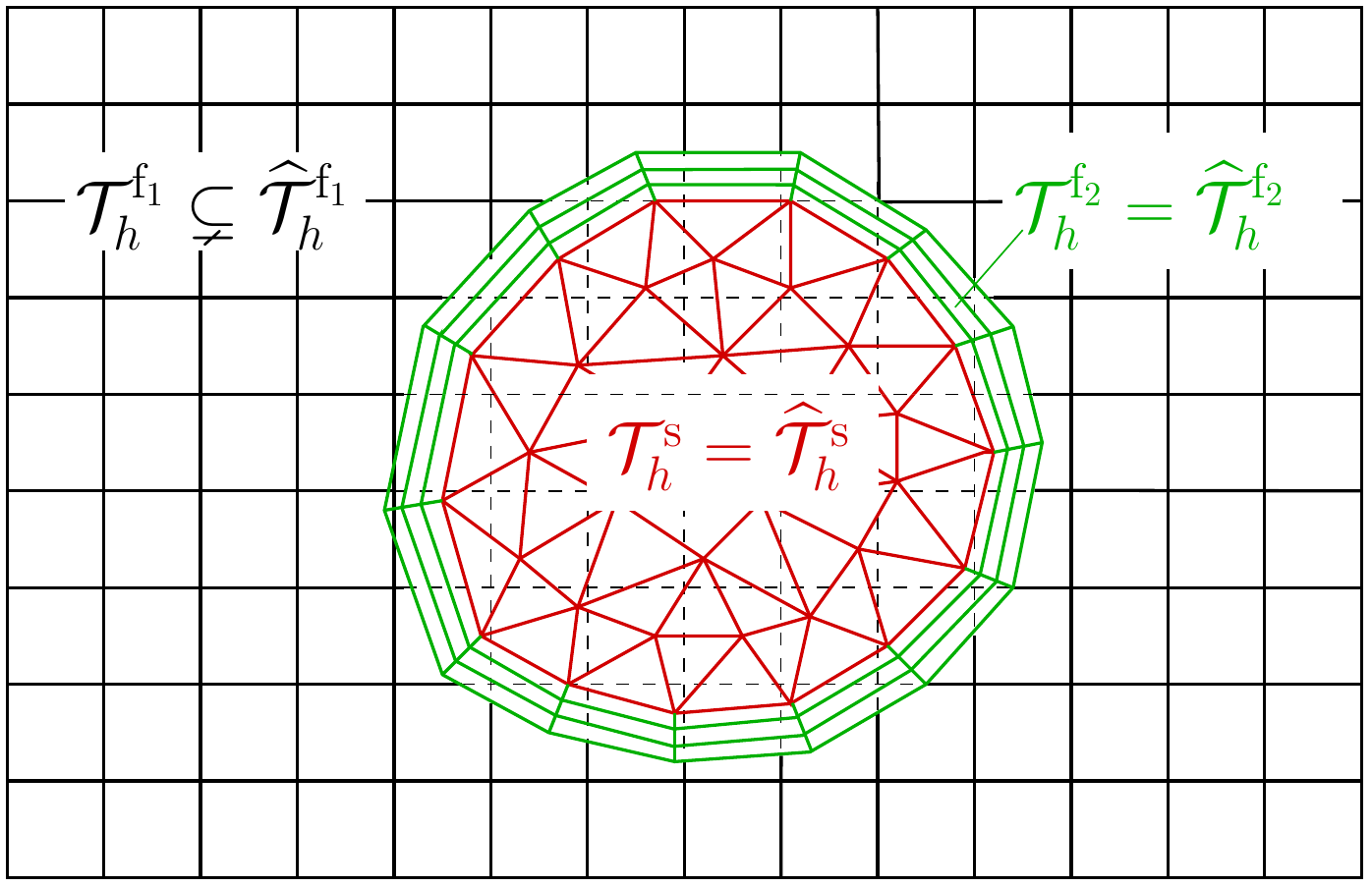}
\includegraphics[width=0.47\textwidth]{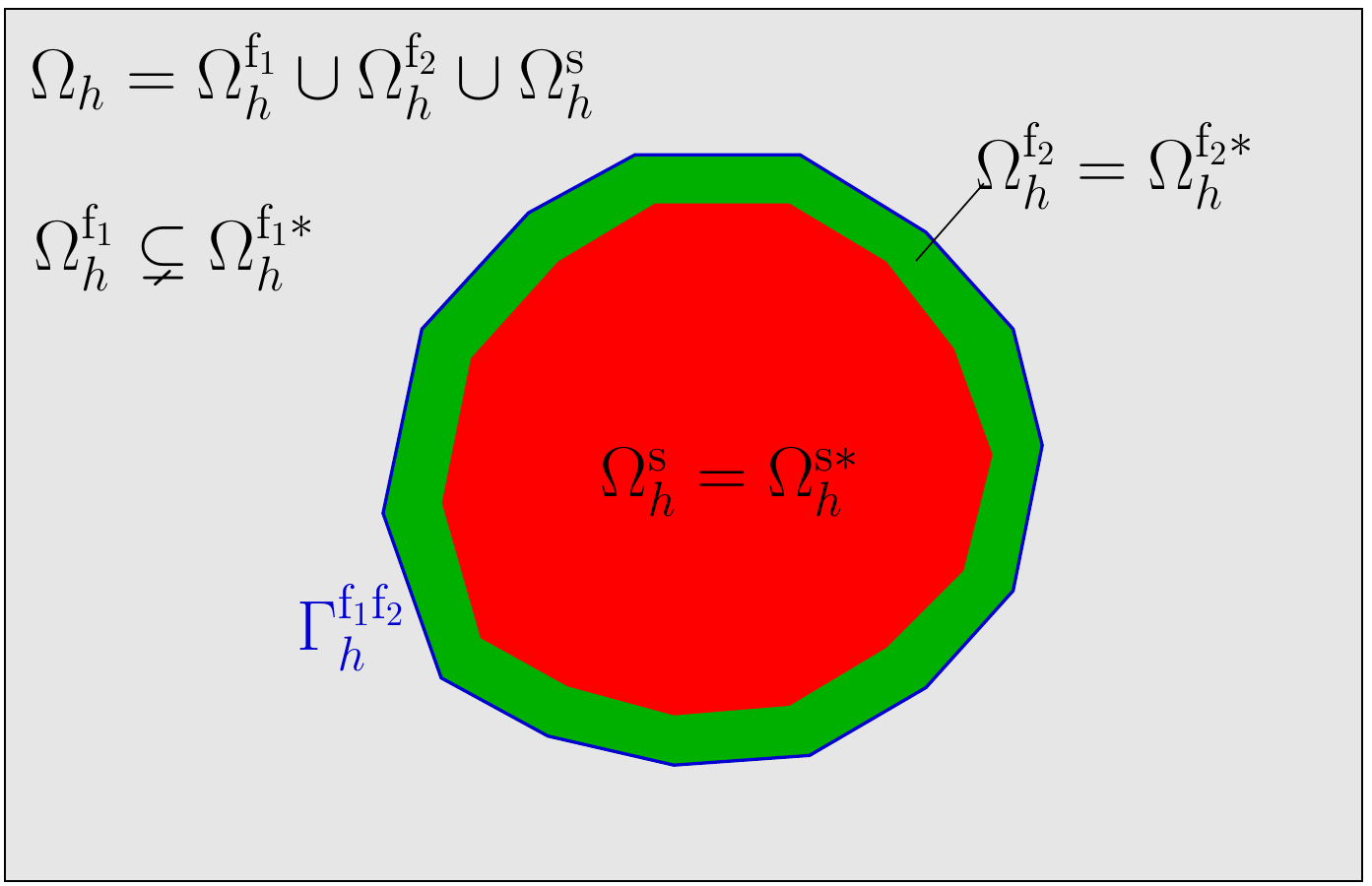}
\caption{Approach~\ref{approach:5_XFFSI}:
Computational meshes (left) and domain partition (right) for the FSI problem.}
\vspace{-12pt}
\label{fig:computational-mesh-unfitted-nonmatching-fsi-xffsi}
\end{figure}
\end{approach}
 
This overview of a various number of novel approaches
indicates a high flexibility when incorporating non-interface-fitted meshes to the overall discretization concept for FSI.
Depending on the FSI problem configuration, the expected structural deformation, the need to account for topological changes
or to even support contact or detachment processes of solids,
the proposed
Approaches~\ref{approach:1_ALE-FSI}--\ref{approach:5_XFFSI}
can exploit their capabilities in different situations.
All unfitted approaches benefit from the fact that meshes can be constructed independent of the respective subdomains.
Suitable inf-sup stable schemes for incompressible low- and high-Reynolds number flows
have been recently provided and will be recalled in Sections~\ref{ssec:stabilized_fitted_unfitted_FEM_fluids} and \ref{ssec:fluid_domain_decomposition_stab_form}.

\section{General Problem Formulation for Fluid-Structure Interaction (FSI)}
\label{sec:problem_formulation_FSI}

\subsection{Review of Frames of References}
\label{ssec:frames_of_reference}

Different kinematic descriptions are used in continuum mechanics when domains under consideration are moving over time.
Depending on the problem configuration and the chosen spatial approximation technique, different frames of reference
are used: the \emph{Lagrangean}, the \emph{Eulerian} and the \emph{Arbitrary-Lagrangean-Eulerian} formalism;
see also the textbook \cite{Donea2003} for an introduction.

\begin{figure}[h!]
  \begin{center}
  \includegraphics[width=0.55\textwidth]{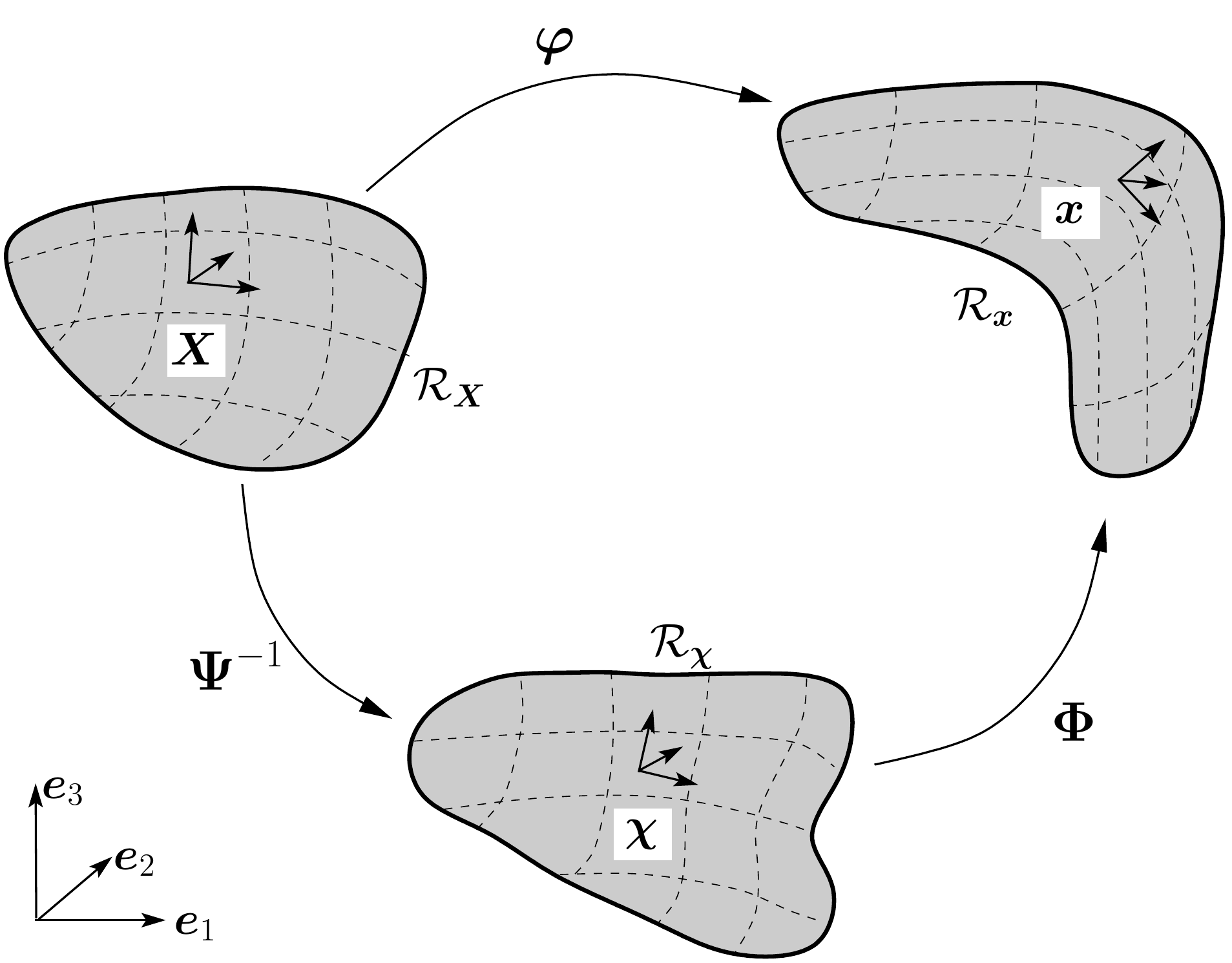} 
  \end{center}
  \caption{Different reference configurations used for FSI: a material domain \mbox{$\mcR_{\bfX}\subset\R^d$}
is mapped to a spatial domain \mbox{$\mcR_{\bfx}\subset\R^d$} at time $t$, whereas
a possibly moved referential configuration \mbox{$\mcR_{\bfChi}\subset\R^d$} is used for an ALE description.
Bijective mappings \mbox{$\bfvarphi,\bfPhi,\bfPsi^{-1}$} allow to transform between the different descriptions.}
  \label{fig:chap_2_frames_of_reference}
\vspace{-12pt}
\end{figure}

The material domain $\mcR_{\bfX}$ is made up of material particles \mbox{$\bfX\in\R^d$} which are under consideration
defined at starting time \mbox{$t=T_0$}.
Following these particles in their motion along time-dependent paths
$\bfx(\bfX,t)$ for times \mbox{$t\in(T_0,T)$}, the \emph{spatial domain} \mbox{$\mcR_{\bfx}(t)$} is formed,
expressed in terms of a bijective mapping 
$\bfvarphi(\bfX,t) : \mcR_{\bfX} \times (T_0,T) \rightarrow \mcR_{\bfx}\times  (T_0,T)$ with $\bfvarphi(\bfX,t)=(\bfx_{\bfX}(\bfX,t),t)$.
As an alternative observer of the region of interest, an independent \emph{referential domain} \mbox{$\mcR_{\bfChi}(t)$} with coordinates~$\bfChi$
is introduced. This is arbitrary in general and allows the observer to move independently of the material or spatial configurations.
It is commonly defined being coinciding with the material configuration and the spatial configuration at the initial time~$T_0$,
\ie~\mbox{$\mcR_{\bfChi}(T_0)=\mcR_{\bfx}(T_0)=\mcR_{\bfX}$}.
A second bijective mapping describes the motion of grid points in the spatial configuration 
$\bfPhi(\bfChi,t) : \mcR_{\bfChi} \times (T_0,T) \rightarrow \mcR_{\bfx}\times  (T_0,T) $ with
$\bfPhi(\bfChi,t)=(\bfx_{\bfChi}(\bfChi,t),t)$.
This mapping combined with $\bfvarphi$ enables to define \mbox{$\bfPsi^{-1} \eqdef \bfPhi^{-1}\circ\bfvarphi$}
relating material and referential configuration as $\bfPsi^{-1}(\bfX,t):\mcR_{\bfX} \times (T_0,T) \rightarrow \mcR_{\bfChi}\times  (T_0,T)$
with $\bfPsi^{-1}(\bfX,t)=(\bfChi_{\bfX}(\bfX,t),t)$.
The different reference configurations \mbox{$\bfX,\bfx,\bfChi$} including the mappings \mbox{$\bfvarphi,\bfPhi, \bfPsi^{-1}$} are visualized in \Figref{fig:chap_2_frames_of_reference}.

\subsection{Governing Equations for the FSI Problem}

\subsubsection{Structural Field}
\label{sssec:strong_formulation_solid_mechanics}

Solid mechanics is in our case described by the non-linear elastodynamics equations stated in a Lagrangean formalism,
in which the observer follows a material particle~$\bfX$ in its motion
from reference configuration~\mbox{$\Dom_0^{\sd}=\Doms(T_0)$} to current configuration~$\Doms(t)$.
This can be mathematically expressed by means
of the mapping~$\bfvarphi$ introduced in~\Secref{ssec:frames_of_reference}, as will be
briefly reviewed in the following. It should be pointed out that the approaches proposed in this paper obviously also work for inelastic behavior of the solid, but for the sake of simplicity of the presentation we only describe the nonlinear elastic case.

For all $\bfX\in\Doms(T_0)$,
kinematics can be formulated in terms of the unknown displacement field~$\bfd(\bfX,t) \eqdef \bfx_{\bfX}(\bfX,t) - \bfX$
between current and initial particle position defined in material description 
and its first- and second-order time derivatives, the velocity and acceleration fields
$\bfu=\dtimedot{\bfd}(\bfX,t)$ and $\bfa=\dsecondtimedot{\bfd}(\bfX,t)$.
The gradient tensor of the spatial part of the invertible mapping $\bfvarphi$, \ie~$\bfx_{\bfX}$,
is referred to as \emph{deformation gradient tensor}
\begin{align}
\label{eq:gov_eq_Solid_deformation_gradient}
\bfF(\bfX,t) &\eqdef \partderiv{\bfx_{\bfX}(\bfX,t)}{\bfX} = (\I + \partderiv{\bfd}{\bfX})(\bfX,t).
\end{align}
A suitable strain-measure for solids subjected to large deformations, but moderate stretching or compressing,
expressed in material coordinates is provided by the \emph{Green-Lagrange strain tensor} 
\begin{align}
\label{eq:gov_eq_Solid_Green_Lagrange_tensor}
\bfE &\eqdef \tfrac{1}{2}(\bfF^T \cdot \bfF - \I).
\end{align}
The corresponding \emph{second Piola--Kirchhoff stress tensor} is defined as
$\bfS = J_{\bfX\mapsto\bfx}\bfF^{-1} \cdot \bfsigma_{\bfx} \cdot \bfF^{-T}$,
where $\bfsigma_{\bfx}$ denotes the Cauchy stresses in spatial coordinates.
For the sake of simplicity, only homogeneous structural bodies that exhibit hyper-elastic material behavior
are considered; for a more comprehensive overview, see, \eg, the textbook \cite{Holzapfel2000}.
Assuming the existence of a so-called strain-energy function~\mbox{$\tilde{\Psi}$},
the second Piola--Kirchhoff stress tensor can be deduced as
$\bfS = 2\partderiv{\tilde{\Psi}}{(\bfF^T\bfF)}$.

In this paper, we exclusively consider the nonlinear \emph{Neo-Hookean} \name{(NH)} material with strain-energy function
\begin{align}
\label{eq:gov_eq_Solid_strain_energy_NH}
\tilde{\Psi}_{\name{NH}}(\bfF^T\bfF) &\eqdef \tfrac{\mu^\sd}{2}(\trace{\bfF^T\bfF} -3) - \mu^\sd \ln (J_{\bfX\mapsto\bfx}) + \tfrac{\lambda^\sd}{2} (\ln (J_{\bfX\mapsto\bfx}))^2 
\end{align}
with \mbox{$J_{\bfX\mapsto\bfx}=(\det(\bfF^T\bfF))^{1/2}$}.
The involved Lam\'{e} parameters $\lambda^\sd$ and $\mu^\sd$ can be expressed in terms of Young's modulus $E^\sd>0$ and Poisson's ratio $\nu^\sd\in(-1,0.5)$ as
\begin{equation}
\label{eq:gov_eq_Solid_Lame_parameters}
\lambda^\sd = E^\sd \nu^\sd ((1+\nu^\sd)(1-2\nu^\sd))^{-1} \quad\text{ and }\quad \mu^\sd= E^\sd (2(1+\nu^\sd))^{-1}.
\end{equation}

The resulting strong form of the non-linear elastodynamics \name{PDE}s can be summarized as:
Find solid displacements $\bfd:\Dom_0^{\sd}\times (T_0,T] \rightarrow \R^d$ defined in the reference configuration
such that
\begin{alignat}{2}
\label{eq:gov_eq_Solid_nonlinear-elastodynamic}
\rho^{\sd}\DSecondTime{\bfd} - \Div(\bfF\cdot\bfS)(\bfd) &= \rho^{\sd}\bff^{\sd} &&\qquad \foralls (\bfX,t) \in \Dom_0^{\sd}\times (T_0,T],\\
\label{eq:gov_eq_Solid_bc_1}
 \bfd &= \bfgD^{\sd}                             &&\qquad \forall (\bfX,t) \in \IntDRef^{\sd}\times (T_0,T], \\ 
\label{eq:gov_eq_Solid_bc_2}
 (\bfF\cdot\bfS)\cdot\bfN &= \bfhN^{\sd}         &&\qquad \forall (\bfX,t) \in \IntNRef^{\sd}\times (T_0,T],\\
\label{eq:gov_eq_Solid_initial_cond_1}
 \bfd(T_0)         &= \bfd_0          &&\qquad \foralls \bfX\in \Dom_0^{\sd}, \\
\label{eq:gov_eq_Solid_initial_cond_2}
 \DTime{\bfd}(T_0) &= \dot{\bfd}_0    &&\qquad \foralls \bfX\in \Dom_0^{\sd},
\end{alignat}
where $\rho^{\sd} = \rho_{\bfX}^{\sd}J_{\bfX\mapsto\bfx}$ denotes the structural material density
in the initial referential configuration,
and $\nabla\cdot(\cdot)=\nabla_{\bfX}\cdot(\cdot)$ is the divergence operator with respect to material referential coordinates.
Appropriate Dirichlet and Neumann boundary data, given by $\bfgD^{\sd},\bfhN^{\sd}$,
and initial values for structural displacements and velocities, defined as $\bfd_0, \dot{\bfd}_0$, complement the second order initial boundary value problem.
More detailed explanations can be found, \eg, in the textbooks \cite{Wriggers2008,Zienkiewicz2000}.

\subsubsection{Fluid Field}
\label{sssec:strong_formulation_fluid_mechanics}

As common in fluid mechanics, the description of the flow kinematics are expressed in spatial coordinates~$\bfx\in\Omega(t)$,
but are here given in a form that allows for a potential temporally moving observed domain.

Principles of conservation of mass and momentum in low Mach number flows,
based on the assumption of non-varying fluid density in space and time (\mbox{$\rho^{\fd}(\bfx,t)=\const$}),
render in the transient incompressible Navier-Stokes equations.
The resulting single phase initial boundary value problem can be written in \emph{Arbitrary-Lagrangean-Eulerian formulation} as:

Find flow velocity $\bfu:\Omega^{\fd}(t)\times t \rightarrow \R^d$ and dynamic pressure $p:\Omega^{\fd}(t)\times t \rightarrow \R$ such that
\begin{alignat}{2}
\label{eq:navier-stokes-strong-momentum}
\rho^{\fd}\dTime{\vel_{\bfChi}}\circ\bfPhi^{-1} + \rho^{\fd}(\bfc\cdot\Grad)\vel + \Grad p - 2\mu^{\fd}\Div\visc{\vel} &= \rho^{\fd}\bodyf^{\fd} \qquad && \foralls (\x,t) \in\Dom^{\fd}(t)\times (T_0,T], \\
\label{eq:navier-stokes-strong-mass}
\Div\vel &= 0 \qquad && \foralls (\x,t) \in \Dom^{\fd}(t)\times (T_0,T], \\
\label{eq:gov_eq_fluid_bc_1}
 \vel           &= \bfgD^{\fd}       \qquad &&\forall (\x,t) \in \IntD^{\fd}\times (T_0,T], \\ 
\label{eq:gov_eq_fluid_bc_2}
 \stress\cdot\n &= \bfhN^{\fd}       \qquad &&\forall (\x,t) \in \IntN^{\fd}\times (T_0,T], \\
\label{eq:gov_eq_fluid_initial}
 \vel(\bfx,0) &= \vel_0(\bfx)  \qquad &&\foralls \bfx\in \Dom^{\fd}(T_0),
\end{alignat}
where $\bodyf^{\fd}$ denotes an external body force load and \mbox{$((\bfc\cdot\Grad)\bfu)_i \eqdef  \sum_{j}{\bfc_j\cdot\partderiv{\bfu_i}{x_j}}$}
the \name{ALE} convective velocity with $\bfc\eqdef\bfu - \hat{\bfu}$, where $\hat{\bfu} \eqdef \partial_t \bfx_{\bfChi}\circ \bfPhi^{-1}$
is the velocity of the referential system.
The mapping therein tracks the deformation of the observed fluid domain $(\Dom^{\fd}(t),t) = \bfPhi(\Dom^{\fd}(T_0),t )$
from its initial configuration.
Further, \mbox{$\visc{\bfu} \eqdef 1/2 \left( \Grad\vel + (\Grad\vel)^T \right)$} denotes the symmetric strain rate tensor,
which is linearly related to the viscous part of the Cauchy stresses $\bfsigma(\bfu,p) = -p \I + 2\mu^{\fd} \visc{\bfu}$
by the dynamic viscosity~\mbox{$\mu^{\fd}>0$};
the latter one can be expressed in terms of a kinematic viscosity~$\nu^{\fd}$ as \mbox{$\mu^{\fd}=\nu^{\fd}\rho^{\fd}$}.
Appropriate Dirichlet and Neumann boundary data are specified at all times~$t$ by functions $\bfgD^{\fd},\bfhN^{\fd}$.
The initial condition for the flow field is specified as $\vel_0(\bfx)$ in $\Dom^{\fd}(T_0)$ which needs being conform with the problem definition.

\subsubsection{ALE Mesh Motion}
\label{sssec:strong_formulation_ALE_mechanics}

Considering a moving mesh ALE formalism for the fluid field,
the change of the fluid domain $\Dom^{\fd}(T_0)\mapsto\Dom^{\fd}(t)$ over time can be tracked in terms of the mapping $\bfPhi(\bfChi,t)$.
Similar to the Lagrangean formalism for structures,
introducing a displacement field~$\bfd^{\fd}(\bfChi,t) \eqdef \bfx_{\bfChi}(\bfChi,t) - \bfChi$
for the computational grid motion,
the ALE time derivative $\partial_t \bfx_{\bfChi}$ can be simply expressed in terms of the fluid domain displacements
resulting in $\hat{\bfu} = \partial_t \bfx_{\bfChi} = \partial_t (\bfd^{\fd}(\bfChi,t) +  \bfChi) = \partial_t \bfd^{\fd}(\bfChi,t)$
for the fluid grid velocity.

Note, for interface-fitted FSI approximations, the fluid domain displacement field is constraint being matching to the solid displacement field,
such that
\begin{alignat}{2}
\label{eq:ALE-constraint_FSI}
\bfd^{\fd}\circ \bfPhi^{-1}(\bfx,t) & = \bfd^{\sd} \circ \bfvarphi^{-1}(\bfx,t) \quad\foralls (\bfx,t) \in (\Gamma^{\fd\sd}(t),t)
\end{alignat}
and potential further essential constraints at $\Gamma^{\fd}$.
Thereby, $\bfd^{\fd}$ can be extended arbitrarily into the interior of the domain $\Omega^{\fd}(T_0)$.
In practice, the mesh quality needs to be preserved properly; see \cite{LeTallec2001, Kloppel2011} for an overview of possible mesh update techniques.
For all computations shown in the present work, a pseudo-structure mesh update algorithm as described, \eg~in \cite{Kloppel2011} has been used for adapting the fluid grid. For an interesting extension of this approach towards improved robustness and efficiency, see \cite{LaSpina}.

\subsubsection{Fluid-Structure Interface}
\label{ssec:strong_formulation_interface_mechanics}

Macroscopic considerations of fluid-structure interaction
are based on conservation laws for mass and momentum.
For viscous fluids, \ie~$\mu^{\fd}>0$, kinematic and dynamic interface constraints emerge as
\begin{alignat}{2}
\label{eq:fsi_interface_condition_u_jump}
\jump{\bfu} &= \bfu^\fd - \frac{\mathrm d{\bfd^\sd}}{\mathrm d{t}}\circ\bfvarphi_t^{-1} = \bfg_\Int^\fs \eqdef \bfzero && \qquad \foralls \bfx\in\Intfs(t), \\
\label{eq:fsi_interface_condition_traction_jump}
\jump{\bfsigma}\cdot\bfn^\fs &= (\bfsigma_{\bfx}(\bfu^\fd,p^\fd) - \bfsigma_{\bfx}(\bfd^\sd\circ\bfvarphi_t^{-1}))\cdot\bfn^\fs = \bfh_\Int^\fs \eqdef \bfzero  &&\qquad \foralls \bfx\in\Intfs(t),
\end{alignat}
where \mbox{$\bfsigma_{\bfx}(\bfu^\fd,p^\fd)$}, \mbox{$\bfsigma_{\bfx}(\bfd^\sd\circ\bfvarphi_t^{-1})$}
denote the Cauchy stresses defined on fluid and structural side of the interface.
Note that in the limit case of \mbox{$\mu^{\fd}\rightarrow 0$}, the constraint \Eqref{eq:fsi_interface_condition_u_jump} on the tangential component
needs to be relaxed and only mass conservation in interface normal direction need to be ensured, \ie~\mbox{$\jump{\bfu}\cdot\bfn^\fs=0$}.
This aspect can be smoothly accounted for via a Nitsche-type weak enforcement discussed in Remark~\ref{rem:Nitsche_weak_DBC}.
More complex physical models that take roughness of the structural surface into consideration
render usually in generalized Robin-type coupling constraints.
As such conditions require special numerical treatment to guarantee stability and optimality
in a discrete fashion, as proposed, \eg, in the works by \cite{Winter2018,JuntunenStenberg2009a},
such models are not treated within the scope of this work.

\subsection{Variational Formulation of the FSI Problem}

The coupled system between an incompressible fluid and a compressible or incompressible structure can be composed of two initial boundary value
problems: a first one for the structural phase in~$\Doms$, see \Secref{sssec:strong_formulation_solid_mechanics},
and a second one for the fluid phase in~$\Domf$, see \Secref{sssec:strong_formulation_fluid_mechanics},
complemented with outer Dirichlet and Neumann boundary conditions at their boundaries, respectively.
Both phases are glued together by interfacial coupling constraints
\Eqref{eq:fsi_interface_condition_u_jump}--\eqref{eq:fsi_interface_condition_traction_jump}.

\subsubsection{Structural Field}
\label{sssec:variational_formulation_solid_mechanics}
For the solid weak formulation the space of admissible displacements
\mbox{$\mcD_{\bfgD} \eqdef [H^1_{\IntD,\bfgD}(\Dom_0^\sd)]^d \subset [H^1(\Dom_0^\sd)]^d$}
satisfies the Dirichlet boundary condition \Eqref{eq:gov_eq_Solid_bc_1},
the space of admissible test functions~$\mcD_{\bfzero}$ exhibits zero trace on~$\IntD^{\sd}$.
Furthermore, \mbox{$\mcD$} denotes the admissible space for structural velocities. 

The weak formulation of the non-linear structural problem \Eqref{eq:gov_eq_Solid_nonlinear-elastodynamic}-\eqref{eq:gov_eq_Solid_initial_cond_2} reads:
for any \mbox{$t\in(T_0,T]$}, find solid displacements \mbox{$\bfd(t)\in\mcD_{\bfgD}$} and velocities \mbox{$\dtimedot{\bfd}(t)=\mathrm d{\bfd}/{\mathrm d t}\in\mcD$} such that for all
$\bfw\in\mcD_{\bfzero}$
\begin{align}
 \label{eq:gov_eq_Solid_weak_form_operator_form}
 (\rho^\sd\DTime{\dtimedot{\bfd}},\bfw)_{\Doms_0} + a^\sd(\bfd,\bfw) = l^\sd(\bfw)
\end{align}
with the following operators
\begin{align}
\label{eq:gov_eq_Solid_operator_ah}
 a^\sd(\bfd,\bfw) &\eqdef ((\bfF\cdot\bfS)(\bfd),\Grad\bfw)_{\Doms_0}, \\
\label{eq:gov_eq_Solid_operator_lh}
 l^\sd(\bfw)         &\eqdef (\rho^\sd\bff^\sd,\bfw)_{\Doms_0} + \scalbound{\bfhN^\sd}{\bfw}{\IntNRef^\sd},
\end{align}
where the elastic form $a^\sd$ is linear in the variable~$\bfw$, however, in general non-linear in the displacement field~$\bfd$.

\subsubsection{Fluid Field}
\label{sssec:variational_formulation_fluid_mechanics}

While the functional space for the admissible fluid velocities is
{$\mcV_{\bfgD} \eqdef [H^1_{\Int_D,\bfgD}(\Dom(t))]^d \subset [H^1(\Dom(t))]^d$}
and satisfies the Dirichlet boundary condition \Eqref{eq:gov_eq_fluid_bc_1},
the related space of admissible test functions~$\mcV_{\bfzero}$ exhibits zero trace on~$\IntD^{\fd}$.
The trial and test function spaces for the pressure are given by
\mbox{$\mcQ=L^2(\Dom(t))$} provided that \mbox{$\IntN^{\fd}=\emptyset$}.
The non-linear variational formulation for incompressible flow then reads as follows:
for all \mbox{$t\in(T_0,T]$}, find fluid velocity and pressure
\mbox{$U(t)=(\bfu(t),p(t))\in \mcV_{\bfgD}\times \mcQ$} such that 
\begin{alignat}{2}
 \label{eq:gov_eq_Fluid_weak_form}
\mcA(U,V) &= \mcL(V) \quad \foralls V=(\bfv,q) \in \mcV_{\bfzero}\times\mcQ,
\end{alignat}
where
\begin{alignat}{2}
\mcA(U,V) & \eqdef \scalLDom{\rho^{\fd}\dTime{\bfu_{\bfChi}}\circ \bfPhi^{-1}}{\bfv}{\Domf(t)} + n^{\fd}(\bfu-\hat{\bfu}; \bfu, \bfv) + a^{\fd}(\bfu,\bfv) + b^{\fd}(p, \bfv) - b^{\fd}(q,\bfu), \\
\mcL(V)   & \eqdef l^{\fd}(\bfv)
\end{alignat}
with the following trilinear, bilinear and linear forms $n,a,b\text{ and }l$ introduced in order to shorten the presentation
\begin{align}
\label{eq:fld:gov_eq_Fluid_operator_cf}
 n^{\fd}(\bfc; \bfu, \bfv) &\eqdef \scalLDom{\rho^{\fd}(\bfc\cdot\nabla)\bfu}{\bfv}{\Domf(t)}, \\
\label{eq:fld:gov_eq_Fluid_operator_af}
 a^{\fd}(\bfu,\bfv)           &\eqdef \scalLDom{2\mu^{\fd}\visc{\bfu}}{\visc{\bfv}}{\Domf(t)}, \\
\label{eq:fld:gov_eq_Fluid_operator_bf}
 b^{\fd}(p,\bfv)              &\eqdef -\scalLDom{p}{\Div{\bfv}}{\Domf(t)}, \\
\label{eq:fld:gov_eq_Fluid_operator_lf}
 l^{\fd}(\bfv)                &\eqdef \scalLDom{\rho^{\fd}\bodyf^{\fd}}{\bfv}{\Domf(t)}  + \scalbound{\bfhN^{\fd}}{\bfv}{\IntN^{\fd}(t)}.
\end{align}
Therein, $\hat{\bfu}$ denotes the grid velocity for an \name{ALE} frame,
as introduced in Sections~\ref{sssec:strong_formulation_fluid_mechanics}--\ref{sssec:strong_formulation_ALE_mechanics}.

\subsubsection{Coupled Variational Fluid-Structure Problem}
\label{sssec:variational_formulation_coupled_problem}

Consider the functional spaces for fluid and solid as given for the single phase problems to
$\mcV_{\bfgD}\times\mcQ, \mcV_{\bfzero}\times\mcQ$ for the fluid and to
$\mcD_{\bfgD}\times\mcD, \mcD_{\bfzero}$
for the solid,
the coupled non-linear variational formulation then reads as follows:
for all \mbox{$t\in(T_0,T]$}, find fluid velocities and pressures as well as solid displacements and velocities
\mbox{$(\bfu(t),p(t),\bfd(t),\dtimedot{\bfd}(t))\in (\mcV_{\bfgD} \times \mcQ) \oplus (\mcD_{\bfgD} \times \mcD)$}
such that
\begin{alignat}{2}
 \label{eq:gov_eq_Fluid_Solid_weak_form}
\begin{split}
& \scalLDom{\rho^\fd \dTime{\bfu_{\bfChi}}\circ \bfPhi^{-1}}{\bfv}{\Domf(t)} + n^\fd(\bfu-\hat{\bfu}; \bfu, \bfv) + a^\fd(\bfu,\bfv) + b^\fd(p, \bfv) - b^\fd(q,\bfu)\\
&\qquad + \scalLDom{\rho^\sd(\DTime{\bfd}-\dtimedot{\bfd})}{\bfz}{\DomsRef}  + \scalLDom{\rho^\sd\DTime{\dtimedot{\bfd}}}{\bfw}{\DomsRef} + a^\sd(\bfd,\bfw)= l^\fd(\bfv) + l^\sd(\bfw), \\
& \jump{\bfu}\big|_{\Intfs(t)} = \bfzero \quad \text{ with } \quad \bfu^\fd=\bfu \quad \text{ and } \quad \bfu^\sd=\dtimedot{\bfd}\circ \bfvarphi_t^{-1},
\end{split}
\end{alignat}
for all \mbox{$(\bfv,q,\bfw) \in (\mcV_{\bfzero}\times\mcQ)\oplus\mcD_{\bfzero}$},
with fluid operators \mbox{$c^\fd,a^\fd,b^\fd,l^\fd$} as defined in equations \Eqref{eq:fld:gov_eq_Fluid_operator_cf}--\eqref{eq:fld:gov_eq_Fluid_operator_lf}
and structural operators \mbox{$a^\sd,l^\sd$} as defined in \Eqref{eq:gov_eq_Solid_operator_ah} and \Eqref{eq:gov_eq_Solid_operator_lh}.

Note that after integration by parts on the two bulk phases, the traction coupling constraint \mbox{$\bfh_\Int^{\fs}=\bfzero$}
has been directly incorporated into the variational formulation as a natural constraint.

Multiplying \eqref{eq:gov_eq_Solid_nonlinear-elastodynamic} and \eqref{eq:navier-stokes-strong-momentum} with functions~$\bfv^i$ (where for the solid phase $\bfv^{\sd}=\bfw$)
and integrating all stress-related bulk terms by parts yields at $\partial \Domi$
\begin{flalign}
   &\sum\nolimits_{1\leqslant i \leqslant N_{\textrm{dom}}}{\left( - \scalbound{\bfsigma_{\bfx}^i\cdot \bfn^i}{\bfv^i}{\partial \Domi}  \right)} \\
 &\quad= \sum\nolimits_{1\leqslant i \leqslant N_{\textrm{dom}}}{\left( - \scalbound{\bfhN^i}{\bfv^i}{\IntN^i} -  \sum\nolimits_{j>i} \int_{\Intij}{\bfh_\Gamma^{ij}\cdot \wavginv{\bfv} + (\wavg{\bfsigma_{\bfx}}\cdot \bfn^{ij})\cdot \jump{\bfv}}\dGamma \right)},
\label{eq:gov_eq_Fluid_Fluid_weak_form:interface_terms}
\end{flalign}
where relation \Eqref{eq:jump-average-relation} allows to incorporate the flux coupling condition
\mbox{$\jump{\bfsigma(\bfu,p)}\cdot \bfn^{ij}=\bfh_\Int^{ij}=\bfzero$} as a natural constraint.
The last term vanishes due to \mbox{$\jump{\bfv}|_{\Intij}=\bfzero$},
provided the coupling \mbox{$\jump{\bfu}=\bfg_\Int^{ij}$} is incorporated in the continuous function space~$\mcV_{\bfgD}\oplus\mcD$.

\begin{remark}
For discrete composed approximation spaces the last term in \Eqref{eq:gov_eq_Fluid_Fluid_weak_form:interface_terms} does not vanish anymore
and remains as a standard consistency term.
Nitsche-type enforcement strategies of essential coupling constraints on discrete subspaces
that do not inherently guarantee continuity as claimed in \Eqref{eq:fsi_interface_condition_u_jump}
across~$\Intfs$ will be discussed in \Secref{ssec:fsi_stab_form}.
Therein, also the importance of the choice of average weights~\mbox{$w^i,w^j$} incorporated in the definition of~\mbox{$\wavg{\cdot},\wavginv{\cdot}$}
will be elucidated.
\end{remark}

\begin{remark} \label{rem:need_for_fluid_domain_decomposition}
Note that similar to the fluid-structure coupling, also the fluid domain $\Domf(t)$ can be split artificially
for domain decomposition discretization purposes,
as required by Approach~\ref{approach:4_FXFSI}.
Operators $c^{\fd_i},a^{\fd_i},b^{\fd_i}$ and respective time derivatives are then evaluated in partitions
$\Dom^{\fd_i}(t)$ of the fluid subdomain $\Domf(t)$ with potential
different ALE or Eulerian descriptions of the time derivatives $\scalLDom{\rho^\fd \partial_t\bfu^i_{\bfChi^i}\circ (\bfPhi^i)^{-1}}{\bfv^i}{\Dom^{\fd_i}(t)}$.
Coupling constraints can be incorporated by analogy to \Eqref{eq:gov_eq_Fluid_Fluid_weak_form:interface_terms}, see further details in
\Secref{ssec:fluid_domain_decomposition_stab_form}.
\end{remark}

\section{Finite Element Formulations - Different Spatial Discretizations}
\label{sec:spatial_discretization}

Within this section, the different spatial discretization techniques used
for Approaches~\ref{approach:1_ALE-FSI}--\ref{approach:4_FXFSI}
will be first introduced
for the single fields and afterwards complemented by Nitsche-type coupling techniques at (artificial) interfaces between the different subdomains.

\subsection{Classical Fitted FEM for Structures}
\label{ssec:fitted_FEM_structures}

For the space semi-discretization of the structural weak formulation~\Eqref{eq:gov_eq_Solid_weak_form_operator_form},
let $\{\mcT_h^\sd\}_h$ be a family of boundary-fitted quasi-uniform meshes with mesh size parameter \mbox{$h>0$}.
Each mesh~$\mcT_h^\sd$ approximates \mbox{$\DomsRef\approx\Dom^\sd_{0,h}=\cup_{T\in\mcT^\sd_h}{T}$} and
consists of possibly curvilinear finite elements \mbox{$T\in\mcT_h^\sd$}
with isoparametric mappings \mbox{$S_T(t):\hat{T}\mapsto T$} to the element parameter space.
Solid displacements~$\bfd$ and velocities~$\dtimedot{\bfd}$ are approximated with standard continuous isoparametric finite element spaces
\begin{equation}
 \mcX_{0,h} = \left\{ x_h \in C^0(\overline{\Dom^\sd_{0,h}}): \restr{x_h}{T}=v_{\hat{T}}\circ S_T^{-1}(t) \text{ with } v_{\hat{T}}\in \mathbb{V}^k(\hat{T})\, \foralls T \in \mcT_h^\sd \right\}
\end{equation}
such that
\mbox{$\mcD_{\bfgD,h} \eqdef [\mcX_{0,h}]^d \cap \mcD_{\bfgD}$}, \mbox{$\mcD_h \eqdef [\mcX_{0,h}]^d \cap \mcD$} and
\mbox{$\mcD_{\bfzero,h}\eqdef [\mcX_{0,h}]^d \cap \mcD_{\bfzero}$}
are the discrete displacement and velocity approximation and test function spaces which take into account the respective trace values.

The space semi-discrete approximation of~\Eqref{eq:gov_eq_Solid_weak_form_operator_form} reads:
for any \mbox{$t\in(T_0,T]$}, find solid displacements \mbox{$\bfd_h(t)\in\mcD_{\bfgD,h}$} and velocities \mbox{$\dtimedot{\bfd}_h(t)\in\mcD_h$},
\ie~$D_h=(\bfd_h,\dtimedot{\bfd}_h)$, such that
\begin{align}
 \label{eq:gov_eq_Solid_weak_form_discrete}
\mcA_h^{\sd}(D_h,W_h) = \mcL_h^{\sd}(W_h) \quad  \foralls W_h=\bfw_h\in\mcD_{\bfzero,h},
\end{align}
where
\begin{align}
 \mcA_h^{\sd}(D_h,W_h) &\eqdef \scalL{\rho^\sd\DTime{\dtimedot{\bfd}_h}}{\bfw_h}_{\Dom_{0,h}^\sd} + a_h^\sd(\bfd_h,\bfw_h), \label{eq:ahs}\\
 \mcL_h^{\sd}(W_h)     &\eqdef l_h^\sd(\bfw_h) \label{eq:lhs}
\end{align}
with \mbox{$a_h^\sd=a^\sd$} and \mbox{$l_h^\sd=l^\sd$} as defined in \Eqref{eq:gov_eq_Solid_operator_ah} and \Eqref{eq:gov_eq_Solid_operator_lh}.
All domain and boundary integrals are then defined on the discrete counterparts \mbox{$\Dom_{0,h}^\sd,\Int_{N,0,h}^\sd$}, respectively.

\subsection{Stabilized Fitted and Unfitted FEMs for Fluids}
\label{ssec:stabilized_fitted_unfitted_FEM_fluids}

In the sequel, finite element approximations for boundary-fitted and non-boundary-fitted flow approximations are recalled
and particular focus is put on stabilization mechanisms in the interior of the fluid domain and the boundary zone
of unfitted meshes.

A spatial semi-discrete cut finite element formulation of \Eqref{eq:gov_eq_Fluid_weak_form},
which utilizes a stabilizing residual-based variational multiscale (\name{RBVM}) concept in the interior of the domain,
\ie~\name{SUPG}/\name{PSPG}/\name{LSIC} terms (see, \eg, in \cite{Gresho2000,HughesScovazziFranca2004}),
and ghost penalty (\name{GP}) terms in the boundary zone, is based on previous works
\cite{MassingSchottWall2016_CMAME_Arxiv_submit, SchottRasthoferGravemeierWall2015, SchottShahmiriKruseWall2015} and reads:

\begin{definition}[Semi-discrete formulation of the \name{RBVM}/\name{GP}-\name{CutFEM}]
\label{def:RBVM_GP_CUTFEM_Navier_Stokes}
For all \mbox{$t\in(T_0,T]$}, find fluid velocity and pressure \mbox{$U_h(t)=(\bfu_h(t),p_h(t))\in\mcV_{\bfgD,h}\times\mcQ_h$}
such that for all \mbox{$V_h=(\bfv_h,q_h)\in\mcV_{\bfzero,h}\times\mcQ_h$}
\begin{equation}
\label{eq:chap_3_Nitsche_RBVM_GP_CUTFEM_form}
   \mcA_h^{\fd,\mathrm{RBVM/GP}}(U_h,V_h) = \mcL_h^{\fd,\mathrm{RBVM/GP}}(U_h,V_h),
\end{equation}
where
\begin{align}
\label{eq:chap_3_Nitsche_RBVM_GP_CUTFEM_form_Ah}
   &\mcA_h^{\fd,\mathrm{RBVM/GP}}(U_h,V_h)
          \eqdef (\rho^{\fd}\dTime{\bfu_{\bfChi,h}}\circ \bfPhi^{-1}, \bfv_h) + (\mcB_h + \mcG_h^{\mathrm{GP}})(\bfu_h-\hat{\bfu}_{h};(\bfu_h,p_h), (\bfv_h,q_h)),\nonumber\\
&\quad
+ \sum_{T\in\mcT_h}{\Big( \rho^{\fd}\dTime{\bfu_{\bfChi,h}}\circ \bfPhi^{-1} + \bfr_{\mathrm{M}}(\bfu_h-\hat{\bfu}_{h};\bfu_h,p_h), \tau_\mathrm{M} ((\rho^{\fd}(\bfu_h-\hat{\bfu}_{h}) \cdot\nabla) \bfv_h + \nabla q_h)\Big)_{T\cap\Domh^\fd}}
\nonumber\\
&\quad
+ \sum_{T\in\mcT_h}{\Big(r_{\mathrm{C}}(\bfu_h),\tau_\mathrm{C} \nabla\cdot\bfv_h\Big)_{T\cap\Domh^\fd}},\\
   &\mcL_h^{\fd,\mathrm{RBVM/GP}}(U_h,V_h)
	  \eqdef \mcL_h(\bfv_h)
+ \sum_{T\in\mcT_h}{\Big(\rho^{\fd}\bodyf^{\fd}, \tau_\mathrm{M} ((\rho^{\fd}(\bfu_h-\hat{\bfu}_{h}) \cdot \nabla \bfv_h) + \nabla q_h)\Big)_{T\cap\Domh^\fd}}
\end{align}
where
\begin{flalign}
\label{eq:chap_3_Nitsche_CIP_GP_CUTFEM_form_Bh}
   \mcB_h(\bfc_h;(\bfu_h,p_h), (\bfv_h,q_h)) &= (n_h + a_h)(\bfc_h;\bfu_h,\bfv_h) + b_h(p_h, \bfv_h) - b_h(q_h,\bfu_h), \\ 
\label{eq:chap_3_Nitsche_CIP_GP_CUTFEM_form_Gh}
   \mcG_h^{\mathrm{GP}} (\bfc_h;(\bfu_h,p_h), (\bfv_h,q_h)) &= (g_c + g_u + g_p )(\bfc_h;(\bfu_h,p_h), (\bfv_h,q_h)), \\
\label{eq:chap_3_Nitsche_CIP_GP_CUTFEM_form_Lh}
   \mcL_h(\bfv_h) &= l_h(\bfv_h).
\end{flalign}
Terms according to the standard Galerkin formulation ($n_h,a_h,b_h,l_h$) are as defined in \Eqref{eq:fld:gov_eq_Fluid_operator_cf}--\eqref{eq:fld:gov_eq_Fluid_operator_lf},
evaluated on the discrete counterparts $\Domh^{\fd},\Int^{\fd}_{\mathrm{N},h}$.
Furthermore, the residual terms are defined as $\bfr_{\mathrm{M}}(\bfc_h;\bfu_h,p_h) = \rho^{\fd}(\bfc_h \cdot \nabla)\bfu_h + \nabla p_h - 2\mu^{\fd}\nabla\cdot\visc{\bfu_h}$
and $r_{\mathrm{C}}(\bfu_h) = \nabla\cdot\bfu_h$.
Appropriate piecewise constant stabilization scaling functions are given as
\begin{align}
\label{eq:residual_based_tau_definition}
 \tau_{\mathrm{M},T}(\bfc_h) &=(\left(\tfrac{2\rho^{\fd}}{\Delta t}\right)^2+(\rho^{\fd} \bfc_h)\cdot \bfG(\rho^{\fd} \bfc_h)+C_{\mathrm{I}}{\mu^{\fd}}^2 \bfG:\bfG)^{-\onehalf}, \quad
 \tau_{\mathrm{C},T} = (\tau_\mathrm{M,T}\mathrm{tr}\left(\bfG\right))^{-1},
\end{align}
as defined in \cite{Taylor1998,Whiting2001} with
the second rank metric tensor $G_{kl}(\bfx)=\sum_{i=1}^{d}(\partial \xi_i / \partial x_k\big|_{\bfx})(\partial \xi_i / \partial x_l\big|_{\bfx})$
and $C_\mathrm{I}$ set to~$36.0$ for linearly and to~$60.0$ for quadratically interpolated hexahedral finite elements.
For further details on the RBVM technique, the interested reader is referred to \cite{Franca1992,Whiting2001},
and for a general overview on different stabilization techniques for incompressible flow to, \eg, \cite{BraackBurmanJohnEtAl2007,Roos2008}.

Defining \mbox{$c_{\infty,F} \eqdef \|\bfc_h \|_{0,\infty,F}$}, interface zone face-jump ghost-penalty terms are given by
\begin{align}
\label{eq:chap_3_Nitsche_CIP_GP_CUTFEM_form:g_beta}
   g_c (\bfc_h;\bfu_h,\bfv_h)           &= \gamma_{c} \sum\nolimits_{F\in\mcF_\Int} \sum\nolimits_{1\leqslant j \leqslant k}{  \rho^{\fd}( \nu^{\fd} + \phi_{c,F} c_{\infty,F}^2 + \sigma h_F^2)  h_F^{2j-1}        \langle \jump{\nablan^j \bfu_h},\jump{\nablan^j \bfv_h} \rangle_F}, \\
   g_u (\bfc_h;\bfu_h,\bfv_h)               &= \gamma_{u}     \sum\nolimits_{F\in\mcF_\Int} \sum\nolimits_{0\leqslant j \leqslant k-1}{ \phi_{u,F}    \rho^{\fd}                 h_F^{2j+1}        \langle \jump{\nabla \cdot \nablan^j\bfu_h},\jump{\nabla\cdot \nablan^j \bfv_h} \rangle_F}, \\
   g_p (\bfc_h;p_h,q_h)                     &= \gamma_{p}     \sum\nolimits_{F\in\mcF_\Int} \sum\nolimits_{1\leqslant j \leqslant k}{   \phi_{p,F}    (\rho^{\fd})^{-1}            h_F^{2j-1}        \langle \jump{\nablan^j p_h},\jump{\nablan^j q_h} \rangle_F}
.
\end{align}
Moreover, $\bfu_h$ has to fulfill the initial condition \mbox{$\bfu_h(T_0)=\bfu_0$} in~$\Domh(T_0)$.
Note, to shorten the presentation of the stabilized formulation the time variable in \mbox{$U_h(t)=(\bfu_h(t),p_h(t))$} has been omitted.
The scaling~$\sigma$ denotes a (pseudo-)reaction which results from temporal discretization
as $\sigma=1/(\theta\Delta t)$ for the one-step-$\theta$ scheme.
Stabilization parameters for the \name{GP} terms are
taken from \cite{SchottWall2014,MassingSchottWall2016_CMAME_Arxiv_submit} and the incorporated scaling functions are
\begin{align}
  \label{eq:cip-s_scalings_recalled_unfitted}
  \phi_T(\bfc_h)= \nu^{\fd} + c_u (\|\bfc_h\|_{0,\infty,T} h_T) + c_{\sigma} (\sigma h_T^2), \quad \phi_{c,T}=\phi_{p,T}=h_T^2 \phi_T^{-1} , \quad \phi_{u,T}=\phi_T.
\end{align}
Further, $\jump{\cdot}$ denotes the jump of quantities across interior facets $F$ and $\partial^j_{\bfn}$ denotes the normal derivative of order~$j$.
The subscript~$(\cdot)_F$ in stabilization scalings indicates to take the mean over quantities from both adjacent elements~$T$.
\end{definition}

\begin{remark}[Notes on fitted approximations]
Note that the classical \name{RBVM} stabilized method
with $\mcG_h^{\mathrm{GP}}\equiv 0$ in \Eqref{eq:chap_3_Nitsche_RBVM_GP_CUTFEM_form_Ah}
is sufficient to control incompressible flow
on boundary-fitted approximations.
\end{remark}

\begin{remark}[Notes on unfitted mesh approximations]
\label{rem:notes_on_unfitted_approximations}
Note that for consistency reasons the RBVM terms in \Eqref{eq:chap_3_Nitsche_RBVM_GP_CUTFEM_form}
are evaluated just on the physical mesh part, \ie~in $T\cap\Domh^\fd$.
\end{remark}

\begin{remark}[Ghost penalty stabilization for unfitted approximations]
It is well-known that a major challenge in translating a fitted finite element formulation into its cut finite element version is to
maintain the stability and approximation properties of the underlying scheme irrespective of how the boundary of the
subdomain cuts the background mesh.
To overcome issues with regards to system conditioning, stability and optimality independent of the mesh intersection,
so-called ghost-penalties \cite{BeckerBurmanHansbo2009, Burman2010, BurmanHansbo2012, Burman2014b}
consisting of weakly consistent jump penalties of order~$k$ are applied;
see the works by Massing~\etal~\cite{MassingSchottWall2016_CMAME_Arxiv_submit} and Schott and Wall~\cite{SchottWall2014} for more detailed
mathematical explanations on the numerical analysis
and studies on the proposed ghost-penalty terms comprised in $\mcG_h^{\mathrm{GP}}$ \Eqref{eq:chap_3_Nitsche_CIP_GP_CUTFEM_form_Gh}.
Applications successfully utilizing this technique can be found, \eg, in \cite{SchottShahmiriKruseWall2015, SchottRasthoferGravemeierWall2015}.
\end{remark}

\begin{remark}[Other types of fluid stabilizations]
The standard Galerkin discrete formulation would suffer from different instabilities in the interior of the approximative meshes.
First, schemes with equal-order interpolation spaces $\mcW_h = \mcV_h\times \mcQ_h$ do not satisfy an inf-sup condition
in the sense of Babuska--Brezzi. Second, spurious oscillations in the numerical solution can arise
yielding sub-optimal error estimates at high Reynolds numbers.
As an alternative to the chosen \name{RBVM} stabilization concept, also other types of fluid stabilizations (see \cite{BraackBurmanJohnEtAl2007} for an overview)
might show advantages in the context of \name{CutFEM},
see discussion in \cite{SchottShahmiriKruseWall2015,SchottRasthoferGravemeierWall2015}.

For a recent analysis of a CutFEM for flow problems based on the continuous interior penalty \name{CIP} method, the interested reader is referred to
\cite{MassingSchottWall2016_CMAME_Arxiv_submit}. The \name{CIP}/\name{GP}-\name{CutFEM} formulation reads
\begin{equation}
\label{eq:chap_3_Nitsche_CIP_GP_CUTFEM_form}
   \mcA_h^{\fd,\mathrm{CIP/GP}}(U_h,V_h) = \mcL_h^{\fd,\mathrm{CIP/GP}}(U_h,V_h) \quad \foralls V_h=(\bfv_h,q_h)\in\mcV_{\bfzero,h}\times\mcQ_h,
\end{equation}
where
\begin{align}
   \mcA_h^{\fd,\mathrm{CIP/GP}}(U_h,V_h)
          &\eqdef (\rho^{\fd}\dTime{\bfu_{\bfChi,h}}\circ \bfPhi^{-1}, \bfv_h) + (\mcB_h + \mcS_h^{\mathrm{CIP}} + \mcG_h^{\mathrm{GP}})(\bfu_h-\hat{\bfu}_{h};(\bfu_h,p_h), (\bfv_h,q_h))
,\nonumber\\
    \mcL_h^{\fd,\mathrm{CIP/GP}}(U_h,V_h)
	  &\eqdef \mcL_h(\bfv_h)
\end{align}
and $\mcS_h^{\mathrm{CIP}}$ a weakly consistent \name{CIP} stabilization operator from, \eg, \cite{BurmanFernandezHansbo2006, MassingSchottWall2016_CMAME_Arxiv_submit}.
\end{remark}

\begin{remark}[Weak constraint enforcement of boundary conditions using Nitsche's method]
\label{rem:Nitsche_weak_DBC}
Note that weak Dirichlet constraint enforcement techniques are often superior over strong enforcements in terms of
automated relaxation of constraints for the benefit of reduced errors in the vicinity of under-resolved boundary layers.
Such schemes are of particular importance for unfitted mesh approximations, where grid nodes do not match the domain boundary.
Even though not presented and utilized in the present work,
additional Nitsche terms can be consistently build in the fluid formulation to incorporate the boundary constraint $\bfu=\bfg_D^{\fd}$ weakly.
For details on this technique, the reader is referred to, \eg, \cite{Bazilevs2007, BurmanFernandezHansbo2006, MassingSchottWall2016_CMAME_Arxiv_submit}.
\end{remark}

\subsection{Nitsche-type Fluid Domain Decomposition}
\label{ssec:fluid_domain_decomposition_stab_form}

Approach~\ref{approach:4_FXFSI} makes use of an artificial decomposition of the fluid domain
into disjoint subregions
\mbox{$\Dom_{h}^{\fd} \eqdef \bigcup_{1\leqslant i \leqslant l}\Dom_h^{\fd_i}$},
which are then approximated with either boundary-fitted or non-interface-fitted finite element approximation $\mcW_{h}^{\fd_i}$ such that
\mbox{$\mcW_{h}^{\fd} \eqdef \oplus_{i=1}^{l}{\mcW_h^{\fd_i}}$},
see also Remark~\ref{rem:need_for_fluid_domain_decomposition}.
In the following, a spatial semi-discrete formulation for $l$ fluid subdomains is recalled.
The method utilizes subregion approximations as proposed in Definition~\ref{def:RBVM_GP_CUTFEM_Navier_Stokes}.
The respective approximations may exhibit possibly different element types, shapes, characteristic mesh sizes and polynomial orders.
The subregions are finally coupled via a Nitsche-type method proposed in \cite{SchottShahmiriKruseWall2015}.

\begin{definition}[Semi-discrete Nitsche-type \name{CutFEM} for coupled flows]
	Let \mbox{$\mcW_{\bfgD,h}^{\fd} \eqdef \oplus_{i=1}^{l}\mcW_{\bfgD,h}^{\fd_i}$} be the composed discrete approximation space for velocity and pressure
associated with the different fluid subregions in possibly time-dependent subdomains~$\Domh^i$
	and $\mcW_{\bfgD,h}^{\fd_i}\eqdef\mcV_{\bfgD,h}^i\times \mcQ_h^i$,
then a Nitsche-type stabilized formulation for
coupling incompressible flows reads as follows:
	for all \mbox{$t\in(T_0,T]$}, find fluid velocity and pressure \mbox{$U_h(t)=(\bfu_h(t),p_h(t))\in\mcW_{\bfgD,h}^{\fd}$} with
\begin{equation}
\label{eq:chap_4_Nitsche_CIP_GP_CUTFEM_form_fluidfluid_1}
	\widetilde{\mcA}_h^{\fd}(U_h,V_h) = \widetilde{\mcL}_h^{\fd}(U_h,V_h) \quad \foralls V_h=(\bfv_h,q_h)\in\mcW_{\bfzero,h}^{\fd},
\end{equation}
where
\begin{align}
 \widetilde{\mcA}_h^{\fd}(U_h,V_h) &\eqdef
 \sum\nolimits_{1\leqslant i \leqslant N_{\textrm{dom}}}
 {
   \left(\mcA_h^{i,\mathrm{RBVM/GP}}(U_h,V_h) + \sum\nolimits_{j>i}{\mcC_h^{ij}(\bfu_h;(\bfu_h,p_h), (\bfv_h,q_h))}\right)
 }, \\
\widetilde{\mcL}_h^{\fd}(U_h,V_h) &\eqdef
\sum\nolimits_{1\leqslant i \leqslant N_{\textrm{dom}}}
 {
   \left(\mcL_h^{i,\mathrm{RBVM/GP}}(U_h,V_h) \right)
 },
\end{align}
consisting of stabilized single-phase formulations \mbox{$\mcA_h^{i,\mathrm{RBVM/GP}}-\mcL_h^{i,\mathrm{RBVM/GP}}$} for each subregion.
The superscript~$(\cdot)^i$ indicates that included quantities like domains, interfaces, integrals, element- and face-sets
and functions \mbox{$(\bfv_h^i,q_h^i)\in\mcW_h^i$} belong to the respective approximation space.

Conserving mass and momentum at the artificial interfaces $\Intij$ requires to weakly enforce
\begin{alignat}{2}
\label{eq:gov_eq_Fluid_Fluid_strong_form_5}
 \jump{\bfu} = \bfu^i -\bfu^j   &= \bfzero                                                             \qquad &&\foralls \bfx \in \Intij(t), \\
\label{eq:gov_eq_Fluid_Fluid_strong_form_6}
 \jump{\bfsigma(\bfu,p)}\cdot\bfn^{ij} = (\bfsigma(\bfu^i,p^i) - \bfsigma(\bfu^j,p^j)) \cdot\bfn^{ij} &= \bfzero   \qquad &&\foralls \bfx \in \Intij(t)
\end{alignat}
by incorporating consistent Nitsche-type coupling terms
\mbox{$\mcC_h^{ij}$} with
\begin{flalign}
   \mcC_h^{ij}(\bfu_h;(\bfu_h,p_h), (\bfv_h,q_h))
         &=
         - \langle \wavg{2\mu^{\fd}\bfepsilon(\bfu_h)}\bfn^{ij}, \jump{\bfv_h} \rangle_{\Intij}
	 + \langle \wavg{p_h}, \jump{\bfv_h}\cdot\bfn^{ij} \rangle_{\Intij}
\label{eq:chap_4_Nitsche_CIP_GP_CUTFEM_form_fluidfluid_form_Chij_1}  \\
	 &\quad
         \mp \langle \jump{\bfu_h}, \wavg{2\mu^{\fd}\bfepsilon(\bfv_h)}\bfn^{ij} \rangle_{\Intij}
         -   \langle \jump{\bfu_h}\cdot\bfn^{ij}, \wavg{q_h} \rangle_{\Intij}  
\label{eq:chap_4_Nitsche_CIP_GP_CUTFEM_form_fluidfluid_form_Chij_2} \\
         & \quad
         + \langle \gamma(\wavg{\varphi}/2)\jump{\bfu_h},\jump{\bfv_h} \rangle_{\Intij}
\label{eq:chap_4_Nitsche_CIP_GP_CUTFEM_form_fluidfluid_form_Chij_3} \\
	 & \quad
         + \langle \gamma(\wavg{\rho^{\fd}\phi/h}/2)\jump{\bfu_h}\cdot\bfn^{ij}, \jump{\bfv_h}\cdot\bfn^{ij} \rangle_{\Intij}
\label{eq:chap_4_Nitsche_CIP_GP_CUTFEM_form_fluidfluid_form_Chij_4}  \\
         & \quad
	 + \langle (\meanavg{\rho^{\fd}\bfu_h}\cdot\bfn^{ij})\jump{\bfu_h},\meanavg{\bfv_h} \rangle_{\Intij}
\label{eq:chap_4_Nitsche_CIP_GP_CUTFEM_form_fluidfluid_form_Chij_5} \\
         & \quad
	 +  \langle 1/2 |(\meanavg{\rho^{\fd}\bfu_h}\cdot\bfn^{ij})|\jump{\bfu_h},\jump{\bfv_h} \rangle_{\Intij}.
\label{eq:chap_4_Nitsche_CIP_GP_CUTFEM_form_fluidfluid_form_Chij_6}
\end{flalign}
Since the coupling constraints \Eqref{eq:gov_eq_Fluid_Fluid_strong_form_5}--\eqref{eq:gov_eq_Fluid_Fluid_strong_form_6} are homogeneous, no right-hand-side contributions are present.
\end{definition}

\begin{remark}[Nitsche-type interface coupling]
\label{rem:Nitsche_type_interface_coupling}
Detailed explanations of Nitsche's formulation can be found in \cite{SchottRasthoferGravemeierWall2015} and \cite{SchottShahmiriKruseWall2015}.
Note that the sign choice~($\mp$) allows to switch between adjoint-consistent~$(-)$
and adjoint-inconsistent~$(+)$ viscous parts -- different variants which come along with substantial changes of the stability properties and restrictions on the choice of~$\gamma$
and potential differences in the $L^2$ error estimates for the velocity.
The adjoint-inconsistent variant is used for all simulations provided in this work.
The scaling~$\phi$ occurring in the latter term is the stabilization scaling from \Eqref{eq:cip-s_scalings_recalled_unfitted} and accounts for the different flow regimes,
see also \cite{MassingSchottWall2016_CMAME_Arxiv_submit}.
All terms together enable to cope with instabilities arising from the convective terms, to control mass transport across interfaces
and to deal with issues due to the discontinuity of the convective velocity~$\bfu_h$.
\end{remark}

\begin{remark}[The role of flux averaging]
\label{rem:Nitsche_type_weighting_strategy}
For details on the importance of properly chosen average weights occurring in the weighted average operators~\mbox{$\wavg{\cdot},\wavginv{\cdot}$},
the interested reader is referred to \cite{SchottShahmiriKruseWall2015}.
Following the latter publication, in this work, we define the average weights
as $w^{\fd_2} = 1- w^{\fd_1} \eqdef 1$, 
resulting in a flux weighting with respect to the uncut fluid mesh~$\mcT_h^{\fd_2}$ at $\Int^{\fd_1\fd_2}$.
This choice is advantageous from a stability point of view, see elaborations in \cite{SchottShahmiriKruseWall2015}.
Moreover, the scaling $\varphi$ is defined as \mbox{$\varphi^k\eqdef\mu^{\fd} (f^k)^2$}, where $f^k$ is obtained from a weakened trace inequality
\begin{equation}
\label{eq:chap_3_trace_inequality_unfitted_mesh_full_triangle}
 \norm{\nabla v_h \cdot \bfn}_{0,\Intij \cap T^k} \lesssim \norm{\nabla v_h \cdot \bfn}_{0, \partial T^k} \leqslant f^k \norm{\nabla v_h}_{0,T^k}
\end{equation}
which scales as \mbox{$(f^k)^2 \approx 1/h^k$} for all interfacial elements $T^k\in\mcT_h^k$ with subdomain index $k\in\{i,j\}$,
however accounts for element distortion, polynomial degree and element type, see \cite{SchottShahmiriKruseWall2015, BurmanErn2007}.
Further, $\meanavg{\cdot}$ in \Eqref{eq:chap_4_Nitsche_CIP_GP_CUTFEM_form_fluidfluid_form_Chij_6} denotes the mean average with \mbox{$w^i=w^j=1/2$}.
\end{remark}

\begin{remark}[Eulerian and ALE fluid mesh approximations]
To enable the different Approaches~\ref{approach:1_ALE-FSI}--\ref{approach:4_FXFSI},
it is possible/required to use ALE and Eulerian techniques for the single fluid meshes which allows them to displace or remain fixed over time, respectively.
As an extension to \cite{SchottShahmiriKruseWall2015},
to allow respective meshes to move, they get equipped with an ALE-field with according changes in the nonlinear convective term,
see Sections~\ref{sssec:strong_formulation_fluid_mechanics}--\ref{sssec:strong_formulation_ALE_mechanics}.
All meshes are allowed to move independently as long as an overlap is ensured.
Due to potential mesh motion, intersection related quantities and the cut-cell integration need to be updated accordingly.
Algorithmic details and temporal discretization steps which address issues arising from changing function spaces will be provided in \Secref{ssec:changing_functions_spaces}.
\end{remark}

\subsection{Nitsche-type Coupling for Fluids and Structures}
\label{ssec:fsi_stab_form}

Depending on the FSI approach introduced in \Secref{ssec:powerful_fitted_and_unfitted_discretization_concepts},
for the fluid function space, either single-mesh approximations (see \Secref{ssec:stabilized_fitted_unfitted_FEM_fluids}) or composite techniques (see \Secref{ssec:fluid_domain_decomposition_stab_form})
can be applied, comprised in an associated function space \mbox{$\mcW_{\bfgD,h}^\fd\eqdef\mcV_{\bfgD,h}\times\mcQ_h$}.
Structural governing equations are formulated in a Lagrangean description
and related approximation spaces for displacement and velocity
\mbox{$\mcW_{\bfgD,h}^\sd\eqdef\mcD_{\bfgD,h} \times \mcD_h $} as well as test spaces
\mbox{$\mcW_{\bfzero,h}^\sd \eqdef \mcD_{\bfzero,h}$}
are based on interface-fitted triangulations~$\mcT_h^\sd$,
as described in \Secref{ssec:fitted_FEM_structures}.
Note, Dirichlet boundary conditions are assumed being imposed strongly via the function space.
Due to the potential unfittedness of the meshes at the fluid-solid interface~$\Intfs$, \ie~when \mbox{$\mcT_h^\fd\subset \widehat{\mcT}_h^\fd$}
with \mbox{$\Domh^\fd\subsetneq \Domh^{\fd\ast}$}, coupling constraints are enforced weakly.
The spatial semi-discrete formulation for the coupled fluid-structure system \Eqref{eq:gov_eq_Fluid_Solid_weak_form}
follows the coupling of fluid subdomains by analogy,
as provided in \Secref{ssec:fluid_domain_decomposition_stab_form}.

We first present the coupling between one solid and one fluid mesh (as used for Approaches~\ref{approach:1_ALE-FSI}--\ref{approach:3_ALE-XFSI}), and afterwards extend the formulation
to take fluid domain decomposition into account
(as used for Approach~\ref{approach:4_FXFSI}). 

\paragraph*{Nitsche-type fluid-structure coupling with one fluid mesh.}

\begin{definition}[Semi-discrete Nitsche-type formulation for fluid-structure interaction]
	Let \mbox{$\mcW_{\bfgD,h} \eqdef \mcW_{\bfgD,h}^\fd \oplus \mcW_{\bfgD,h}^\sd$} be the admissible space for discrete FSI solutions,
then a Nitsche-type stabilized formulation for the FSI problem setting reads as follows:
	for all \mbox{$t\in(T_0,T]$}, find fluid velocity and pressure \mbox{$U_h(t)=(\bfu_h(t),p_h(t))\in\mcW_{\bfgD,h}^\fd$}
as well as solid displacement and velocity \mbox{$D_h(t)=(\bfd_h(t),\dtimedot{\bfd}_h(t))\in\mcW_{\bfgD,h}^\sd$}
	such that \mbox{$\foralls(V_h,W_h)=(\bfv_h,q_h,\bfw_h)\in\mcW_{\bfzero,h}^\fd \oplus \mcW_{\bfzero,h}^\sd$}
\begin{flalign}
\label{eq:chap_4_Nitsche_CIP_GP_CUTFEM_form_fluidsolid_1}
   \mcA_h^{\fd,\mathrm{RBVM/GP}}(U_h,V_h) + \mcA_h^{\sd}(D_h,W_h) + \mcC_h^{\fs}((U_h,D_h),(V_h,W_h)) \nonumber\\
        = \mcL_h^{\fd,\mathrm{RBVM/GP}}(U_h,V_h) + \mcL_h^{\sd}(W_h),
\end{flalign}
where the coupled formulation includes the stabilized single-mesh fluid formulation
\mbox{$\mcA_h^{\fd,\mathrm{RBVM/GP}}-\mcL_h^{\fd,\mathrm{RBVM/GP}}$}.
The incorporated form \mbox{$\mcA_h^{\sd}-\mcL_h^{\sd}$} denotes the structural discrete formulation proposed in \Eqref{eq:gov_eq_Solid_weak_form_discrete}
evaluated in the structural subdomain.

Nitsche-type fluid-solid coupling terms at~$\Intfs$ are comprised in the operator~$\mcC_h^{\fs}$,
similar to the couplings \Eqref{eq:chap_4_Nitsche_CIP_GP_CUTFEM_form_fluidfluid_form_Chij_1}--\eqref{eq:chap_4_Nitsche_CIP_GP_CUTFEM_form_fluidfluid_form_Chij_6}
between two fluid subdomains,
\begin{flalign}
 \mcC_h^{\fs}((U_h,D_h),(V_h,W_h))
         &=
         - \langle 2\mu^{\fd}\bfepsilon(\bfu_h)\bfn^{\fs}, \jump{\bfv_h} \rangle_{\Intfs}
	 + \langle p_h, \jump{\bfv_h}\cdot\bfn^{\fs} \rangle_{\Intfs}
\label{eq:chap_5_Nitsche_CIP_GP_CUTFEM_form_fluidsolid_form_Chfs_1}  \\
	 &\quad
         \mp \langle \jump{\bfu_h}, 2\mu^{\fd}\bfepsilon(\bfv_h)\bfn^{\fs} \rangle_{\Intfs}
         -   \langle \jump{\bfu_h}\cdot\bfn^{\fs}, q_h \rangle_{\Intfs}  
\label{eq:chap_5_Nitsche_CIP_GP_CUTFEM_form_fluidsolid_form_Chfs_2} \\
         & \quad
         + \langle \gamma(\mu/h)\jump{\bfu_h},\jump{\bfv_h} \rangle_{\Intfs}
\label{eq:chap_5_Nitsche_CIP_GP_CUTFEM_form_fluidsolid_form_Chfs_3}
         + \langle \gamma(\rho^\fd\phi/h)\jump{\bfu_h}\cdot\bfn^{\fs}, \jump{\bfv_h}\cdot\bfn^{\fs} \rangle_{\Intfs}
\end{flalign}
with \mbox{$\jump{\bfu_h} = \bfu_h^\fd - \bfu_h^\sd =  \bfu_h - \dtimedot{\bfd}_h \circ \bfvarphi_t^{-1}$} and
\mbox{$\jump{\bfv_h} = \bfv_h^\fd - \bfv_h^\sd = \bfv_h - \bfw_h\circ \bfvarphi_t^{-1}$}.
It needs to be pointed out that for the homogeneous constraints \mbox{$\bfg_\Int^{\fs}=\bfzero$} and \mbox{$\bfh_\Int^{\fs}=\bfzero$}, all related right-hand-side terms vanish.
\end{definition}

\begin{remark}[Fluid-sided average weighting]
In contrast to the proposed Nitsche coupling between two fluids \Eqref{eq:chap_4_Nitsche_CIP_GP_CUTFEM_form_fluidfluid_form_Chij_1}--\eqref{eq:chap_4_Nitsche_CIP_GP_CUTFEM_form_fluidfluid_form_Chij_6}, at fluid-solid interfaces the average weights are chosen as \mbox{$w^\fd=1-w^\sd=1$},
which renders in a fluid-sided flux weighting strategy. This choice simplifies ensuring stability for couplings with non-linear
elastic structural materials. Since throughout this work fluids are assumed being less viscous/stiff than solids, these weights conform to the
optimality of harmonic weights when high contrast in material properties are present,
see discussion in, \eg, \cite{Burman2012b}.
As a result of this choice, material parameters occurring in stabilization scalings of \Eqref{eq:chap_5_Nitsche_CIP_GP_CUTFEM_form_fluidsolid_form_Chfs_3}
belong to the fluid phase only.
\end{remark}

\begin{remark}
Note that for unfitted mesh couplings, ghost penalties \eqref{eq:chap_3_Nitsche_CIP_GP_CUTFEM_form_Gh} are required also for facets located next to fluid elements
$T\in\mcT_{\Intfs}^{\fd}$ that are crossed by $\Intfs$, see definition~\eqref{notation:p2:eq:define-cutting-cell-mesh}.
\end{remark}

\paragraph*{Extension of Nitsche-type fluid-structure coupling to a composite of fluid meshes.}

Let the space of admissible discrete solutions be denoted with
\mbox{$\mcW_{\bfgD,h} \eqdef (\oplus_{i=1}^{l}{\mcW_{\bfgD,h}^{\fd_i}}) \oplus \mcW_{\bfgD,h}^\sd$},
consisting of \mbox{$l= N_{\textrm{dom}}-1$} fluid subdomain spaces and,
without loss of generality, one structural function space.
The subspaces are as defined in Sections~\ref{ssec:fluid_domain_decomposition_stab_form}
and \ref{ssec:fsi_stab_form}.
The Nitsche-type coupled stabilized formulation for this multidomain \name{FSI} problem setting reads as follows:
for all \mbox{$t\in(T_0,T]$}, find fluid velocity and pressure approximations
\mbox{$U_h(t)=(\bfu_h(t),p_h(t))\in (\oplus_{i=1}^{l}{\mcW_{\bfgD,h}^{\fd_i}})$},
where \mbox{$U_h(t)|_{\Dom^i} = U_h^i(t)\in \mcW_{\bfgD,h}^{\fd_i}$},
and solid displacement and velocity \mbox{$D_h(t)=(\bfd_h(t),\dtimedot{\bfd}_h(t))\in\mcW_{\bfgD,h}^\sd$}
such that for all \mbox{$(V_h,W_h)\in(\oplus_{i=1}^{l}\mcW_{\bfzero,h}^{\fd_i}) \oplus \mcW_{\bfzero,h}^\sd$}
\begin{equation}
\label{eq:chap_5_Nitsche_CIP_GP_CUTFEM_form_multiplefluidsolid_1}
   \widetilde{\mcA}_h^{\textrm{FSI}}((U_h,D_h),(V_h,W_h)) = \widetilde{\mcL}_h^{\textrm{FSI}}(U_h,(V_h,W_h)),
\end{equation}
where
\begin{align}
 &\widetilde{\mcA}_h^{\textrm{FSI}}((U_h,D_h),(V_h,W_h)) \eqdef \nonumber\\
 &\sum_{1\leqslant i \leqslant l} 
 {
  \Big(\mcA_h^{i,\mathrm{RBVM/GP}}(U_h,V_h) + \sum_{l\geqslant j>i}{\mcC_h^{ij}(U_h, V_h)}  + \mcC_h^{i\textrm{s}}((U_h,D_h),(V_h,W_h)) \Big)
 }  + \mcA_h^{\sd}(D_h,W_h), \\
&\widetilde{\mcL}_h^{\textrm{FSI}}(U_h,V_h) \eqdef
\sum_{1\leqslant i \leqslant l}
 {
   \Big(\mcL_h^{i,\mathrm{RBVM/GP}}(U_h,(V_h,W_h)) + \sum_{l\geqslant j>i}{\mcL_h^{ij}(U_h, V_h)} \Big)
 }
+ \mcL_h^{\sd}(W_h)
\end{align}
with operators \mbox{$\mcA_h^{i,\mathrm{RBVM/GP}}-\mcL_h^{i,\mathrm{RBVM/GP}}$} for the single fluid phases,
see \Secref{ssec:stabilized_fitted_unfitted_FEM_fluids},
respective fluid-fluid Nitsche-type couplings~\mbox{$\mcC_h^{ij}$}
from \Eqref{eq:chap_4_Nitsche_CIP_GP_CUTFEM_form_fluidfluid_form_Chij_1} 
and the structural variational form \mbox{$\mcA_h^{\sd}-\mcL_h^{\sd}$} \Eqref{eq:gov_eq_Solid_weak_form_discrete}.
The latter is coupled to the fluid phases with Nitsche-type couplings~$\mcC_h^{i\textrm{s}}$
\Eqref{eq:chap_5_Nitsche_CIP_GP_CUTFEM_form_fluidsolid_form_Chfs_1}--\eqref{eq:chap_5_Nitsche_CIP_GP_CUTFEM_form_fluidsolid_form_Chfs_3}.

\begin{remark}
Note that for unfitted fluid domain decomposition, ghost penalties \eqref{eq:chap_3_Nitsche_CIP_GP_CUTFEM_form_Gh} have to be applied for facets located next to fluid elements
$T\in\mcT_{\Int^{12}}^{\fd_1}$ that are crossed by $\Int^{12}$, see definition~\eqref{notation:p2:eq:define-cutting-cell-mesh}.
More detailed elaborations on their need depending on the Nitsche weighting strategy can be found in \cite{SchottShahmiriKruseWall2015}.
\end{remark}

\section{Monolithic Solution Strategies for Time-Discrete Coupled FSI Systems}
\label{sec:fsi:monolithic}

This section aims at formulating a fully discrete finite-dimensional system of non-linear equations for the Nitsche-coupled FSI
systems \Eqref{eq:chap_4_Nitsche_CIP_GP_CUTFEM_form_fluidsolid_1} and \Eqref{eq:chap_5_Nitsche_CIP_GP_CUTFEM_form_multiplefluidsolid_1}
which is to be solved for all discrete time levels.
Note that, if unmistakable, in the following we omit superscripts $(\cdot)^{\fd},(\cdot)^{\sd}$ specifying fluid and solid subdomain quantities to shorten the presentation.

\subsection{Time Stepping for Nitsche-Coupled Fluid-Structure Systems}
\label{ssec:fsi:stabilized_Nitsche_formulation:timestepping}

Let the time domain~\mbox{$(T_0, T]$} be partitioned into~$N$ equal-sized time step intervals \mbox{$J^n=(t^{n-1},t^n]$}
of size $\Delta t$ with discrete time levels \mbox{$t^n = T_0 + n\Delta t$} and \mbox{$t^N = T$}.
For the temporal discretization of the coupled system, different single-field time-stepping schemes can be utilized.
Subsequently, a discrete coupled formulation is exemplarily provided for the combination of a Generalized\hbox{-}$\alpha$~method for the structural elastodynamics
equations \mbox{$\mcA_h^{\sd}-\mcL_h^{\sd}$} and a one-step-$\theta$ scheme for the stabilized fluid formulation
\mbox{$\mcA_h^{\fd,\mathrm{RBVM/GP}}-\mcL_h^{\fd,\mathrm{RBVM/GP}}$}.
Neglecting the Nitsche couplings~$\mcC_h^{\fs}$ in \Eqref{eq:chap_4_Nitsche_CIP_GP_CUTFEM_form_fluidsolid_1} for a moment,
the temporally discretized non-linear decoupled systems for fluids and solids, \mbox{$\bfR^\fd(\bfU^n,\bfP^n)$} and \mbox{$\bfR^\sd(\bfD^n)$},
can be written as
\begin{align}
\label{eq:chap_5_fsi_fluid_residual}
 \bfR^\fd(\bfU^n,\bfP^n) & = \bfM^{\fd,n} (\bfU^n,\bfP^n) + \sigma^{-1} \bfF^{\fd,n}(\bfU^n,\bfP^n)  
- \bfH^{\fd,n-1}(\tilde{\bfU}^{n-1},\tilde{\bfA}^{n-1}), \\ 
\label{eq:chap_5_fsi_structural_residual}
 \bfR^\sd(\bfD^n)        & 
 = \bfM^\sd \frac{1-\alpha_m}{\beta \Delta t^2}  \bfD^{n} + (1-\alpha_f)(\bfF^{\sd,n}_{\textrm{int}}(\bfD^{n}) - \bfF^{\sd,n}_{\textrm{ext}}) \nonumber\\
 & \quad - \bfH^{\sd,n-1}(\bfD^{n-1},\bfU^{n-1},\bfA^{n-1}).
\end{align}

Following \cite{Forster2009} and \cite{Dettmer2003}, the fluid residual $\bfR^\fd$ \Eqref{eq:chap_5_fsi_fluid_residual} contains the matrix~$\bfM^{\fd,n}$ resulting from the time derivative term $(\rho^{\fd}\bfu_h^{n},\Upsilon(\bfv_h,q_h))$
with $\Upsilon$ depending on the type of fluid stabilization.
For the \name{RBVM/GP}-\name{CutFEM} this function is given by $\Upsilon(\bfv_h,q_h)=\bfv_h + \tau_\mathrm{M} (\rho^{\fd}((\bfu_h-\hat{\bfu}_{\bfChi,h}) \cdot\nabla) \bfv_h + \nabla q_h)$
and \mbox{$\Upsilon(\bfv_h,q_h)=\bfv_h$} for the \name{CIP/GP}-\name{CutFEM} formulation.
The function~$\bfF^{\fd,n}$ comprises all operators to be evaluated at time level~$t^n$, \ie, all standard Galerkin terms,
stabilization operators, external loads and Nitsche terms according to the weak imposition of boundary conditions,
\ie~\mbox{$\mcB_h,\mcG_h^{\textrm{GP}},\mcS_h,-\mcL_h$} and does not include further time derivatives of~$\bfu_h$.
Terms belonging to the previous time levels are comprised in~$\bfH^{\fd,n-1}$.
For the OST-scheme it is defined
\begin{equation}
\label{eq:chap_3_OST_pseudo_reaction}
 \sigma_\theta = (\theta \Delta t)^{-1} \quad \text{ and } \quad \bfH^{\fd,n-1}(\bfv_h) \Leftrightarrow (\rho^{\fd} \tilde{\bfu}_h^{n-1} ,\Upsilon(\bfv_h,q_h)) + (1-\theta)\Delta t (\rho^{\fd}\tilde{\bfa}_h^{n-1}, \bfv_h).
\end{equation}
For computed $\bfu_h^{n}$, accelerations can be updated by utilizing a \name{OST}-scheme for \mbox{$\partial_t{\bfu_h(t)} = \bfa_h(t)$}
\begin{equation}
\label{eq:chap_3_Nitsche_CIP_GP_CUTFEM_form_OST_moving_domain_2}
 \bfa_h^{n} = (\bfu_h^{n} - \tilde{\bfu}_h^{n-1})(\theta\Delta t)^{-1} - (1-\theta)\theta^{-1} \tilde{\bfa}_h^{n-1}.
\end{equation}
Note that using this technique, the stabilized formulations to the incompressible Navier-Stokes equations hidden in~$\bfF^{\fd,n}$
have to be evaluated only for the current time step~$t^{n}$.

\begin{remark}
\label{rem:changing_functions_spaces}
While for fitted mesh approximation, the finite-dimensional function space remains unchanged over time, which allows to define
$\tilde{\bfa}_h^{n-1}={\bfa}_h^{n-1}$ and $\tilde{\bfu}_h^{n-1}={\bfu}_h^{n-1}$ in \Eqref{eq:chap_3_Nitsche_CIP_GP_CUTFEM_form_OST_moving_domain_2},
for unfitted methods with moving domains, the function spaces might differ between two consecutive time levels,
\ie~$\mcW_h^n\neq\mcW_h^{n-1}$ and thus requires additional projection methods.
An algorithmic treatment of such issues will be addressed in \Secref{ssec:fsi:stabilized_Nitsche_formulation:monolithic_solution_algorithm}.
\end{remark}

The structural residual \Eqref{eq:chap_5_fsi_structural_residual} results from applying the family of \emph{Generalized-$\alpha$} (G\hbox{-}$\alpha$) time-stepping schemes as
established in \cite{Chung1993,Erlicher2002}.
Rewriting the discrete weak formulation of structural dynamics
\Eqref{eq:gov_eq_Solid_weak_form_discrete} in terms of global finite-dimensional vector-valued displacement
and velocity fields \mbox{$\bfD(t)\in\R^\ndof$} and \mbox{$\bfU(t)\in\R^\ndof$} yields
\mbox{$2\cdot\ndof$} non-linear ordinary differential equations (\name{ODE}s) of first order
\begin{align}
 \label{eq:gov_eq_Solid_weak_form_discrete_matrix_form_1}
 \bfM\dtimedot{\bfU} + \bfF_{\textrm{int}}(\bfD) - \bfF_{\textrm{ext}} &= \bfzero \qquad
 \text{and}\qquad\dtimedot{\bfD} - \bfU = \bfzero,
\end{align}
where~$\bfM$ is the global mass matrix, $\bfF_{\textrm{int}}$ the vector of non-linear internal forces resulting from~$a_h^\sd$ \Eqref{eq:ahs} and
$\bfF_{\textrm{ext}}$ external forces resulting from $l_h^\sd$ \Eqref{eq:lhs}.
Further, $\ndof$ denotes the total number of structural nodal degrees of freedom.

For the approximation of internal and external forces,
a generalized trapezoidal rule (\name{GTR}) is utilized and $\bfH^{\sd,n-1}$ comprises all terms evaluated at the previous time level~$t^{n-1}$.
Note that the system of first-order structural \name{ODE}s can be expressed solely in the unknown displacement vector~$\bfD^n$.
Once $\bfD^{n}$ is computed, approximations to $\bfU^{n}$ and $\bfA^{n}$ can be recovered, \viz
\begin{align}
 \label{eq:genalpha_recover_acceleration}
 \bfA^{n}(\bfD^{n}) &= [\bfD^{n} - (\bfD^{n-1} + \Delta t \bfU^{n-1} + (1/2-\beta)(\Delta t)^2 \bfA^{n-1})] \cdot (\beta (\Delta t)^2)^{-1}, \\
 \label{eq:genalpha_recover_velocity}
 \bfU^{n}(\bfA^{n}) &= \bfU^{n-1} + (1-\gamma)\Delta t \bfA^{n-1} + \gamma \Delta t \bfA^{n},
\end{align}
where
$\gamma=1/2-\alpha_m+\alpha_f$,
$\beta=1/4(1-\alpha_m+\alpha_f)^2$ with $\alpha_f=\rho_\infty/(\rho_\infty+1)$ and $\alpha_m=(2\rho_\infty-1)/(\rho_\infty+1)$
depending on the user-specified spectral radius $\rho_\infty\in[0,1]$ controlling the numerical high frequency dissipation of the second-order accurate scheme.

Expressing structural interface velocities \mbox{$\bfu_h^\sd = \dtimedot{\bfd}_h \circ \bfvarphi_t^{-1}$}
in terms of global structural displacements vectors~$\bfD_\Int\subset\bfD$,
an independent temporal discretization at the fluid-solid interface can be introduced.
Applying an independent one-step-$\theta_\Int$ scheme to this \name{ODE} yields
\begin{equation}
\label{eq:chap_5_fsi_displacement_velocity_conversion}
 \bfU_\Int^{\sd,n}(\bfD_\Int^n) = (\bfD_\Int^n-\bfD_\Int^{n-1})(\theta_\Int \Delta t)^{-1} - (1-\theta_\Int)\theta_\Int^{-1} \bfU_\Int^{\sd,n-1},
\end{equation}
where~$\bfU_\Int^{\sd,n}$ is a vector-valued approximation on the structural interface velocities
at time level~$t^n$, \ie~\mbox{$\bfu_h^\sd \circ \bfvarphi_{t^n} = \dtimedot{\bfd}_h(t^n)$}.
Denoting internal forces, which have to be in equilibrium at the fluid-solid interface~$\Intfs$,
with $\bfF^{\sd,n}_{\Intfs}$ and $\bfF^{\fd,n}_{\Intfs}$
for the solid and fluid phases, respectively, residuals can be formulated as
\begin{align}
\label{eq:chap_5_fsi_final_residuals}
 &\bfR^\fd(\bfU^n,\bfP^n) - \sigma^{-1} \bfF^{\fd,n}_{\Intfs} \quad\text{ and }\quad
\bfR^\sd(\bfD^n) - ( (1-\alpha_f)\bfF^{\sd,n}_{\Intfs} +\alpha_f \bfF^{\sd,n-1}_{\Intfs})
\end{align}
with $\sigma$ the characteristic temporal factor from \Eqref{eq:chap_3_OST_pseudo_reaction}.
The interfacial forces are replaced by numerical forces in a discrete setting.
These are given by the strongly consistent Nitsche-coupling terms
\Eqref{eq:chap_5_Nitsche_CIP_GP_CUTFEM_form_fluidsolid_form_Chfs_1}--\eqref{eq:chap_5_Nitsche_CIP_GP_CUTFEM_form_fluidsolid_form_Chfs_3},

 holds \mbox{$\bfF^{\fd,n}_{\Intfs} = -\bfC^{\fs}((\bfU^n,\bfP^n),\bfD^n)$}
and \mbox{$\bfF^{\sd,n}_{\Intfs} = -\bfC^{\mathrm{sf}}((\bfU^n,\bfP^n),\bfD^n)$}, where matrix notations correspond to
\begin{align}
\label{eq:chap_5_fsi_linearized_Nitsche_couplings_1}
 \bfC^{\fs}((\bfU^n,\bfP^n),\bfD^n)
	\quad\Leftrightarrow \quad
	  &- \langle 2\mu^{\fd}\bfepsilon(\bfu_h)\bfn^{\fs}, \bfv_h \rangle_{\Intfs} 
	   + \langle p_h, \bfv_h\cdot\bfn^{\fs} \rangle_{\Intfs} \nonumber \\
          &\mp \langle \jump{\bfu_h}, 2\mu^{\fd}\bfepsilon(\bfv_h)\bfn^{\fs} \rangle_{\Intfs} 
           -   \langle \jump{\bfu_h}\cdot\bfn^{\fs}, q_h \rangle_{\Intfs} \nonumber \\
          &+ \langle \gamma(\mu/h)\jump{\bfu_h},\bfv_h \rangle_{\Intfs} \nonumber \\
          &+ \langle \gamma(\rho^\fd\sigma h + \rho^\fd|\bfu_h^\fd| + \mu/h)\jump{\bfu_h}\cdot\bfn^{\fs}, \bfv_h\cdot\bfn^{\fs} \rangle_{\Intfs}, \\
\label{eq:chap_5_fsi_linearized_Nitsche_couplings_2}
 \bfC^{\mathrm{sf}}((\bfU^n,\bfP^n),\bfD^n)
	\quad\Leftrightarrow \quad
	  &- \langle 2\mu^{\fd}\bfepsilon(\bfu_h)\bfn^{\fs}, (-\bfw_h) \rangle_{\Intfs} 
	   + \langle p_h, (-\bfw_h)\cdot\bfn^{\fs} \rangle_{\Intfs}  \nonumber\\
          &+ \langle \gamma(\mu/h)\jump{\bfu_h},(-\bfw_h) \rangle_{\Intfs} \nonumber\\
          &+ \langle \gamma(\rho^\fd\sigma h + \rho^\fd|\bfu_h^\fd| + \mu/h)\jump{\bfu_h}\cdot\bfn^{\fs}, (-\bfw_h)\cdot\bfn^{\fs} \rangle_{\Intfs}.
\end{align}
These can be identified by splitting contributions from \Eqref{eq:chap_5_Nitsche_CIP_GP_CUTFEM_form_fluidsolid_form_Chfs_1}--\eqref{eq:chap_5_Nitsche_CIP_GP_CUTFEM_form_fluidsolid_form_Chfs_3} into fluid and structural residuals, \ie, with respect to $\bfv_h$ and $\bfw_h$.
Furthermore, the structural interface force from the previous time level~$t^{n-1}$ occurring in \Eqref{eq:chap_5_fsi_final_residuals}
can be recovered from the respective structural coupling matrix-vector product
\begin{align}
\label{eq:chap_5_fsi_structural_force_previous_time_step}
 \bfF^{\sd,n-1}_{\Intfs(t^{n-1})} = -\bfC^{\mathrm{sf},n-1}((\bfU^{n-1},\bfP^{n-1}),\bfD^{n-1}).
\end{align}

After rescaling the single non-linear fluid and solid residuals \Eqref{eq:chap_5_fsi_final_residuals},
the final Nitsche-coupled finite-dimensional fluid-structure system of equations, which is to be solved for the current time level~$t^n$, reads as follows:
find discrete finite-dimensional vectors \mbox{$\bfU^n, \bfP^n,\bfD^n$} such that
\begin{equation}
\label{eq:chap_5_fsi_nonlinear_residual}
\left[
 \begin{array}{c}
  \bfR_{(U,P)}\\
  \bfR_D
 \end{array}
\right]^n
=
\left[
 \begin{array}{c}
  \bfC^{\fs}((\bfU^n,\bfP^n),\bfD^n) + \sigma\bfR^\fd(\bfU^n,\bfP^n) \\
  \frac{1}{1-\alpha_f}\bfR^\sd(\bfD^n) + \bfC^{\mathrm{sf}}((\bfU^n,\bfP^n),\bfD^n) - \tfrac{\alpha_f}{1-\alpha_f} \bfF^{\sd,n-1}_{\Intfs(t^{n-1})}
 \end{array}
\right]^n
=
\left[
 \begin{array}{c}
  \bfzero \\
  \bfzero
 \end{array}
\right]
\end{equation}
with given solid approximations \mbox{$\bfD^{n-1},\bfU^{n-1},\bfA^{n-1}$} and fluid approximations \mbox{$\bfU^{n-1},\bfP^{n-1}$}
from the previous time level~$t^{n-1}$
as well as projected fluid solutions \mbox{$\tilde{\bfU}^{n-1}(\bfU^{n-1}), \tilde{\bfA}^{n-1}(\bfA^{n-1})$} with respect to the current
interface location~$\Intfs(t^{n})$, see Remark~\ref{rem:changing_functions_spaces} and subsequent elaborations in \Secref{ssec:changing_functions_spaces}.
For the sake of clarity, the fluid residual \mbox{$\bfR_{(U,P)}=[\bfR_{U},\bfR_{P}]^T$} can be further split into velocity and pressure residuals
corresponding to the momentum and continuity equations, respectively.

In analogy to \Eqref{eq:chap_5_fsi_nonlinear_residual},
the final Nitsche-type coupled finite-dimensional system of equations consisting of $l$~fluid blocks and one structure block
for a discrete time level~$t^n$ reads:
find discrete finite-dimensional vectors \mbox{$((\bfU, \bfP)^1, \ldots, (\bfU, \bfP)^l, \bfD)^n$} such that
\begin{align}
\label{eq:chap_5_fsi_nonlinear_residual_multifluid}
\left[
 \begin{array}{c}
  \bfR_{(U,P)^1} \\
  \vdots \\
  \bfR_{(U,P)^{l}}\\
  \bfR_D
 \end{array}
\right]^n
&=
\left[
 \begin{array}{c}
   \sigma\bfR^{\fd}((\bfU,\bfP)^1) + \sum\limits_{i=1}^l\bfC^{1 i}((\bfU,\bfP)^1,(\bfU,\bfP)^i) +  \bfC^{1 \sd}((\bfU,\bfP)^1,\bfD)\\
  \vdots \\
  \sigma\bfR^{{\fd}}((\bfU,\bfP)^l) + \sum\limits_{i=1}^l\bfC^{l {i}}((\bfU,\bfP)^i,(\bfU,\bfP)^l) + \bfC^{{l} \sd}((\bfU,\bfP)^l,\bfD)\\
  \frac{1}{1-\alpha_f}\bfR^\sd(\bfD) + \sum\limits_{i=1}^l\bfC^{\mathrm{s}i}((\bfU,\bfP)^i,\bfD) - \tfrac{\alpha_f}{1-\alpha_f} \bfF^{\sd,n-1}_{\Intfs(t^{n-1})}
 \end{array}
\right]^n =\bfzero
\end{align}
where \mbox{$\bfC^{i \sd},\bfC^{\sd i}$} denote the splits of fluid-structure Nitsche couplings
between fluid subdomain~$\Dom^{\fd_i}$ and the solid subdomain $\Dom^{\sd}$, 
as defined in \Eqref{eq:chap_5_fsi_linearized_Nitsche_couplings_1}--\eqref{eq:chap_5_fsi_linearized_Nitsche_couplings_2}.
Similar splits for the Nitsche coupling terms
\Eqref{eq:chap_4_Nitsche_CIP_GP_CUTFEM_form_fluidfluid_form_Chij_1}--\eqref{eq:chap_4_Nitsche_CIP_GP_CUTFEM_form_fluidfluid_form_Chij_6}
between fluid phase $\Dom^{\fd_i},\Dom^{\fd_j}$
as used for Approach~\ref{approach:4_FXFSI}
are denoted with $\bfC^{i j}$, $\bfC^{j i}$.
Furthermore, $\bfR^\sd$ denotes the structural residual and $\bfR^i$, \mbox{$i=1,\ldots,l$}, the $l$ fluid subdomain residuals without interface coupling terms.

\subsection{Treatment of Changing Function Spaces}\label{ssec:changing_functions_spaces}

As a particular challenge of unfitted approximations in moving domain flow problems, 
temporal time-stepping schemes from \mbox{$t^{n-1}\rightarrow t^n$} which are based on Rothe's technique,
consist of approximating solutions to discrete time levels $t^{n-1}$ and $t^n$ with different function spaces,
\ie~\mbox{$U^{n-1}\in\mcW^{n-1}$} and \mbox{$U^{n}\in\mcW^{n}$}, where $\mcW^{n-1}\neq\mcW^{n}$ in general. Similar issues might arise during a monolithic solution procedure
between two succeeding Newton-Raphson iteration steps.
To compensate this mismatch between two approximation spaces, projected approximations \mbox{$\tilde{\bfu}_h^{n-1}, \tilde{\bfa}_h^{n-1}\in\mcV_h^n$}
adapting \mbox{${\bfu}_h^{n-1}, {\bfa}_h^{n-1}$} to the current domain~$\Domh^n$ are necessary.
Possible projection operators \mbox{$P^n:\mcV_h^{n-1}\rightarrow\mcV_h^n$} between function spaces applicable to velocity and acceleration approximation,
\ie~\mbox{$\tilde{\bfu}_h^{n-1} \eqdef P^n {\bfu}_h^{n-1}$} and \mbox{$\tilde{\bfa}_h^{n-1} \eqdef P^n {\bfa}_h^{n-1}$}, will be introduced subsequently.

\paragraph*{Algorithmic Procedure for Function Space Projections from $\mcX_h^{n-1}$ onto $\mcX_h^{n}$.}

\begin{figure}
\centering
\includegraphics[width=8.0cm, trim=0.5cm 1cm 1.1cm 0cm, clip=true]{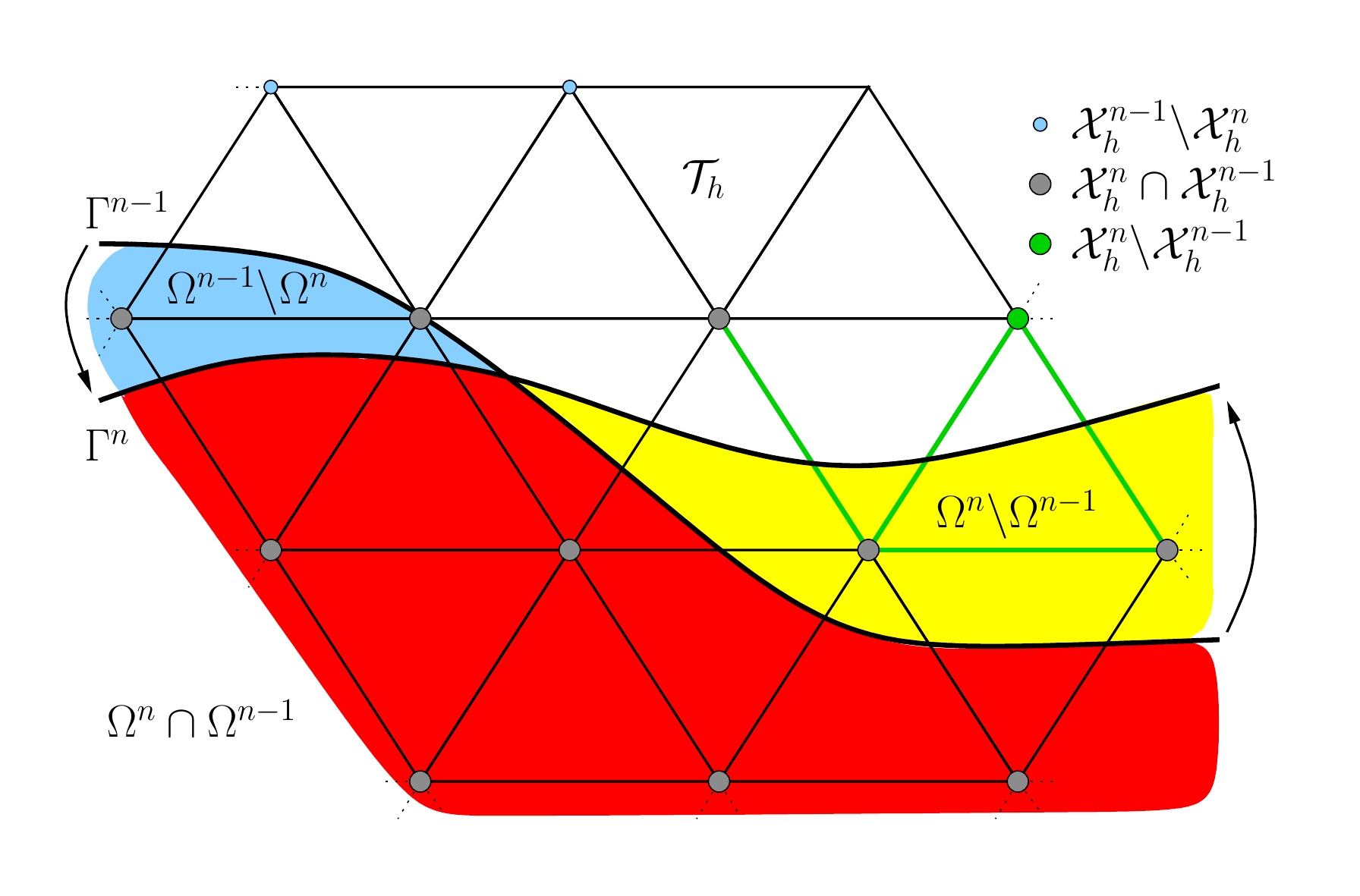}
\caption{Projecting solution functions between discrete function spaces of different time levels
with interfaces moving from~$\Int^{n-1}$ to $\Int^{n}$:
For \name{DOF}s in $\mcX_h^{n}$, pairs with \name{DOF}s of the previous time level can be identified and transcribed between $\mcX_h^{n-1}$ and $\mcX_h^{n}$~(gray markers),
some \name{DOF}s get deactivated~(blue markers) and one ghost-\name{DOF} gets newly activated whose value~$\tilde{u}$ will be determined in the \name{Extension-Phase}~(green marker).
 physically meaningful values can be copied from $\mcX_h^{n-1}$ are marked in gray color.
Faces evaluated for face-jump-penalty extension operators \Eqref{eq:chap_3_moving_domain_extension_system} are colored in green.
}
  \label{fig:moving_domain_2D_transfer_copy_phase}
  \vspace{-12pt}
\end{figure}

At each active node \mbox{$s\in \mcN$} at the current time level~$t^n$
an appropriate nodal value $\tilde{u}^{n-1}$ has to be constructed on basis of the scalar solution \mbox{$u_h^{n-1}\in\mcX_h^{n-1}$}.
The final procedure suggested for this task is summarized in \Algoref{algo:moving_domain_solution_projection} and is split into two phases:
\begin{itemize}\setlength{\itemsep}{0pt}
 \item a first \name{Transfer/Copy-Phase} and
 \item a second \name{Extension-Phase}.
\end{itemize}
\begin{algorithm}
 \caption{Function space projection from $\mcX_h^{n-1}$ to $\mcX_h^{n}$ for \name{CutFEM}s with moving domains}
 \begin{algorithmic}[1]
  \STATE \name{Input}: function spaces~$\mcX_h^{n-1}$ and $\mcX_h^{n}$ including cut-related entities, discrete scalar field \mbox{$u_h\in\mcX_h^{n-1}$} to be projected onto \mbox{$\mcX_h^{n}$}.
  \STATE \name{Transfer/Copy-Phase} (identify related node-wise \name{DOF} pairs \mbox{$(\tilde{u},u)$} associated with the function spaces \mbox{$\mcX_h^{n},\mcX_h^{n-1}$}
which are reasonable to get transcribed/copied, \ie~\mbox{$u\mapsto\tilde{u}$})
    \FOR{each active node~$s\in\mcN$ with associated \name{DOF} $\tilde{u}$ 
at current time level~$t^n$}
	    \STATE Identify \name{DOF} \mbox{$u$} from \mbox{$u_h\in\mcX_h^{n-1}$}.\\
	    \IF{(identification successful)}
		\STATE Set \name{DOF} pair \mbox{$(\tilde{u},u)$} and copy value \mbox{$u\mapsto \tilde{u}$}.\\
	    \ELSIF{(identification \NOT successful for ghost-\name{DOF}~$\tilde{u}$)}
		\STATE Mark \name{DOF}~$\tilde{u}$ for \name{Extension-Phase}.\\
	    \ELSE
		\STATE Throw exception (``the \name{CFL}-like condition is not satisfied!'').\\
	    \ENDIF
    \ENDFOR
  \STATE \name{Extension-Phase} (extend solution into the interface zone)
  \STATE Build constraint function space $\bar{\mcX}_h^n$ with Dirichlet values for all \name{DOF}s successfully identified in the \name{Transfer/Copy-Phase}.
  \STATE Solve global extension system \Eqref{eq:chap_3_moving_domain_extension_system}: Find \mbox{$\tilde{u}_h\in\bar{\mcX}_h^n$} such that
  \mbox{$\mcE_h^n(\tilde{u}_h,\tilde{v}_h) = 0\, \foralls \tilde{v}_h \in \bar{\mcX}_{h,0}^n$}.
  \STATE \name{Output}: final projection \mbox{$\bfP^n:\mcX_h^{n-1}\rightarrow \mcX_h^n$} with solution \mbox{$\tilde{u}_h^{n-1} = \bfP^n u_h^{n-1}$}. \\
 \end{algorithmic}
 \label{algo:moving_domain_solution_projection} 
\end{algorithm}
The \name{Transfer/Copy-Phase} aims at \emph{directly copying} degrees of freedom (\name{DOF}s) from the previous configuration to the current configuration.
Nodal \name{DOF}s can be kept unchanged in the following two situations:
\begin{itemize}
 \item interface motion within a cut element, non-crossing a grid node,
 \item interface motion across nodes with reduced influence region of basis functions $N^s$ and deactivated ghost nodes.
\end{itemize}
Reasonable \name{DOF}-pairs \mbox{$(\tilde{u}^{n-1},u^{n-1})$}
have to exhibit a related meaning for the solution approximation in their respective physical domains.
This decision usually relies on information about \name{DOF}-associated cut-cells located within the support of its shape function~$N_s$.
Thereby, it often occurs that \name{DOF}s change their role from standard to ghost-\name{DOF}s
and vice versa.
Common scenarios where the \name{Transfer/Copy-Phase} can be successfully applied yielding reasonable \name{DOF}-pairs are visualized in \Figref{fig:moving_domain_2D_transfer_copy_phase}.
Note that identifying \name{DOF}-pairs should be possible for all \name{DOF}s at the current time level~$t^n$
whose node is located within the physical domain~$\Domh^n$.
If the support \mbox{$\supp{(N_s)}$} for a given \name{DOF} is intersected at the current time level,
it was either fully covered by $\Omega_h^{n-1}$ or also intersected at the previous time level.
Otherwise, the $\CFL$-like condition on the interface motion has been violated.
Mathematically, the first phase covers strategies for unchanged nodes $s\in\mcX_h^{n-1}\cap\mcX_h^{n}$ (gray markers),
and nodes which get deactivated at the current configuration (blue markers).

After performing the \name{Transfer/Copy-Phase} there might remain some ghost-\name{DOF}s which have been newly activated compared to
the previous intersection configuration, \ie~such \name{DOF}s belong to $\mcX_h^{n}\backslash\mcX_h^{n-1}$ (green markers).
Missing ghost-\name{DOF}s can then be computed in a second step, the so-called \name{Extension-Phase},
which provides a localized and more accurate variant of the technique suggested in \cite{Burman2014a}.
Fixing values which have been already determined during the \name{Transfer/Copy-Phase} by setting Dirichlet constraints to $\mcX_h^n$,
whose resulting trial and test function spaces are then denoted as $\bar{\mcX}_h^n$ and $\bar{\mcX}_{h,0}^n$,
a global linear extension system can be set up as follows:
find \mbox{$\tilde{u}_h\in\bar{\mcX}_h^n$} such that
\begin{equation}
\label{eq:chap_3_moving_domain_extension_system}
  \mcE_h^n(\tilde{u}_h,\tilde{v}_h) = \sum_{F \in  \mcF_{\Int}^n}{ \sum_{0\leqslant j \leqslant k} h_F^{2j+1} \langle \jump{\nablan^j \tilde{u}_h}, \jump{\nablan^j \tilde{v}_h} \rangle_F } = 0 \qquad \foralls \tilde{v}_h \in \bar{\mcX}_{h,0}^n,
\end{equation}
which is a face-jump ghost-penalty operator similar to \eqref{eq:chap_3_Nitsche_CIP_GP_CUTFEM_form_Gh} and
thus provides a smooth continuation of the previous time step solution into the interface zone.

\begin{remark}
Even though the linear system to be solved is defined globally,
in average the number of non-zero entries per matrix row is much less
compared to those which result from the non-linear discrete systems to be solved in each time step for the incompressible Navier-Stokes equations.
Note that in practical situations only a few ghost-\name{DOF}s have to be determined via this extension technique.
All standard and ghost \name{DOF}s determined previously through the \name{Transfer/Copy-Phase} can be completely removed from the global linear system.
\end{remark}

Combining the \name{Transfer/Copy-Phase} with a consecutive \name{Extension-Phase} allows to define
the final scalar-valued projection \mbox{$\bfP^n:\mcX_h^{n-1}\rightarrow \mcX_h^n$} with \mbox{$\tilde{u}_h^{n-1} = \bfP^n u_h^{n-1}$}.
Adaption to vector-valued function spaces, \ie~sets of \name{DOF}s as required for the velocity and acceleration fields, is straightforward.
Note that splitting this procedure into two phases guarantees that the change of the solution in the interior of the domain is kept at a minimum
and is conform with the case when the interface does not move, \ie~\mbox{$u_\Int=0$}. Then the spaces $\mcX_h^{n-1}$ and $\mcX_h^n$ are identical and
the proposed projection operator simplifies to an identity mapping \mbox{$\bfP^n=\boldsymbol{I}$},
which preserves the convergence properties of a classical Newton-Raphson-like scheme and that of classical ODE based temporal discretization methods.

\subsection{Monolithic Solution Algorithm for Unfitted Non-Linear Systems}
\label{ssec:fsi:stabilized_Nitsche_formulation:monolithic_solution_algorithm}

A monolithic solution procedure for the non-linear coupled fluid-structure system is proposed subsequently.
The final algorithm is split into two nested subroutines.
An extension of the classical time loop to the unfitted mesh case, which even allows to adapt the function space
during the iterative procedure at a fixed time level~$t^n$, is summarized in \Algoref{algo:nonlinear_solution_algo_monolithic_FSI}.
The substantial solution procedure for non-linear \name{FSI} residuals consists of a Newton-Raphson-like iterative method
presented in \Algoref{algo:nonlinear_solution_algo_monolithic_FSI_Newton_Raphson_Loop}.
This main sub-algorithm is applicable to interface-fitted as well as unfitted mesh configurations
as long as the fluid function space does not change while iterating.
Note that, to shorten the presentation, the subsequent algorithms are presented for the approximation of one structural and one fluid domain,
with the corresponding non-linear residual \Eqref{eq:chap_5_fsi_nonlinear_residual}. Extension to several fluid meshes (see Approach~\ref{approach:4_FXFSI}
with non-linear residual \Eqref{eq:chap_5_fsi_nonlinear_residual_multifluid})
is straightforward by applying the fluid/ALE solver steps to the respective subdomain approximations simultaneously.

\begin{algorithm}[t]
 \caption{Monolithic solution algorithm for non-linear unfitted \name{FSI}-systems}
 \begin{algorithmic}[1]
  \STATE \name{Input}: initial conditions \mbox{$\bfd_0^\sd,\dtimedot{\bfd}_0^\sd,\bfu_0^\fd$} at $T_0$.
  \STATE Initialize structural \mbox{$\bfA^\sd_0$} for G\hbox{-}$\alpha$ scheme and set fluid \mbox{$\bfA^\fd_0$} for \name{OST} or \mbox{$\theta=1.0$} for \mbox{$n=1$}.
  \FOR{time steps~\mbox{$1\leqslant n \leqslant N$}}
     \STATE Update time level \mbox{$t^n=T_0+n\Delta t$}. Reset cycle counter to \mbox{$c=1$}.
     \STATE \COMMENT{cycle over fluid function space changes}
     \WHILE{(\NOT converged)}
      \IF{($c>C_{\textrm{max}}$)}
	 \STATE Throw exception (``maximum number of function space changes at~$t^n$ exceeded!'').\\
      \ENDIF
      \IF{($c=1$)} 
	\STATE \textbf{Structural Solver:}
	\STATE Perform predictor \mbox{$(\bfD,\bfU,\bfA)_{c=1}^n \leftarrow P(\bfD,\bfU,\bfA)^{n-1}$}.
	\STATE \textbf{ALE Solver:}
	\STATE Predict fluid grid displacement constraints.
	\STATE Relax fluid grid, update interior displacements $\hat{\bfd}_{\bfChi}$, compute grid velocities $\hat{\bfu}_{\bfChi}$.
	\STATE \textbf{Fluid Solver:}
	\STATE Update interface position~$(\Int_h)_{c=1}^{n}$ according to predicted~$\bfD_{c=1}^n$.
	\STATE Intersect~$\widehat{\mcT}_h^{\fd}$ and obtain the active computational mesh \mbox{$(\mcT_{h}^{\fd})_{c=1}^n$} and 
               update face and element sets \mbox{$(\mcF_{\Int}^{\fd})_{c=1}^n$, $(\mcT_{\Int}^{\fd})_{c=1}^n$}.
	\STATE Allocate \name{DOF}s and set up new fluid function space \mbox{$(\mcW_h^\fd)_{c=1}^n$}, see \cite{SchottWall2014}.
	\STATE Perform entire \Algoref{algo:moving_domain_solution_projection} and transcribe solution vectors\\
		between function spaces \mbox{$(\mcW_h^\fd)^{n-1} \rightarrow (\mcW_h^\fd)_{c=1}^n$} between time levels
                \st~\mbox{$(\tilde{\bfU},\tilde{\bfP},\tilde{\bfA})_{c=1}^{n-1} \in (\mcV_h^\fd\times\mcQ_h^\fd\times\mcV_h)_{c=1}^n$}
	\STATE Perform fluid predictor, \eg, \mbox{$(\bfU,\bfP)_{c=1,i=1}^n = (\tilde{\bfU},\tilde{\bfP})^{n-1} \in (\mcW_h^\fd)_{c=1,i=1}^n$}.
      \ELSE[The fluid function space has changed within the last cycle \mbox{$c-1$}]
	\STATE Perform \Algoref{algo:moving_domain_solution_projection} and transcribe solution vectors
	       between function spaces \mbox{$(\mcW_h^\fd)^{n-1} \rightarrow (\mcW_h^\fd)_{c}^n$} between time levels
	       \st~\mbox{$(\tilde{\bfU},\tilde{\bfP},\tilde{\bfA})_{c}^{n-1} \in (\mcV_h^\fd\times\mcQ_h^\fd\times\mcV_h^\fd)_{c}^n$}.
	\STATE Use recent solution approximation from interrupted pass~\mbox{$c-1$} as initial guess for the following Newton-Raphson procedure. Note that
	       \mbox{$(\bfU,\bfP)_{c,i=1}^n \in (\mcW_h^\fd)_{c}^n$}.
      \ENDIF
      \STATE \COMMENT{previous time level solution available as: \mbox{$(\bfD,\bfU,\bfA)_c^{n-1}$} (solid), \mbox{$(\tilde{\bfU},\tilde{\bfA})_{c}^{n-1}$} (fluid)}
      \STATE \COMMENT{prediction for current time level available as \mbox{$((\bfU,\bfP),\bfD)_{c,i=1}^n$} }
      \STATE Perform \Algoref{algo:nonlinear_solution_algo_monolithic_FSI_Newton_Raphson_Loop}. \COMMENT{(Re-)start NEWTON-RAPHSON-like iterations}
      \IF{(Newton-Raphson converged)}
	\STATE \algorithmicbreakwhile
      \ENDIF
      \STATE $c\leftarrow c+1$
     \ENDWHILE
     \STATE\COMMENT{Inner Newton-Raphson loop converged with non-changing fluid function space}
     \STATE Update structural vectors \mbox{$\bfA^n,\bfU^n$} based on $\bfD^n$ for G\hbox{-}$\alpha$ scheme via \Eqref{eq:genalpha_recover_acceleration}--\eqref{eq:genalpha_recover_velocity}.
     \STATE Update fluid acceleration approximation~$\bfA^n$ for \name{OST}-scheme via \Eqref{eq:chap_3_Nitsche_CIP_GP_CUTFEM_form_OST_moving_domain_2},
	    \ie, \mbox{${\bfa}_h^{n}(\bfu_h^n,\tilde{\bfu}^{n-1}_c,\tilde{\bfa}_c^{n-1})$} with \mbox{$\tilde{\bfu}^{n-1}_c, \tilde{\bfa}^{n-1}_c$} the recent
	    projections of \mbox{$\bfu^{n-1}_c,\bfa^{n-1}_c\in\mcV_h^{n-1}$} to the fluid function space~$\mcV_c^n$ belonging to the converged state.
     \STATE Store the structural force vector \mbox{$\bfF_{\Intfs(t^{n})}^{\sd,n}$} \Eqref{eq:chap_5_fsi_structural_force_previous_time_step} for next time level.
     \STATE Store final approximations: $(\bfD,\bfU,\bfA)^n$ for the solid
            and $(\bfU,\bfP,\bfA)^n$ for the fluid.
  \ENDFOR
  \STATE \name{Output}: solid displacement and fluid velocity and pressure solution approximations \mbox{$\left\{((\bfU,\bfP),\bfD)^n\right\}_{1\leqslant n \leqslant N}$}
for discrete time levels \mbox{$\left\{t^{n}\right\}_{1\leqslant n \leqslant N}$}. Note the possibly time-varying fluid approximation spaces \mbox{$\{\mcW_h^\fd\}_{1\leqslant n \leqslant N}$}.
 \end{algorithmic}
 \label{algo:nonlinear_solution_algo_monolithic_FSI} 
\end{algorithm}

\begin{algorithm}[t]
 \caption{Newton-Raphson-like solution procedure for non-linear \name{FSI}-systems}
 \begin{algorithmic}[1]
  \STATE \name{Input}: initial guess of structural displacements, fluid velocity and pressure \mbox{$((\bfU,\bfP),\bfD)_{i=1}^n$} for iterative solution procedure.
Previous time-step solutions for structure \mbox{$(\bfD,\bfU,\bfA)^{n-1}$},
for fluid \mbox{$(\tilde{\bfU},\tilde{\bfA})^{n-1}$}, interfacial velocities \mbox{$\bfU_\Int^{\sd,n-1}$} and structural forces~\mbox{$\bfF_{\Intfs(t^{n-1})}^{\sd,n-1}$}.
     \FOR{Newton-Raphson-like iterations~$1\leqslant i \leqslant N_{\textrm{max}}$}
	\STATE \textbf{\name{FSI} Solver:}
  	\STATE Update structural interface velocity \mbox{$(\bfU_\Int^\sd)_i^n$} based on $(\bfD_\Int)_i^n$ with \name{OST}-scheme \Eqref{eq:chap_5_fsi_displacement_velocity_conversion}.
	\STATE Apply current iterations to structural solver and to \name{CutFEM} (\name{ALE}-)fluid solver.
	\STATE \textbf{Structural Solver:}
	\STATE Evaluate structural contribution to matrix~\mbox{$\bfL_{DD}|_{\bfD_i^n}$} and to residual~\mbox{$\bfR_D|_{\bfD_i^n}$}, see \Eqref{eq:chap_5_fsi_linearized_system}.
	\STATE \textbf{ALE Solver:}
	\STATE Update fluid grid displacement boundary constraints.
	\STATE Relax fluid grid and obtain updated displacements $\hat{\bfd}_{\bfChi}$, compute grid velocities $\hat{\bfu}_{\bfChi}$.
	\STATE \textbf{Fluid Solver:}
	\IF{($i>1$ and unfitted approximation)} 
	\STATE Intersect~$\widehat{\mcT}_h^{\fd}$ and obtain the active computational mesh $(\mcT_{h}^{\fd})_i^n$ and
               update face and element sets \mbox{$(\mcF_{\Int}^{\fd})_i^n$}, \mbox{$(\mcT_{\Int}^{\fd})_i^n$}.
	\STATE Allocate \name{DOF}s and set up new fluid function space \mbox{$(\mcW_h^\fd)_i^n$}, see \cite{SchottWall2014}.
	\STATE Perform \name{Transfer/Copy-Phase} of \Algoref{algo:moving_domain_solution_projection} and transcribe solution vectors\\
		between function spaces \mbox{$(\mcW_h^\fd)_{i-1}^n \rightarrow (\mcW_h^\fd)_{i}^n$} of previous and current iteration.
	\IF{(\name{Transfer/Copy-Phase} \NOT successful)}
	    \STATE \COMMENT{Change in fluid function space occurred, \ie, \mbox{$(\mcW_h^\fd)_{i-1}^n \neq (\mcW_h^\fd)_{i}^n$}.}
	    \STATE Perform \name{Extension-Phase} of \Algoref{algo:moving_domain_solution_projection} and
                   utilize the resulting\\approximation as initial guess for the next Newton-Raphson cycle \mbox{$c+1$}:\\ \mbox{$((\bfU,\bfP),\bfD)^n \in (\mcW_h^\fd)_{c+1,i=1}^n \times \mcD_h$}
	    \RETURN \FALSE.
	\ENDIF
	\ENDIF
	\STATE Evaluate fluid boundary conditions \mbox{$\bfgD^\fd({t^n})$}, \mbox{$\bfhN^\fd({t^n})$} with respect to \mbox{$(\Int_h^{\fd})_i^n$}.
	\STATE Evaluate stabilized fluid system, coupling matrices contributing to \mbox{$\bfL_{xy}$} and to residuals \\
		$\bfR_U,\bfR_P,\bfR_D$, see \Eqref{eq:chap_5_fsi_linearized_system}.
	\STATE \textbf{FSI Solver:}
	\STATE Set up final linear fluid-structure system \mbox{$\bfL^\fs \cdot \Delta ((\bfU,\bfP),\bfD)^n_i = -\bfR^\fs$} \Eqref{eq:chap_5_fsi_linearized_system}.
	\STATE Apply strong Dirichlet boundary conditions from function spaces $\mcW_{\bfgD,h}^\fd$ and $\mcD_{\bfgD,h}^\sd$.
	\STATE Build block preconditioner and apply to \Eqref{eq:chap_5_fsi_linearized_system}.
	\STATE Solve preconditioned linearized fixed-point like system \Eqref{eq:chap_5_fsi_linearized_system} for $\Delta((\bfU,\bfP),\bfD)_i^n$.
	\STATE Check convergence: $\norm{\Delta((\bfU,\bfP),\bfD)_{i}^n}<TOL$ \AND $\norm{(\bfR^\fs)_{i}^n}<TOL$.
	\IF{(\name{FSI} system converged)}
	  \RETURN \TRUE
	\ENDIF
	\STATE Update Newton-Raphson iteration via \Eqref{eq:chap_5_fsi_linearized_system_update} to $((\bfU,\bfP),\bfD)_{i+1}^n$.
      \ENDFOR
      \STATE Throw exception (``maximum number of Newton-Raphson iterations reached!'').\\
  \STATE \name{Output}:
\RETURN \FALSE~ with iteration \mbox{$((\bfU,\bfP),\bfD)_i^n \in (\mcW_h^\fd)_i^n\times \mcD_h^\sd$} based on changed fluid function\\space $(\mcW_h^\fd)_i^n\neq(\mcW_h^\fd)_{i-1}^n$, or
\RETURN \TRUE~ with converged velocity, pressure and displacement solution approximations\\ \mbox{$((\bfU^n,\bfP^n),\bfD^n)$} for time level~$t^n$.
 \end{algorithmic}
 \label{algo:nonlinear_solution_algo_monolithic_FSI_Newton_Raphson_Loop} 
\end{algorithm}
\paragraph*{Newton-Raphson-like approximation of non-linear \name{FSI} residuals.}
The solution approximation to the coupled fluid-solid system at a discrete time level~$t^n$
is given in terms of the finite-dimensional system of non-linear equations \Eqref{eq:chap_5_fsi_nonlinear_residual}
for fluid velocities and pressure $(\bfU,\bfP)^{n}$ and for structural displacements~$\bfD^{n}$.
Therein, non-linearities emanate from geometrical and material non-linear relations in the structural dynamics \name{PDE}
on the one hand,
and from the non-linear convective fluid term together with its occurrence in related fluid stabilizations
on the other hand. Moreover, while structural equations are integrated in referential configuration, further non-obvious non-linearities are
hidden in the change of the integration volume and surface area associated with the fluid subdomain,
which depend non-linearly on the interface motion.

Neglecting for a short term that due to the unfittedness of fluid and solid computational meshes
in Approaches~\ref{approach:2_XFSI}-\ref{approach:4_FXFSI}
the discrete fluid function space
can change when the structure moves, then the solution of the non-linear system \Eqref{eq:chap_5_fsi_nonlinear_residual} can be 
approximated iteratively by performing a Newton-Raphson-like method.

The solution \mbox{$((\bfU,\bfP),\bfD)^{n}$} is approximated iteratively for \mbox{$i\geqslant 1$} by solving linear systems for increments
\mbox{$\Delta((\bfU,\bfP),\bfD)_i^{n}$} satisfying
\begin{equation}
\label{eq:chap_5_fsi_linearized_system}
\underbrace{
\ADLdrawingmode{3}
\ADLactivate
\left[
\begin{array}{c | c}
\begin{array}{c;{2pt/2pt} c}
\bfL_{UU} & \bfL_{UP}\\ \hdashline[2pt/2pt]
\bfL_{PU} & \bfL_{PP}
\end{array}
&
\begin{array}{c}
\bfL_{UD} \\ \hdashline[2pt/2pt]
\bfL_{PD}
\end{array}\\
\cline{1-2}
\begin{array}{c;{2pt/2pt} c}
\bfL_{DU}  & \bfL_{DP}
\end{array}
&
\bfL_{DD}
\end{array}
\right]^{n}_{i}
}
_{\bfL^{\fs}}
\cdot
\left[
\begin{array}{c}
\ADLdrawingmode{3}
\ADLactivate
  \Delta \bfU \\ \hdashline[2pt/2pt]
  \Delta \bfP \\ \cline{1-1}
  \Delta \bfD
\end{array}
\right]^{n}_{i}
=
-
\underbrace{
\left[
\begin{array}{c}
  \bfR_U \\ \hdashline[2pt/2pt]
  \bfR_P \\ \cline{1-1}
  \bfR_D
\end{array}
\right]^{n}_{i}
}
_{\bfR^{\fs}}
\end{equation}
followed by a Newton-Raphson incremental update step for the next iteration
\begin{align}
\label{eq:chap_5_fsi_linearized_system_update}
\left[
\begin{array}{c}
\bfU \\ \hdashline[2pt/2pt]
\bfP \\ \cline{1-1}
\bfD
\end{array}
\right]^{n}_{i+1}
=
\left[
\begin{array}{c}
\bfU \\ \hdashline[2pt/2pt]
\bfP \\ \cline{1-1}
\bfD
\end{array}
\right]^{n}_{i}
+
\left[
\begin{array}{c}
\Delta \bfU \\ \hdashline[2pt/2pt]
\Delta \bfP \\ \cline{1-1}
\Delta \bfD
\end{array}
\right]^{n}_{i}.
\end{align}

\newpage

The subscript~\mbox{$()_{i}$} in matrix and right-hand side of the linear system \Eqref{eq:chap_5_fsi_linearized_system} indicates sub-matrices and residuals being
evaluated on basis of the current approximation \mbox{$((\bfU,\bfP),\bfD)_{i}^{n}$}.
Therein, let \mbox{$\bfL_{xy} = \frac{\partial{\bfR_x}}{\partial y}$} denote (pseudo)-directional derivatives of residuals~$\bfR_x$ from \Eqref{eq:chap_5_fsi_nonlinear_residual}
with respect to the finite dimensional solution approximation~$y$, where \mbox{$x,y\in\{\bfU,\bfP,\bfD\}$}.
Exemplarily, approximation $\bfL_{DD}$
is given as
\begin{equation}
\label{eq:chap_5_fsi_linearized_system_structure_DD}
  \bfL_{DD}|_{((\bfU,\bfP),\bfD)_{i}^n}=\frac{1}{1-\alpha_f}\left(\bfM^\sd \frac{1-\alpha_m}{\beta \Delta t^2} + \frac{\partial \bfF^{\sd,n}_{\textrm{int}}(\bfD)}{\partial\bfD}\bigg|_{\bfD=\bfD_{i}^n}\right) + \bfG|_{((\bfU,\bfP),\bfD)_{i}^n},
\end{equation}
where
\begin{align}
  \bfG|_{((\bfU, \bfP),\bfD)^n_i}   &\eqdef \dfrac{\partial \bfC^{\textrm{sf}}((\bfU, \bfP),\bfD)}{\partial \bfD}\bigg|_{((\bfU, \bfP),\bfD) = ((\bfU, \bfP),\bfD)_i^n} \\
  \Leftrightarrow
	  &\quad  + \langle \gamma(\mu/h)(-\partial_{\bfD}(\bfu_h^\sd)|_{\bfD_i^n},(-\bfw_h) \rangle_{\Intfs(\bfD_i^n)} \nonumber\\
          &\quad  + \langle \gamma(\rho^\fd\sigma h + \rho^\fd|\bfu_h^\fd| + \mu/h) (-\partial_{\bfD}(\bfu_h^\sd)|_{\bfD_i^n}) \cdot\bfn^{\fs}(\bfD_i^n), (-\bfw_h)\cdot\bfn^{\fs}(\bfD_i^n) \rangle_{\Intfs(\bfD_i^n)}, \nonumber
\end{align}
with \mbox{$\partial_{\bfD}(\bfu_h^\sd)|_{\bfD_i^n}=\frac{\partial\bfu_h^\sd(\bfD)}{\partial \bfD}\big|_{\bfD=\bfD_i^n} \Leftrightarrow \frac{1}{\theta_\Int \Delta t} \bfD_i^n $}.
Note that changes of integration area and interface unit normal vectors are treated in a fixed-point fashion.
All other approximations~$\bfL_{xy}$ can be derived by analogy, but are not presented to shorten the presentation.

The Newton-Raphson-like scheme for residual \Eqref{eq:chap_5_fsi_nonlinear_residual_multifluid}
according to Approach~\ref{approach:4_FXFSI} becomes

\begin{equation}
\label{eq:chap_5_fsi_linearized_system_multifluid}
\underbrace{
\ADLdrawingmode{3}
\ADLactivate
\left[
\begin{array}{c | c | c | c}
\begin{array}{c;{2pt/2pt} c}
\bfL_{UU}^{11} & \bfL_{UP}^{11}\\ \hdashline[2pt/2pt]
\bfL_{PU}^{11} & \bfL_{PP}^{11}
\end{array}
&
\cdots
&
\begin{array}{c;{2pt/2pt} c}
\bfL_{UU}^{1l} & \bfL_{UP}^{1l}\\ \hdashline[2pt/2pt]
\bfL_{PU}^{1l} & \bfL_{PP}^{1l}
\end{array}
&
\begin{array}{c}
\bfL_{UD}^{1 \sd} \\ \hdashline[2pt/2pt]
\bfL_{PD}^{1 \sd}
\end{array}
\\
\cline{1-4}
\vdots & \ddots & \vdots & \vdots
\\
\cline{1-4}
\begin{array}{c;{2pt/2pt} c}
\bfL_{UU}^{l1} & \bfL_{UP}^{l1}\\ \hdashline[2pt/2pt]
\bfL_{PU}^{l1} & \bfL_{PP}^{l1}
\end{array}
&
\cdots
&
\begin{array}{c;{2pt/2pt} c}
\bfL_{UU}^{ll} & \bfL_{UP}^{ll}\\ \hdashline[2pt/2pt]
\bfL_{PU}^{ll} & \bfL_{PP}^{ll}
\end{array}
&
\begin{array}{c}
\bfL_{UD}^{l \sd} \\ \hdashline[2pt/2pt]
\bfL_{PD}^{l \sd}
\end{array}
\\
\cline{1-4}
\begin{array}{c ;{2pt/2pt}c} \bfL_{DU}^{\sd 1} & \bfL_{DP}^{\sd 1} \end{array} & \cdots & \begin{array}{c ;{2pt/2pt}c} \bfL_{DU}^{\sd l} & \bfL_{DP}^{\sd l} \end{array} & \bfL_{DD}^{\sd\sd} \\
\end{array}
\right]^{n}_{i}
}
_{\bfL}
\cdot
\left[
\begin{array}{c}
\ADLdrawingmode{3}
\ADLactivate
  \Delta\bfU^1 \\ \hdashline[2pt/2pt]
  \Delta\bfP^1 \\ \cline{1-1}
  \vdots \\ \cline{1-1}
  \Delta\bfU^l \\ \hdashline[2pt/2pt]
  \Delta\bfP^l  \\ \cline{1-1}
  \Delta\bfD
\end{array}
\right]^{n}_{i}
=
-
\underbrace{
\left[
\begin{array}{c}
  \bfR_{U^1} \\ \hdashline[2pt/2pt]
  \bfR_{P^1} \\ \cline{1-1}
  \vdots \\ \cline{1-1}
  \bfR_{U^l} \\ \hdashline[2pt/2pt]
  \bfR_{P^l}  \\ \cline{1-1}
  \bfR_D
\end{array}
\right]^{n}_{i}
}
_{\bfR}
\end{equation}
with incremental update step for the next iteration
\begin{align}
\label{eq:chap_5_fsi_linearized_system_update_multifluid}
\left[
\begin{array}{c}
  \bfU^1 \\ \hdashline[2pt/2pt]
  \bfP^1 \\ \cline{1-1}
  \vdots \\ \cline{1-1}
  \bfU^l \\ \hdashline[2pt/2pt]
  \bfP^l  \\ \cline{1-1}
  \bfD
\end{array}
\right]^{n}_{i+1}
=
\left[
\begin{array}{c}
  \bfU^1 \\ \hdashline[2pt/2pt]
  \bfP^1 \\ \cline{1-1}
  \vdots \\ \cline{1-1}
  \bfU^l \\ \hdashline[2pt/2pt]
  \bfP^l  \\ \cline{1-1}
  \bfD
\end{array}
\right]^{n}_{i}
+
\left[
\begin{array}{c}
  \Delta\bfU^1 \\ \hdashline[2pt/2pt]
  \Delta\bfP^1 \\ \cline{1-1}
  \vdots \\ \cline{1-1}
  \Delta\bfU^l \\ \hdashline[2pt/2pt]
  \Delta\bfP^l  \\ \cline{1-1}
  \Delta\bfD
\end{array}
\right]^{n}_{i}.
\end{align}

\paragraph*{Newton-Raphson-like algorithmic procedure.}

Assuming non-changing fluid function spaces, which is guaranteed for Approach~\ref{approach:1_ALE-FSI}, but only partially for
Approaches~\ref{approach:2_XFSI}--\ref{approach:4_FXFSI},
the solution procedure for the non-linear \name{FSI} residual \Eqref{eq:chap_5_fsi_nonlinear_residual} is summarized in
\Algoref{algo:nonlinear_solution_algo_monolithic_FSI_Newton_Raphson_Loop}.

For starting the iterative solution procedure, an initial guess \mbox{$((\bfU,\bfP),\bfD)^n_{i=1}$} for the solution fields is required.
Following \cite{Forster2007addedmass,Mayr2015},
these may be given in terms of a predicted displacement field and the velocity solution of the previous time level.
Evaluating \name{FSI} residuals requires history from the previous time level
for structural fields \mbox{$(\bfD,\bfU,\bfA)^{n-1}$} and
structural forces \mbox{$\bfF_{\Intfs(t^{n-1})}^{\sd,n-1}$} \Eqref{eq:chap_5_fsi_structural_force_previous_time_step}.

\begin{remark}
Note that for unfitted fluid approximations their function spaces might change due to displacing the fluid-structure interface.
This demands to project previous time step quantities onto the current function space~\mbox{$(\mcW_h^{\fd})_{i=1}^n$}
associated with the time level~$t^n$ by applying \Algoref{algo:moving_domain_solution_projection}.
This provides approximations \mbox{$(\tilde{\bfU},\tilde{\bfA})^{n-1}\in (\mcV_h)^n_{i=1}\times(\mcV_h)^n_{i=1}$}.
\end{remark}

For a current displacement iteration~$\bfD_i^{n}$, structural interface velocities~$(\bfU_\Int^{\sd})_i^n$ can be updated
via \Eqref{eq:chap_5_fsi_displacement_velocity_conversion}.
If the fluid grid is expected to move in an \name{ALE} fashion
(see Approaches~\ref{approach:1_ALE-FSI} and \ref{approach:3_ALE-XFSI}),
the grid displacements need to be updated (for instance interface-constrained by \eqref{eq:ALE-constraint_FSI}),
the grid velocities $\hat{\bfu}_{\bfX,h}$ computed 
and afterwards accounted for in the flow solver,
see Sections~\ref{sssec:strong_formulation_fluid_mechanics}--\ref{sssec:strong_formulation_ALE_mechanics}.
More detailed information can be found in, \eg~\cite{Forster2006,Forster2009}. 

Having set the current approximations to the structural and the (\name{ALE}-)fluid solver, their contributions to matrices~$(\bfL_{xy})_i^n$ and
right-hand-side blocks~$(\bfR_x)_i^n$ of the linearized \name{FSI} system \Eqref{eq:chap_5_fsi_linearized_system}
can be evaluated and finally assembled into the global system
\mbox{$\bfL^\fs \cdot \Delta ((\bfU,\bfP),\bfD)^n_i = -\bfR^\fs$}.
After applying strong Dirichlet constraints from \mbox{$\mcW_{\bfgD,h}^{\fd}$} and \mbox{$\mcW_{\bfgD,h}^{\sd}$}
and setting up a block preconditioner,
the linear \name{FSI} system can be solved for the Newton-Raphson increments.
Finally, the approximations can be updated to~\mbox{$((\bfU,\bfP),\bfD)_{i+1}^n$} via \Eqref{eq:chap_5_fsi_linearized_system_update}.
Newton-Raphson-like iterations are performed until convergence of residuals~\mbox{$(\bfR^{\fs})_{i}^n$} and
of increments \mbox{$\Delta((\bfU,\bfP),\bfD)_i^n$} is achieved.

\begin{remark}[Solving Linear Matrix Systems]
For approximating the incremental solution of each linear \name{FSI} iteration \Eqref{eq:chap_5_fsi_linearized_system},
if not indicated otherwise,
a preconditioned \name{GMRES} solver is utilized.
Building specific preconditioners is based upon on a unified framework
for monolithically coupled $n$-field problems
proposed in \cite{Verdugo2016}.
In this work, generic block Gauss--Seidel iterations are used for decoupling the subproblems.
For the respective structural, fluid and \name{ALE} blocks,
any combination of either one-level domain decomposition preconditioners of incomplete factorization type,
see \cite{IFPACK2005}, or algebraic multigrid (\name{AMG}) methods, as proposed in \cite{MueLU2014}, is applicable.
\end{remark}

\paragraph*{Cycle over function space changes for unfitted fluid approximations.}

Recalling the issue of applicability of iterative solution techniques for unfitted mesh methods,
which consists in potentially varying fluid approximation spaces during the iterative procedure,
the following generalization of the Newton-Raphson scheme is suggested:

As described in \Algoref{algo:nonlinear_solution_algo_monolithic_FSI_Newton_Raphson_Loop},
for each iteration the updated interface displacements are set to the \name{CutFEM} fluid solver
on which basis the computational mesh~$\widehat{\mcT}_h^\fd$ is intersected, associated face and element sets $(\mcF_\Int^{\fd})_i^n$ and $(\mcT_\Int^{\fd})_i^n$
are built and the fluid function space $(\mcW_h^\fd)_i^{n}$ is constructed.
Having updated the discrete function space,
previous fluid iterations \mbox{$(\bfU,\bfP)_{i-1}^n\in(\mcW_h^\fd)^n_{i-1}$} have to be transcribed to $(\mcW_h^\fd)^n_{i}$.
This can be carried out node-wise using the \name{Transfer/Copy-Phase} of \Algoref{algo:moving_domain_solution_projection}
if \mbox{$(\mcW_h^\fd)^n_{i-1}\equiv(\mcW_h^\fd)^n_{i}$}.
However, when \mbox{$(\mcW_h^\fd)^n_{i-1}\neq(\mcW_h^\fd)^n_{i}$}, the current Newton-Raphson cycle~$c$ is interrupted.
\Algoref{algo:moving_domain_solution_projection}, that is including the \name{Extension-Phase},
can be then applied to \mbox{$(\bfU,\bfP)_{c,i}^{n}\in (\mcW_h^\fd)_{c,i-1}^n$}
to obtain a predicted approximation \mbox{$((\bfU,\bfP),\bfD)^n_{c+1,i=1} = (P^n_i (\bfU,\bfP)_{c,i}^{n}, \bfD)$}
for the next pass~\mbox{$c+1$}.
These predicting steps for restarting the Newton-Raphson procedure in a new cycle
are comprised in \Algoref{algo:nonlinear_solution_algo_monolithic_FSI_Newton_Raphson_Loop}.

As summarized in \Algoref{algo:nonlinear_solution_algo_monolithic_FSI},
within a subsequent run, \ie~\mbox{$c>1$}, the previous time level fluid solution has to be adapted according to the changed function space.
For this purpose, a projection step as provided by \Algoref{algo:moving_domain_solution_projection} can be performed such that
$(\tilde{\bfU},\tilde{\bfP},\tilde{\bfA})_{c}^{n-1} \in (\mcV_h^\fd\times\mcQ_h^\fd\times\mcV_h^\fd)_{c}^n$.
Note that for unchanged \name{DOF}s the Newton-Raphson is interrupted just formally and the solution approximation remains unmodified
due to the \name{DOF}-wise copying-technique.
As it might happen that \name{DOF}s get repeatedly activated and deactivated
for subsequent cycle runs \mbox{$c,c+1,c+2,\ldots$}, 
situations, which could worsen or even totally destroy convergence,
straightforward techniques like freezing the fluid function space, \ie~\mbox{$(\mcW_h^\fd)_i^n=(\mcW_h^\fd)_{i+1}^n=(\mcW_h^\fd)_{i+2}^n=\ldots$},
can be utilized.
When doing so, \name{DOF}s which are located outside of the integrated fluid domain
can be sufficiently controlled by simply extending the interface zone, \ie~the set of interface zone facets $\widetilde{\mcF}_\Int^{\fd}\supsetneq\mcF_\Int^{\fd}$ for the ghost-penalty-stabilization operators.

After achieving a converged solution from the proposed nested Newton-Raphson procedure,
further time-stepping related quantities need to be updated.
For an \name{OST} scheme in the fluid solver,
an acceleration approximation~$\bfA^n$ is obtained from \Eqref{eq:chap_3_Nitsche_CIP_GP_CUTFEM_form_OST_moving_domain_2},
where \mbox{${\bfa}_h^{n}(\bfu_h^n,\tilde{\bfu}^{n-1}_c,\tilde{\bfa}_c^{n-1})$} incorporates
projected velocity and acceleration approximations \mbox{$\tilde{\bfu}^{n-1}_c,\tilde{\bfa}_c^{n-1}$}
belonging to the converged cycle run~$c$ of the previous time level.
For the structural field, acceleration and velocity fields~$\bfA^n$ and~$\bfU^n$ according to the
G\hbox{-}$\alpha$ scheme are refreshed via \Eqref{eq:genalpha_recover_acceleration}--\eqref{eq:genalpha_recover_velocity}.
After storing the structural forces for the next time step as specified in \Eqref{eq:chap_5_fsi_structural_force_previous_time_step},
all quantities required for proceeding with the next time level~$t^{n+1}$ are available,
\ie~\mbox{$(\bfU,\bfP,\bfA)^n$} for the fluid and \mbox{$(\bfD,\bfU,\bfA)^n$} and \mbox{$\bfF_{\Intfs(t^{n})}^{\sd,n}$} for the solid.

\section{Different Nitsche-based FSI approaches - Numerical Examples}
\label{sec:numerical_examples}

In the following, the proposed unfitted \name{FSI} approaches are validated for
challenging two- and three-dimensional test cases, which exhibit highly dynamic transient fluid-structure interaction.
Their discussed advantages for different \name{FSI} problem settings will be demonstrated
and the importance of the proposed discretization concepts, even beyond \name{FSI}, will be highlighted.
All non-linear \name{FSI} systems are solved fully monolithic based on the techniques provided in \Secref{sec:fsi:monolithic}.
If not indicated otherwise, convergence checks for increments and residuals are performed separately for the distinct solution approximations,
\ie, for \mbox{$\bfR_U,\bfR_P, \bfR_D$} and \mbox{$\Delta \bfU, \Delta \bfP, \Delta \bfD$},
based on relative $l^2$- and $l^\infty$-vector-norms with a uniform tolerance of \mbox{$TOL=10^{-8}$}.
The subsequent results have been computed with the fully parallelized code environment \name{BACI} (see \cite{WalletBaciCommittee2017}).

For the structural temporal discretization the G\hbox{-}$\alpha$ scheme without damping, \ie~\mbox{$\rho_\infty=1.0$},
is used, if not indicated otherwise.
For the temporal approximation of interface quantities it is chosen \mbox{$\theta_\Int=1.0$}, while
the fluid temporal approximation is based on an one-step-$\theta$ scheme and uses different parameters~$\theta$, as will be specified below.
For all involved Nitsche couplings, an adjoint-inconsistent variant is utilized.

Approach~\ref{approach:2_XFSI} will be validated by the simulation of a pulsating flow over a bending, highly flexible three-dimensional flap.
In Approach~\ref{approach:3_ALE-XFSI}, a classical fitting mesh \name{ALE}-based Approach~\ref{approach:1_ALE-FSI} will be combined with the unfitted mesh Approach~\ref{approach:2_XFSI}, where a moving background fluid mesh is involved.
For highlighting the advantages of this approach a
flow excited impermeable fluid container in a flexible pipe serves as test example.
The advantages of Approach~\ref{approach:4_FXFSI} with regards to local mesh refinement based on a fluid domain decomposition technique
will be demonstrated by means of the simulation of two setups of a vibrating flag.

\subsection{Approach~\ref{approach:2_XFSI}: Pulsating Flow over a Bending Flexible Flap}
\label{ssec:fsi:fluid_structure_unfitted_numerical_results:bending_flexible_flap}

\paragraph*{Problem Setup.}
This test case is inspired from steady-state examples presented in \cite{Gerstenberger2009}.
Here, in contrast, the material properties of the fluid and the solid are weakened
to obtain highly dynamic interactions between a mainly convective-dominated flow and a flexible structure.

Initially inlet-driven fluid flow in a cuboid-shaped domain \mbox{$\Dom=[0,1.8]\times [0,0.6]\times [-0.6,0.6]$}
around a rubbery flexible flap of initial dimensions \mbox{$\Dom^\sd=[0.49,0.56]\times [0,0.35] \times [-0.3,0.3]$},
which is clamped by the bottom wall of~$\Domh$, \ie, at \mbox{$x_2=0$}, is considered.
The top and the four side walls of the flap defines the interface $\Int^{\fs}$.
Fluid is periodically pushed into the domain~$\Dom$ at the inlet \mbox{$x_1=0$} with
\mbox{$\bfu_{\mathrm{in}}=(u_1,0,0) \cdot g(t)$},
where \mbox{$u_1(\bfx)= u^{\mathrm{max}} ( 81/2500 ) x_2(x_2-0.6)(x_3+0.6)(x_3-0.6)$}
is parabolic in directions $x_2$ and $x_3$.
The peak velocity is \mbox{$u^{\mathrm{max}}=2.0$}, varied by a temporal factor
$g(t)=\tfrac{1}{2}(1-\cos(\pi t))~\foralls t\in [0,T_1=10.0]$ and $g(t)=0~\foralls t\in(T_1,T=30.0]$.
The flow entering the domain excites the structural flap to initially bend and deform, while, later, its periodicity
causes highly dynamic mutual stimulations of fluid and solid.
No-slip wall boundary conditions at the four sides perpendicular to the inlet prevent the flow to escape
which is pushed outwards at the outlet at \mbox{$x_1=1.8$}, where \mbox{$\bfhN=\bfzero$} defines the pressure level.
The incompressible flow is characterized by \mbox{$\nu^\fd=0.01$} and \mbox{$\rho^\fd = 1.0$}.
Based on the maximum inflow velocity and the width of the flap, the Reynold's number~$\RE$ ranges from \mbox{$0$--$120$}.
The structure exhibits a Neo-Hookean material with \mbox{$E= 500$}, \mbox{$\nu^\sd=0.4$}
and \mbox{$\rho^\sd=250$} for which large and dynamic deformations are expected.
External volume loads are not present in this setup.

\paragraph*{Computational Approach.}
The spatial approximation of the flow field utilizes
an unfitted fluid mesh which remains fixed over time according to Approach~\ref{approach:2_XFSI}.
For the structural subdomain~$\Dom^\sd$
an fluid-solid-interface fitted computational grid~\mbox{$\widehat{\mcT}_h^\sd$} is used.
It consists of \mbox{$8\times 15 \times 31$} $8$-node $\mathbb{Q}^1$-elements
and overlaps with a background fluid grid \mbox{$\widehat{\mcT}_h^\fd$}, which fits to the outer boundaries, however, does not fit to the fluid-solid interface $\Int_h^{\fs}$,
neither in the initial state nor when the structure displaces, \ie~\mbox{$\Domh^\fd\subsetneq\Domh^{\fd \ast}$}.
The mesh \mbox{$\widehat{\mcT}_h^\fd$} is constructed as follows: a Cartesian grid consisting of \mbox{$30\times 13 \times 26$} elements covers
the domain $\Dom$ and is subsequently refined in a middle block dimensioned by \mbox{$[0.3,0.9]\times[0,0.6]\times[-0.6,0.6]$}
with one level of element splitting in $x_1$- and $x_2$-directions.
This increases the flow resolution in the major region of interest, where the structure is expected to move and deform.
The final entire fluid mesh \mbox{$\widehat{\mcT}_h^\fd$} contains $21632$ $8$-node $\mathbb{Q}^1$-elements,
whereas its active computational mesh $\mcT_h^\fd$ changes over time due to the displacing interface $\Int_h^\fs$.
For the temporal discretization it is chosen \mbox{$\Delta t = 0.01$}
with \mbox{$\theta=1.0$} for the fluid field.
The Nitsche-penalty parameter in \Eqref{eq:chap_5_Nitsche_CIP_GP_CUTFEM_form_fluidsolid_form_Chfs_3}
is chosen as \mbox{$\gamma=10.0$}.
The fluid formulation relies on the proposed \name{RBVM/GP} formulation (see elaborations in \Secref{sec:spatial_discretization} for details) for the cut, unfitted mesh \mbox{$\widehat{\mcT}_h^\fd$}.

\begin{figure}
  \begin{center}
  \includegraphics[trim=0 100 0 100, clip, width=0.475\textwidth]{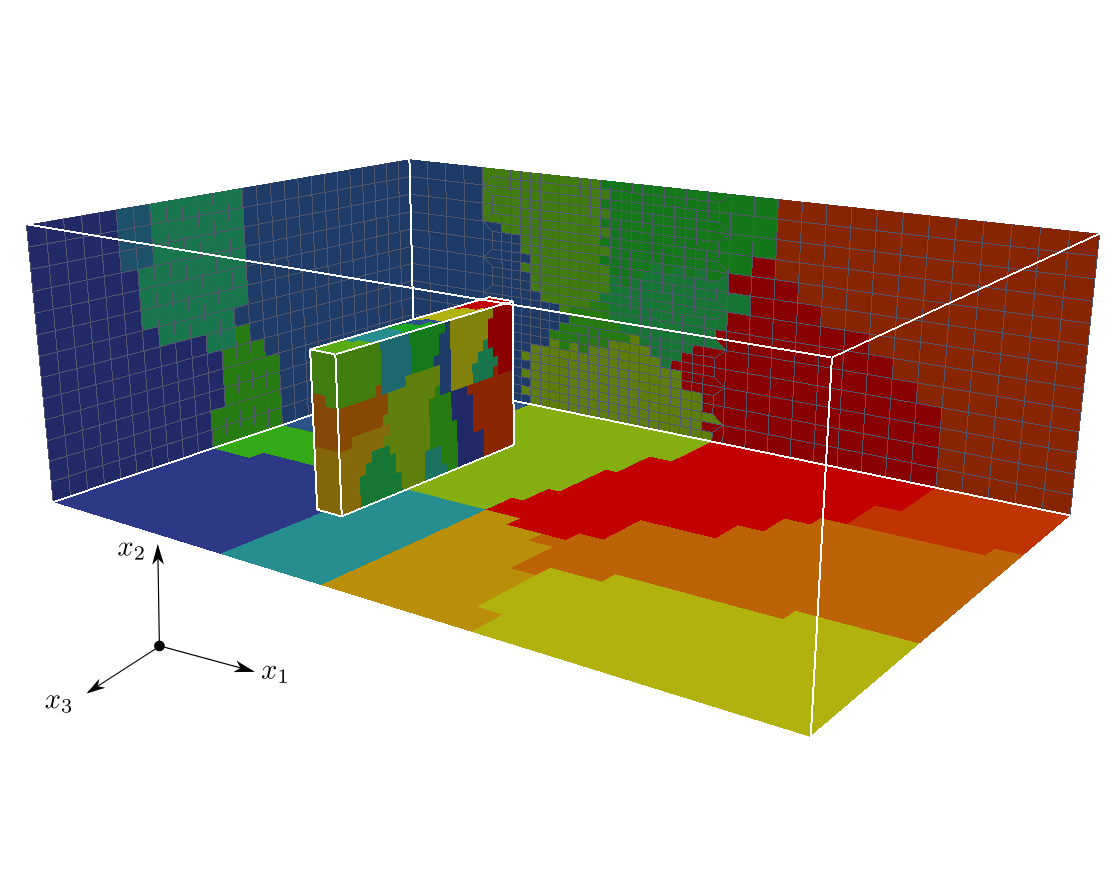}
  \includegraphics[trim=0 100 0 100, clip, width=0.475\textwidth]{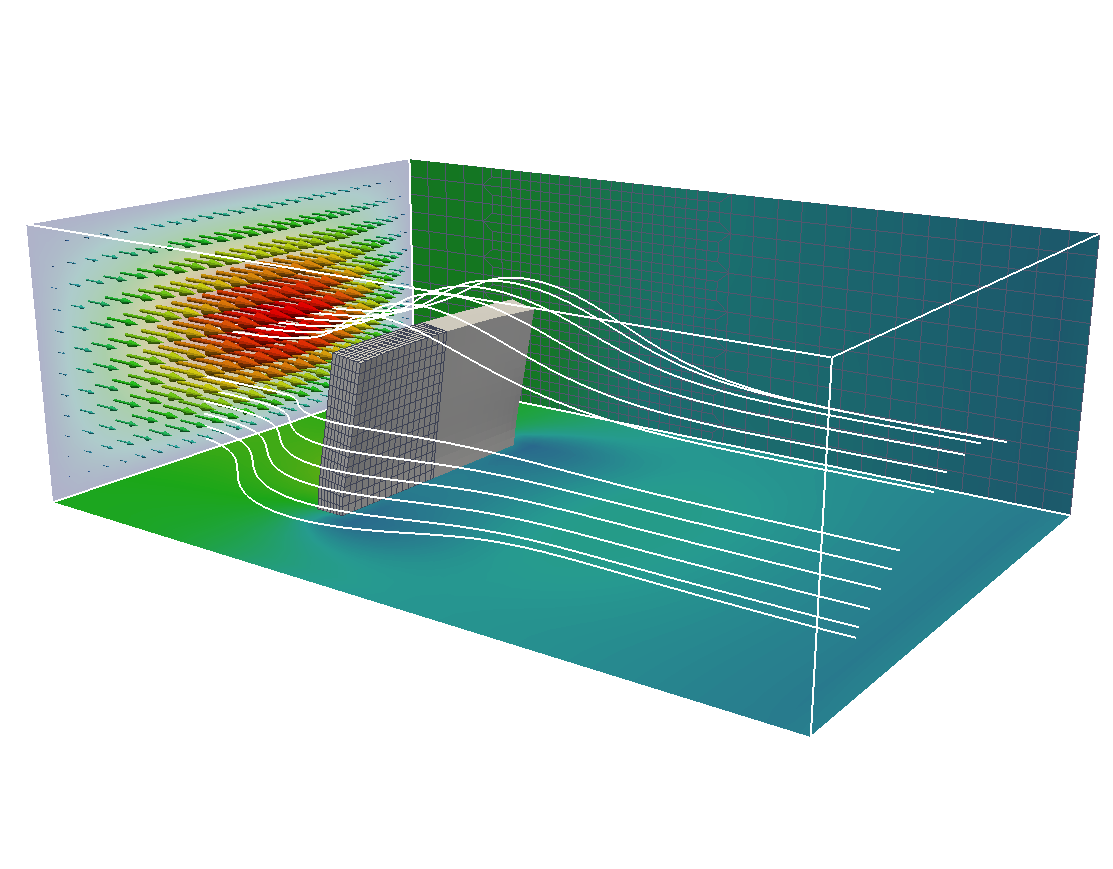}\\
  \includegraphics[trim=0 100 0 100, clip, width=0.475\textwidth]{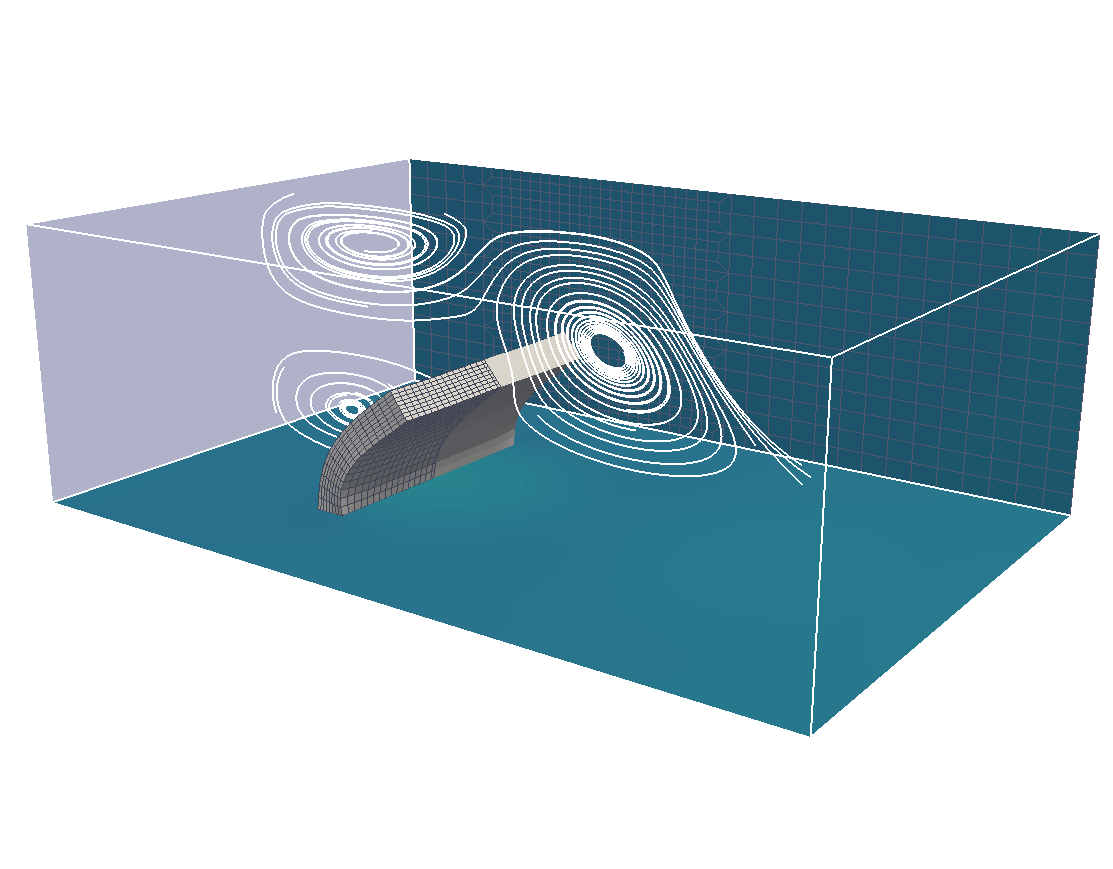}
  \includegraphics[trim=0 100 0 100, clip, width=0.475\textwidth]{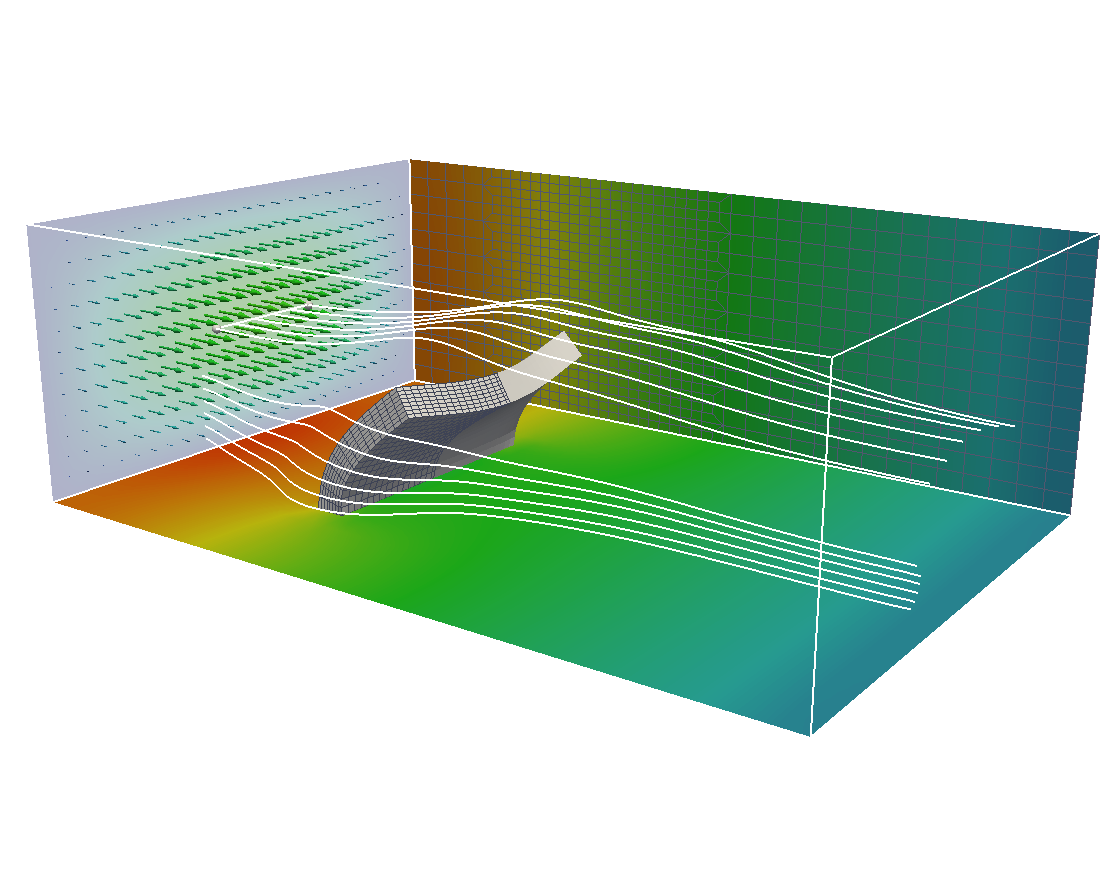}\\
  \includegraphics[trim=0 100 0 100, clip, width=0.475\textwidth]{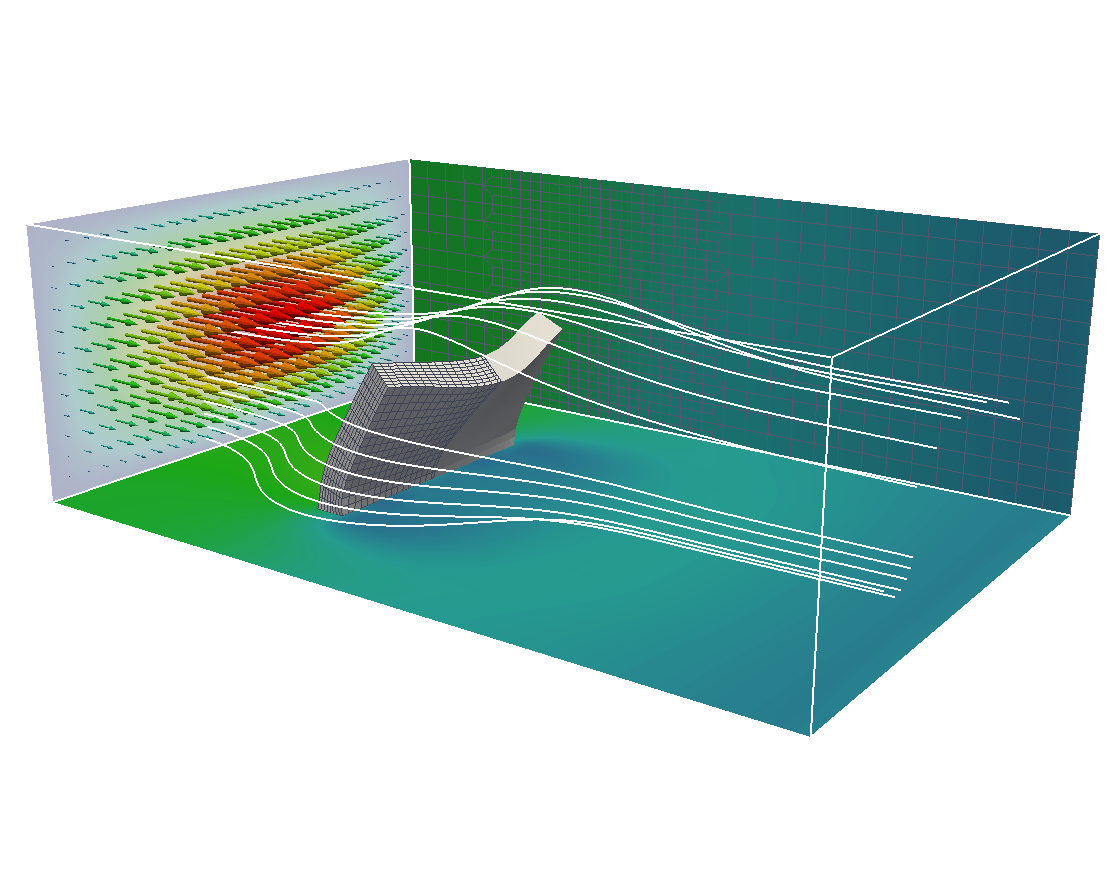}
  \includegraphics[trim=0 100 0 100, clip, width=0.475\textwidth]{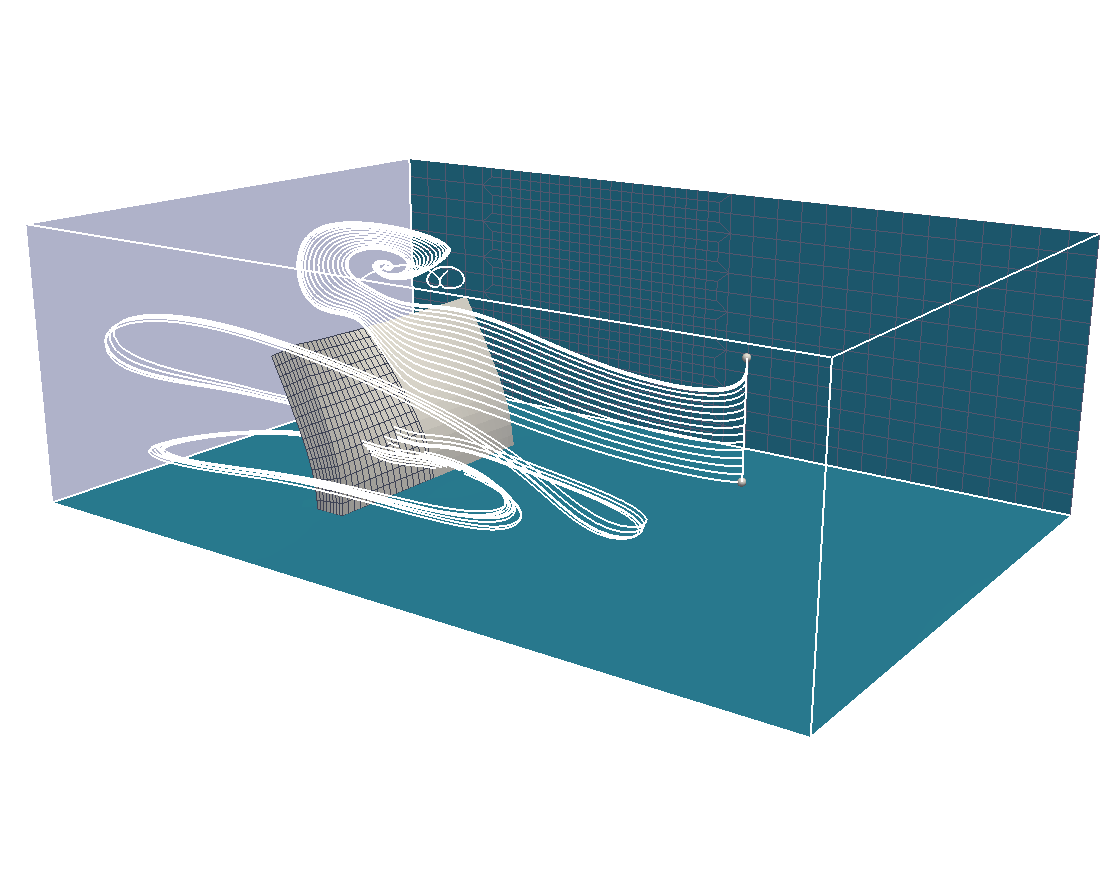}
  \end{center}
  \vspace{-12pt}
  \caption{Pulsating flow over a bending flexible flap: visualization of the surrounding fluid flow and deformation of the bending rubbery flap
at different times (from top left to bottom right) \mbox{$T_1=0.0$}, \mbox{$T_2=1.0$}, \mbox{$T_3=2.0$}, \mbox{$T_4=2.5$}, 
 \mbox{$T_5=9.0$} during the excitation phase when the pulsating inflow strongly affects the structural motion.
When no fluid is entering the inlet anymore, flow and flap motion strongly influence each other shown at \mbox{$T_6=12.5$}. 
Pulsating parabolic inlet indicated by colored arrows.
Side and bottom walls are colored by pressure distribution, whereas for the initial positioning (top left) the processor distribution for
the parallel performed computation is visualized.
}
  \label{fig:chap_5_Pulsating_Flow_Bending_Flexible_Flap:time_evolution}
\end{figure}

\begin{figure}
\centering
\includegraphics[width=6.0cm, trim=0cm 0cm 0cm 0cm, clip=true]{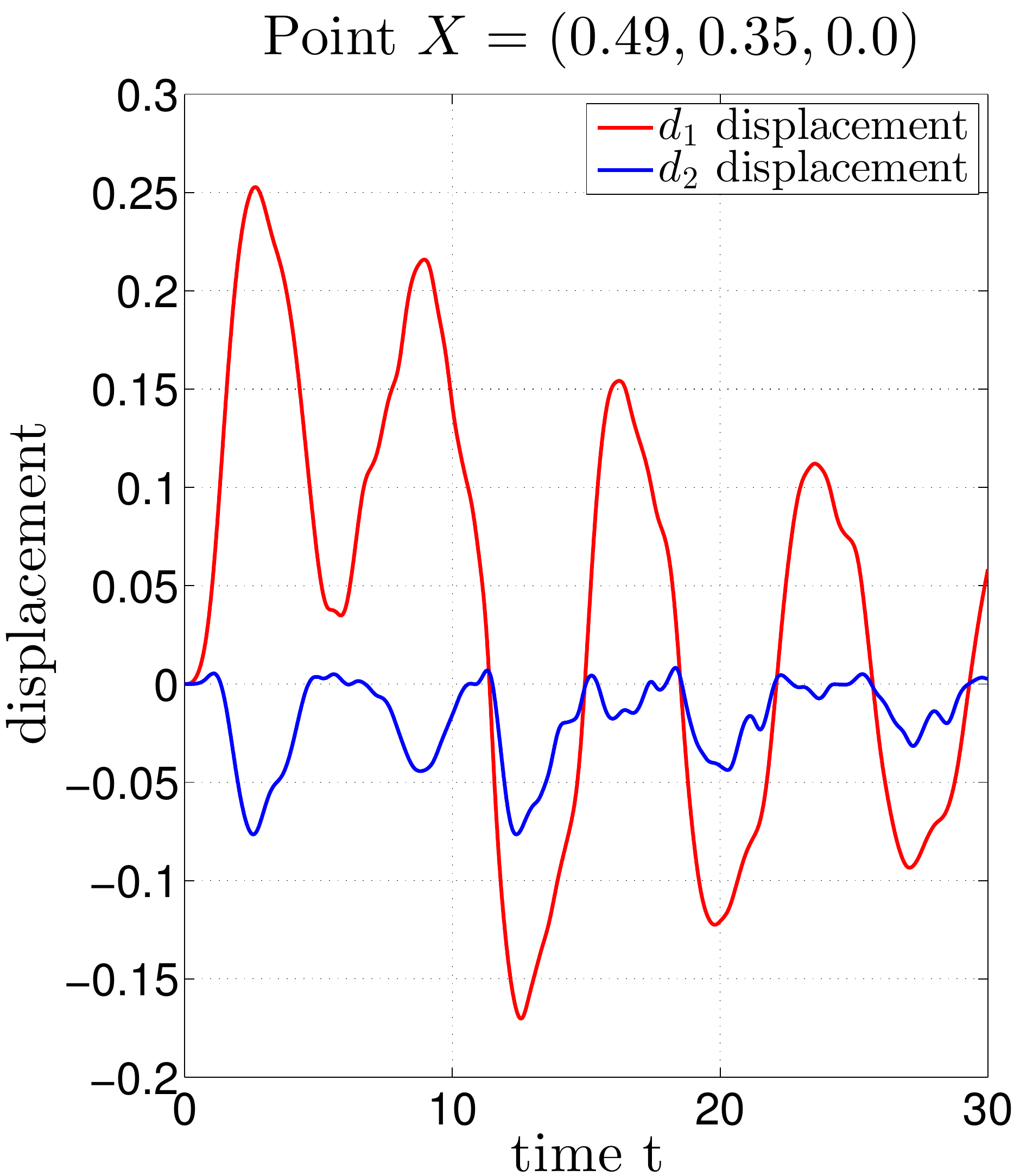}
\qquad
\includegraphics[width=6.0cm, trim=0cm 0cm 0cm 0cm, clip=true]{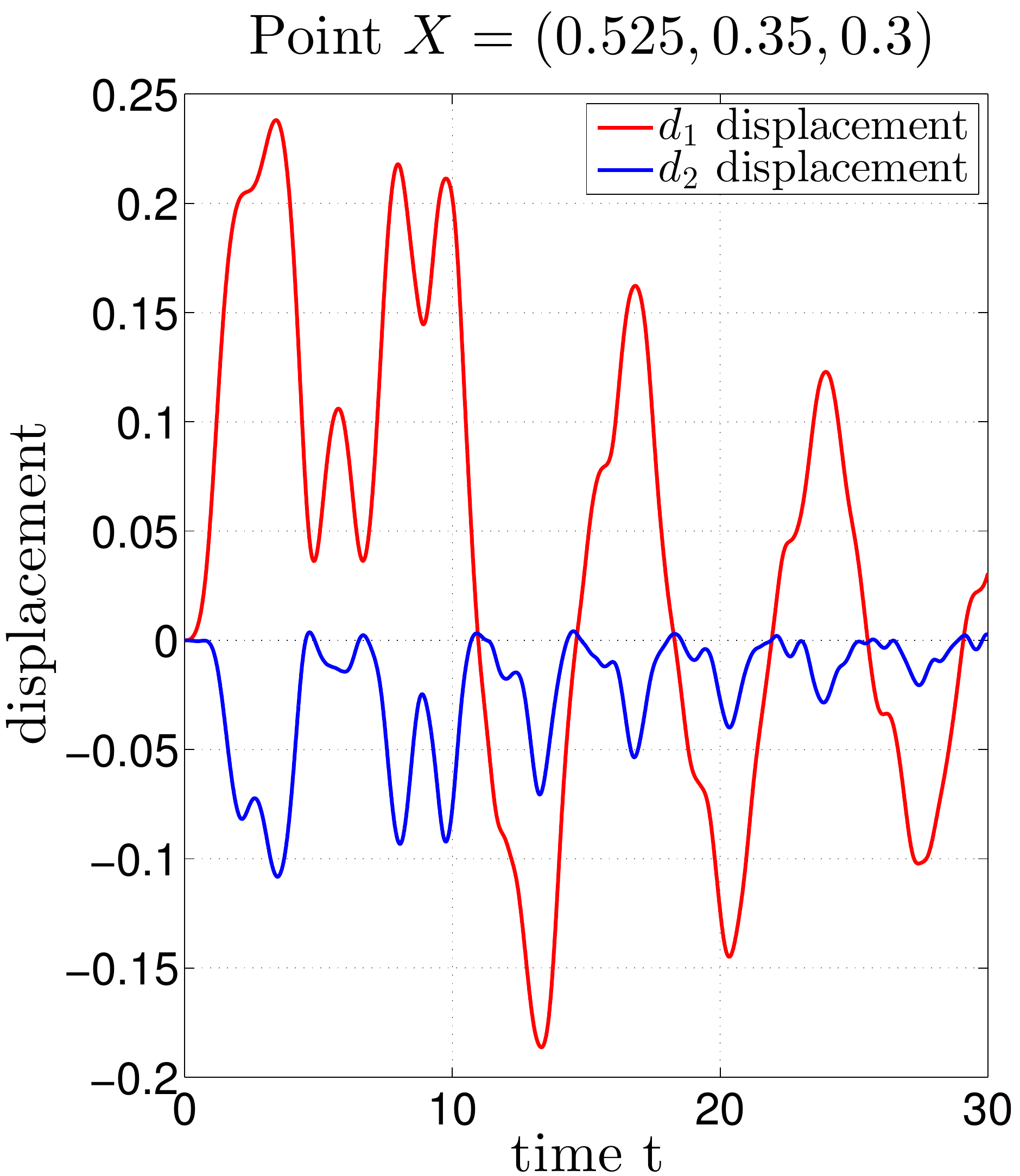}
  \caption{Pulsating flow over a bending flexible flap: history of displacement components \mbox{$d_1,d_2$} of
selected points located at the top surface of the bending flap.
Computed displacements for point \mbox{$\bfX=(0.49,0.35,0.0)$} (left) and for point \mbox{$\bfX=(0.525,0.35,0.3)$} (right)
defined in referential configuration at \mbox{$T_0=0$}.
}
  \label{fig:chap_5_Pulsating_Flow_Bending_Flexible_Flap:displacements}
\vspace{-12pt}
\end{figure}

\paragraph*{Results.} Simulation results on the temporal evolution of the flap bending and its surrounding flow are
shown in \Figref{fig:chap_5_Pulsating_Flow_Bending_Flexible_Flap:time_evolution}.
Flow entering the fluid domain at the inlet streams around the flap and forces acting on structural front-surface
cause bending in mean flow direction as depicted at \mbox{$T_2=1.0$}.
When inflow is decreased to zero for the first time at \mbox{$T_3=2.0$},
the deformed structure tends to turn back to its undeformed initial state.
At this stage, the flow is dominated by the structural motion and different swirls are induced.
Increasing again the mass flow at the inlet reaching its peak at \mbox{$T_4=2.5$}, due to strong induced viscous and pressure forces acting on the flap
it again undergoes deformations mainly in stream direction.
Whenever inflow is strong enough, the structural motion is dominated by the
flow and tries to minimize resistance in the surrounding flow.
In contrast, when the structure is deformed and no flow is entering through the inlet anymore,
its motion towards its undeformed configuration dominates the flow.
States at later times during the excitation phase depict this highly dynamic fluid-structure interaction
and the periodically recurring dominance of pulsating inflow and of strongly deformed structural state,
as exemplarily shown at \mbox{$T_5=9.0$}.
After the excitation phase, different low-velocity vortices are present in the fluid domain and
continuously interact with the rubbery structure.
Also negative deflections in $x_1$-direction occur as indicated at times \mbox{$T_6=12.5$}.
Due to the higher Poisson's ratio, even larger deformation perpendicular to the main stream direction arise.
The amplitude of motion decays slowly over time.
For two selected points located at the flap's top surface,
history of displacements in $x_1$- and $x_2$-directions are reported in \Figref{fig:chap_5_Pulsating_Flow_Bending_Flexible_Flap:displacements}.
Their coordinates are specified in initial configuration at \mbox{$\bfX=(0.49, 0.35, 0.0)$} for the front top point and
at \mbox{$\bfX=(0.525, 0.35, 0.3)$} for the right top point.

The present test case could serve as a useful example in three spatial dimensions (as many cases in literature are only 2D)
that is not too difficult to set up and
demonstrates the robustness of our proposed monolithic \name{FSI} solver based on the discretization Approach~\ref{approach:2_XFSI}.
In addition to the \cite{FernandezLandajuela2016}, further advancements
have been made with regards to temporal discretization of the unfitted \name{FSI} system, monolithic solution strategies
and efficient iterative solution procedures for the resulting linear matrix systems, as proposed in \Secref{sec:fsi:monolithic}.
All these algorithmic peculiarities of the fixed-grid \name{FSI} approach are crucial for this challenging transient test case and
so show the robustness of all involved techniques.
As indicated by this example, moderate-to-large deformations of the structural body in a convective dominated surrounding flow
exhibiting complex flow patterns can be accurately dealt with this discretization concept.
The independent motion of the solid mesh within the background mesh,
accurately captured and realized by the \name{CutFEM}, allows for complex \name{FSI} subjected to large structural motions and deformations.
Such a technique will come to its full extent when extending it to fluid-structure-contact interaction.
This will be addressed in a future publication.

\subsection{Approach~\ref{approach:1_ALE-FSI}+\ref{approach:3_ALE-XFSI}: Flow Excited Impermeable Fluid Container in Flexible Pipe}
To depict the capabilities of an unfitted mixed Eulerian-Lagrangean technique a configuration, where an elastic, 
impermeable fluid container is excited by dynamic flow threw a flexible tube is analyzed.
Such a problem configuration includes large relative motion of the fluid container and the outer flow domain.
On the other hand the relative motion between flexible tube and the outer flow domain as well as between container and the embedded fluid domain is limited.
For this configuration applying classical fitted solid and fluid discretizations for the interfaces between flexible tube and outer flow and also between container and the embedded fluid domain,
allows to benefit from the advantages such as simple standard Gaussian integration quadrature, 
vanishing requirement for ghost-penalty stabilization and superior solution approximation close to the interface.
On the other hand, the large relative motion of the outer flow domain and the fluid container claims the use of unfitted solid and fluid discretizations which is enabled by the \name{CutFEM}.

\paragraph*{Problem Setup.}
The initial setup for the flexible tube consists of a solid domain  \mbox{$\Dom^{\sd_1} = [-0.75,0.75]\times[0.4,0.5]\cup[-0.75,0.75]\times[-0.5,-0.4]$}, 
which is clamped on inlet and outlet boundaries \mbox{$x_1=-0.75$} and \mbox{$x_1=0.75$}, respectively. The outer boundaries of \mbox{$\Dom^{\sd_1}$}, \mbox{$x_2=0.5$} and \mbox{$x_2=-0.5$} remain traction free.
The solid domain of the elastic fluid container \mbox{$\Dom^{\sd_2}$} has an hollow, elliptical shape with inner semi-axis diameter in $x_1$-direction  \mbox{$d_{x_1} = 0.325$},
inner semi-axis diameter in $x_2$-direction \mbox{$d_{x_2} = 0.525$}, thickness of the elastic container \mbox{$a=0.05$} and center position of the ellipsoid \mbox{$\bfx_c=[-0.45,0]$}.
The inner fluid domain \mbox{$\Dom^{\fd_2}$} is inside the inner ellipsoid of  \mbox{$\Dom^{\sd_2}$}.
Finally, the fluid domain of the surrounding dynamic flow \mbox{$\Dom^{\fd_1}$} is given by the remaining part of the overall domain \mbox{$\Dom = [-0.75,0.75]\times[-0.5,0.5]$} with
\mbox{$\Dom^{\fd_1} = \Dom \setminus \left( \Dom^{\sd_1} \cup \Dom^{\sd_2} \cup \Dom^{\fd_2}\right)$}.
On the inlet boundary of domain \mbox{$\Dom^{\fd_1}$} (\mbox{$x_1=-0.75$}) a parabolic time depend velocity profile 
\mbox{$\bfu_{in} = \left[0.4^{-2}\left(0.4^2-x_2^2\right)\left(\sin(\pi t) + 0.1\right),0.0\right]\, \forall \,t \,\in \,(0,10.0]$} is enforced strongly, whereas zero-traction Neumann boundary condition
is applied on the outlet boundary (\mbox{$x_1=0.75$}).
The three occurring fluid-structure interfaces are denoted by \mbox{$\Int^{f_1s_1} = \overline{\Dom^{\fd_1}} \cap \overline{\Dom^{\sd_1}}$},
\mbox{$\Int^{f_1s_2} = \overline{\Dom^{\fd_1}} \cap \overline{\Dom^{\sd_2}}$} and 
\mbox{$\Int^{f_2s_2} = \overline{\Dom^{\fd_2}} \cap \overline{\Dom^{\sd_2}}$}.
The constitutive behavior for both solid bodies, the flexible tube as well as the fluid container, is given by the Neo-Hookean material, with \mbox{$E_1=500, \nu^s_1=0.3, \rho^s_1=100$}
in domain \mbox{$\Dom^{\sd_1}$} and \mbox{$E_2=250, \nu^s_2=0.3, \rho^s_2=100$} for domain  \mbox{$\Dom^{\sd_2}$}.
The properties of the incompressible fluid are given by \mbox{$\nu^f_1=1.0, \rho^f_1=1.0$} for the outer flow in domain \mbox{$\Dom^{\fd_1}$} 
and \mbox{$\nu^f_2=0.01, \rho^f_2=1.0$} for enclosed fluid in domain \mbox{$\Dom^{\fd_2}$}.
No volume loads are present.

\paragraph*{Computational Approach.}
The spatial discretization to solve this problem is generated in a way to benefit from advantages of both methods for boundary or interface fitted and unfitted computational meshes.
All solid and fluid domains in this example are discretized by two-dimensional bilinearly-interpolated elements. 
The fluid domain \mbox{$\Dom^{\fd_2}$} is discretized by an unstructured mesh with 699 elements and 747 computational nodes and is fitted to the fluid-structure interface \mbox{$\Int^{f_2s_2}$}.
Also the the solid domain \mbox{$\Dom^{\sd_2}$} is discretized by an unstructured mesh with 315 elements and 420 computational nodes and is fitted to the interfaces \mbox{$\Int^{f_1s_2}$} and \mbox{$\Int^{f_2s_2}$}.
Both meshes for the domains \mbox{$\Dom^{\fd_2}$} and \mbox{$\Dom^{\sd_2}$} have matching computational nodes on the interfaces \mbox{$\Int^{f_2s_2}$}, which enables a simple transfer of solid motion
into the fluid domain.
Further, solid domain \mbox{$\Dom^{\sd_1}$} is discretized by a structured mesh, with 4 x 60 elements for the upper and lower part of the domain.
The mesh is fitted to the outer boundaries as well as the fluid-structure interface \mbox{$\Int^{f_1s_1}$}.
Finally, the fluid domain  \mbox{$\Dom^{\fd_1}$} is discretized by a structured mesh, with 32 x 60 elements.
It is fitted only to the fluid-structure interface \mbox{$\Int^{f_1s_1}$} and the inflow and outflow boundaries, but not to the interface \mbox{$\Int^{f_1s_2}$} as large motion is expected.
Temporal discretization is performed by the backward Euler scheme with time-step length $\Delta t = 0.005$ for both solid and fluid domains.
The outer flow field in domain \mbox{$\Dom^{\fd_1}$} is approximated with the CIP/GP method and the fluid flow in domain \mbox{$\Dom^{\fd_2}$} by the CIP formulation.
On all fluid-structure interfaces an adjoint inconsistent Nitsche-type coupling with Nitsche-penalty $\gamma = 45.0$ is applied.
A direct solver for the global linear system is favorable, due to the overall number of degrees of freedom. 
The relative residual vector-norms are limited to~$10^{-6}$.

\begin{figure}
\centering
\includegraphics[trim=0 0 100 80,clip,width=0.32\textwidth]{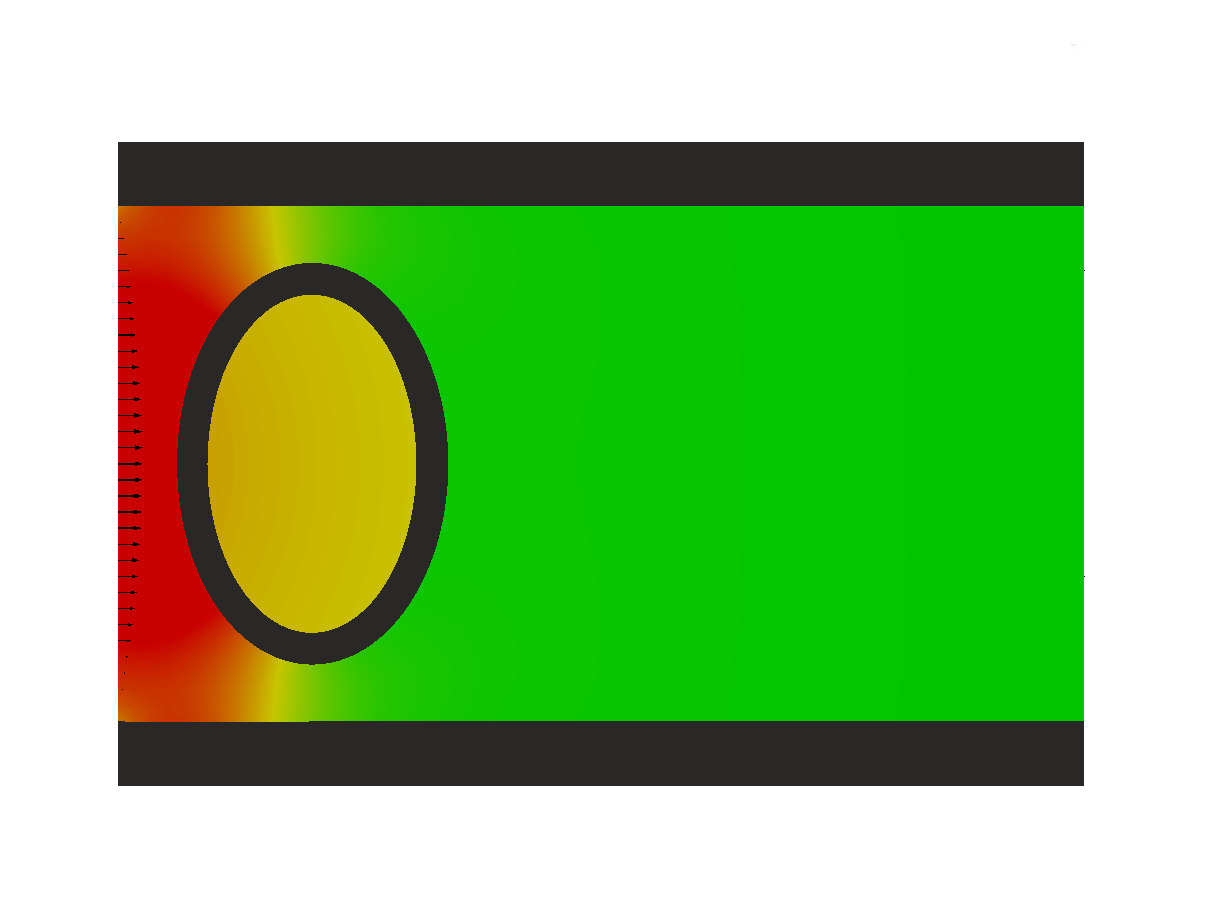}
\includegraphics[trim=0 0 100 80,clip,width=0.32\textwidth]{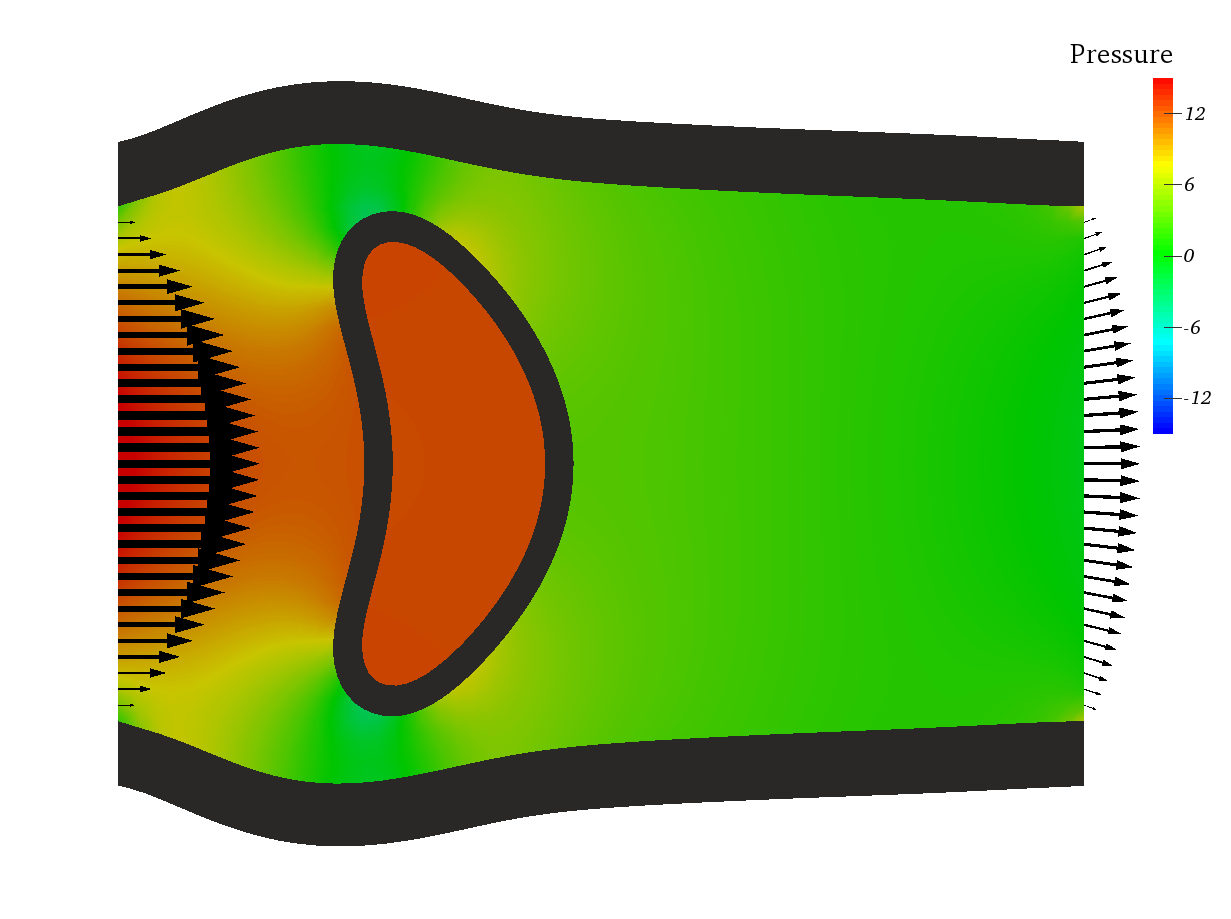}
\includegraphics[trim=0 0 100 0,clip,width=0.32\textwidth]{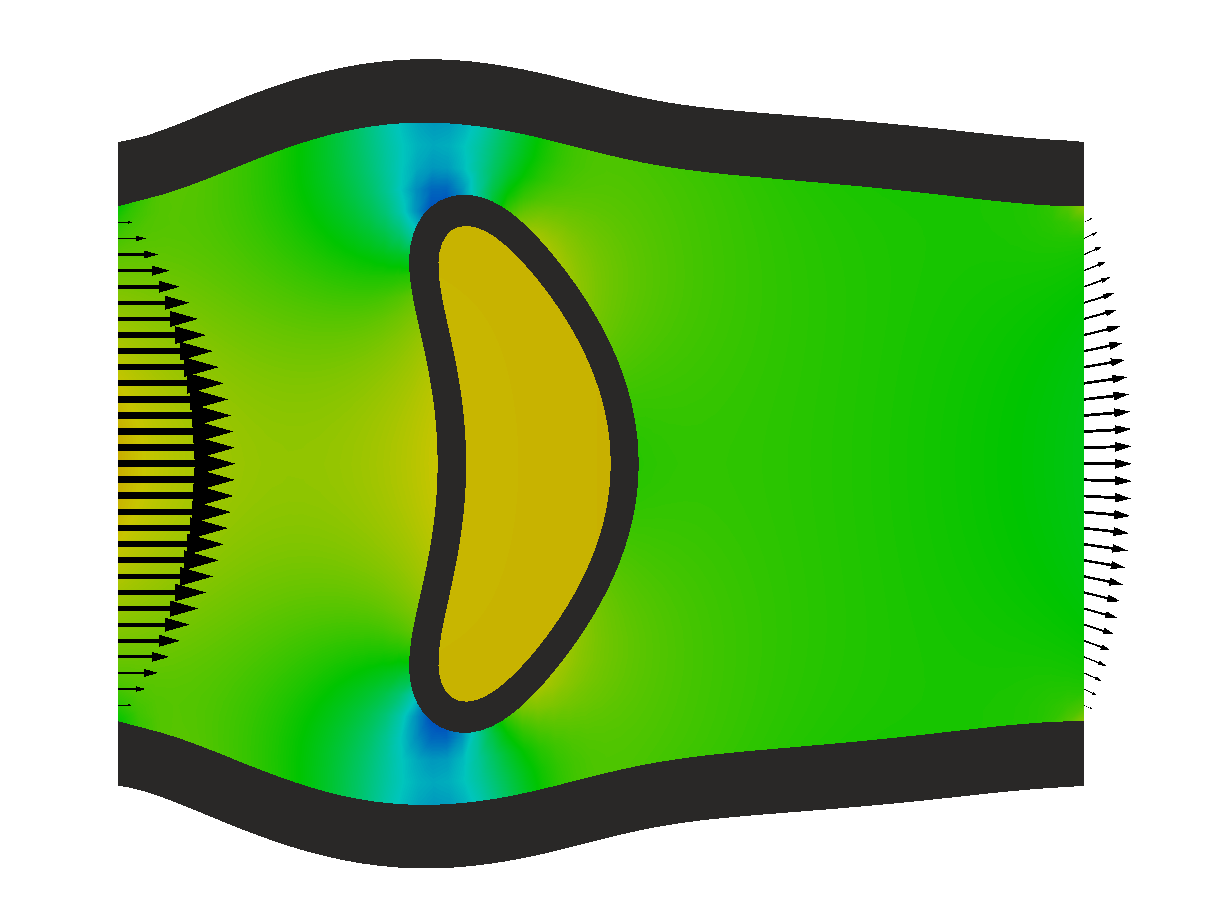}
\includegraphics[trim=0 0 100 80,clip,width=0.32\textwidth]{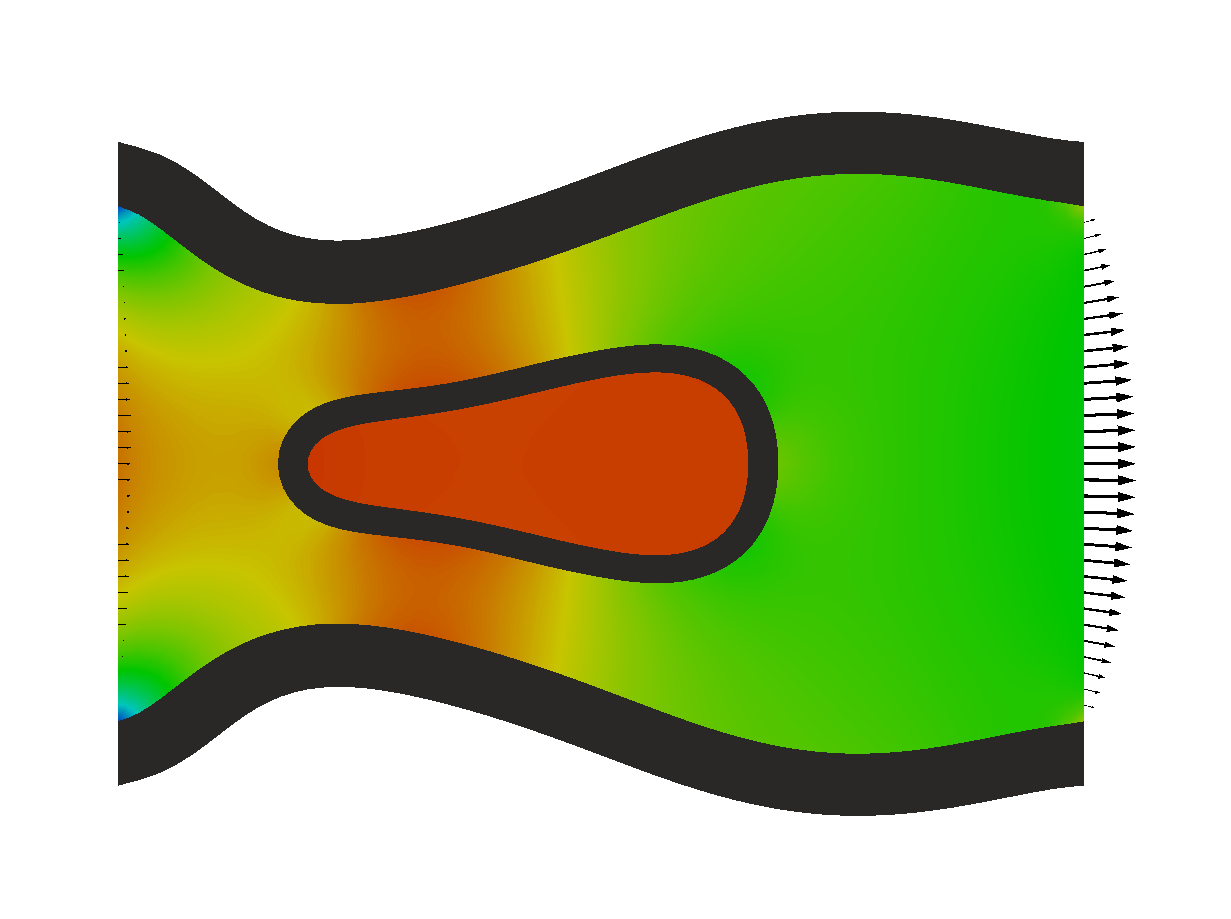}
\includegraphics[trim=0 0 100 0,clip,width=0.32\textwidth]{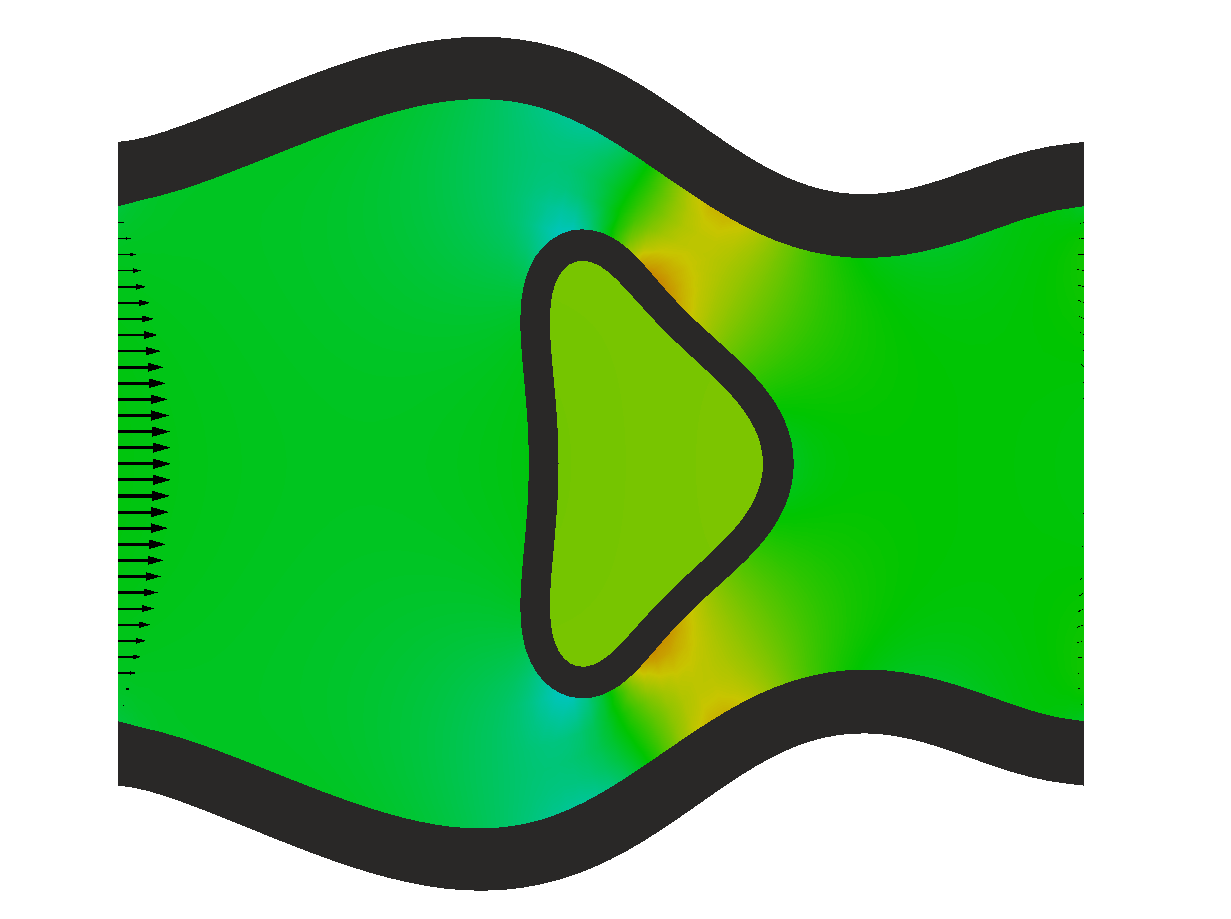}
\includegraphics[trim=0 0 100 0,clip,width=0.32\textwidth]{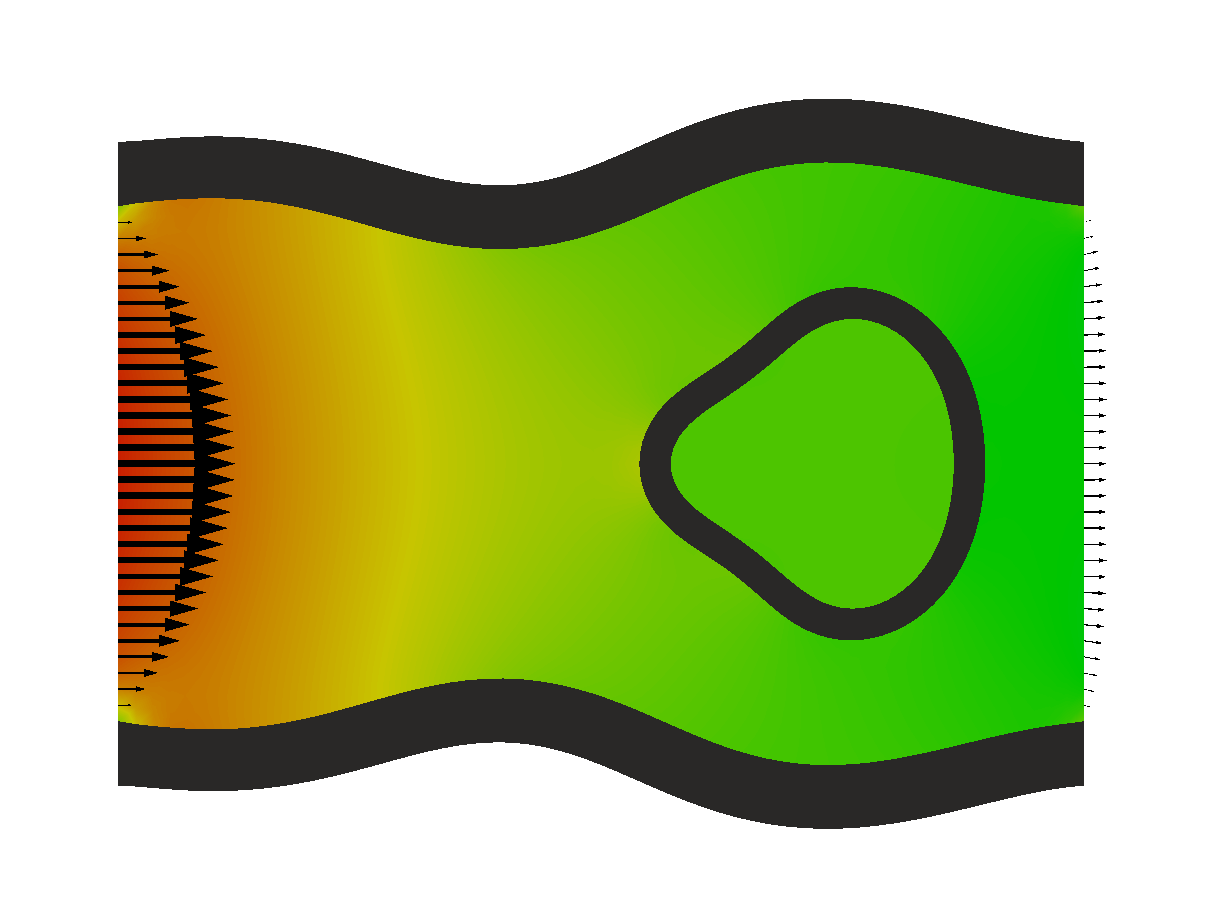}
\includegraphics[trim=300 100 300 730,clip,width=0.5\textwidth]{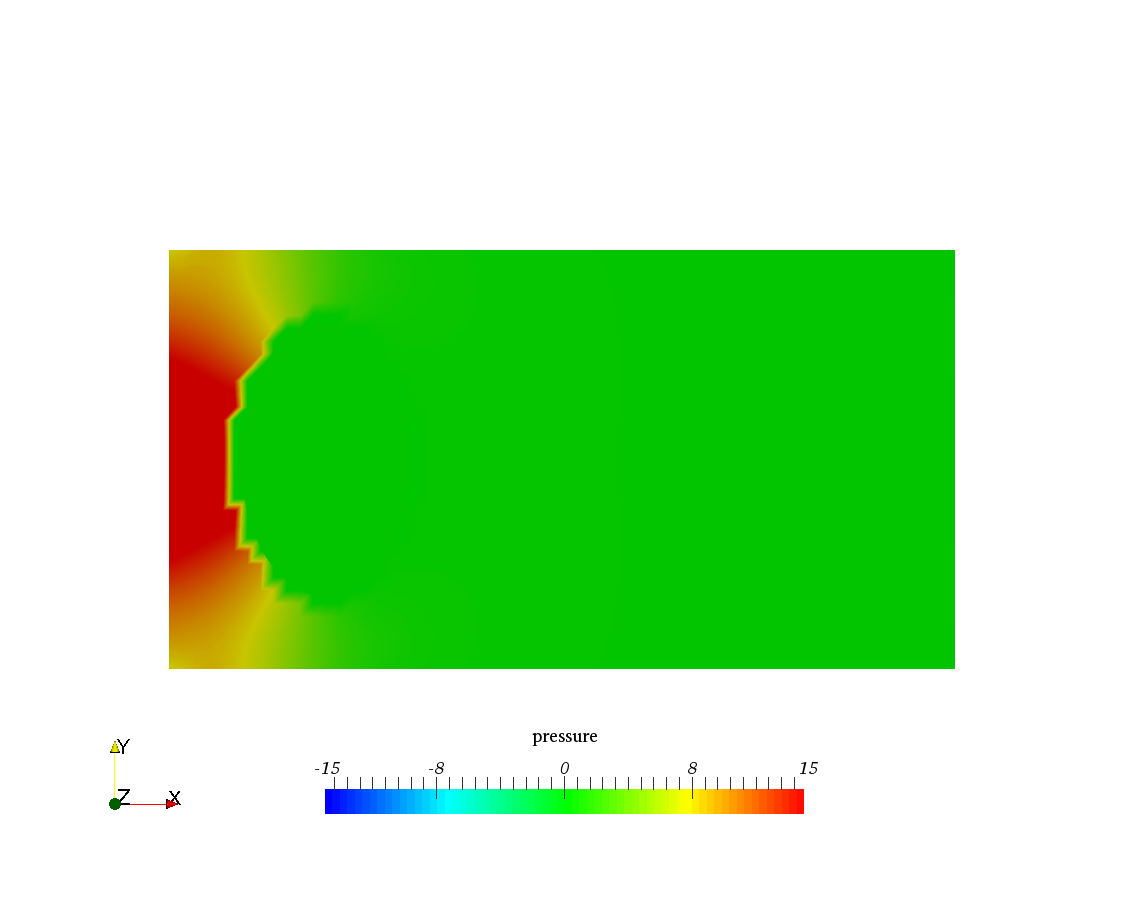}
\caption{Visualization of the pressure solution in fluid domains by color map, inflow and outflow velocities by black arrows and deformation of the domains for specific points in time.
The time-steps from top-left to bottom-right are: $T_1=0.03, T_2=0.5, T_3=0.7, T_4=2.0, T_5=2.9,T_6=8.3$.}
\label{fig:chap_5_Filled_Ballon:Pres}
\end{figure}

\begin{figure}
\centering
\includegraphics[trim=0 0 200 0,clip,width=0.42\textwidth]{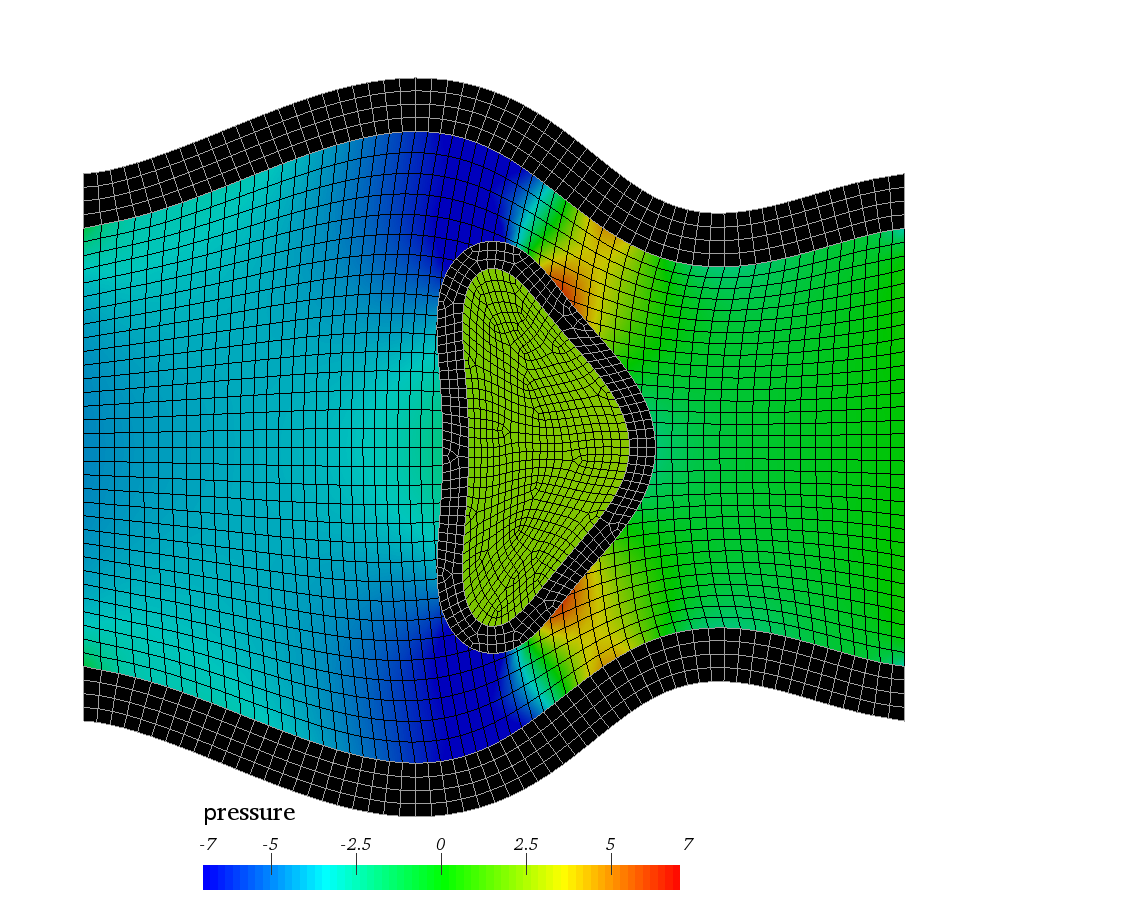}
\includegraphics[trim=0 0 0 0,clip,width=0.57\textwidth]{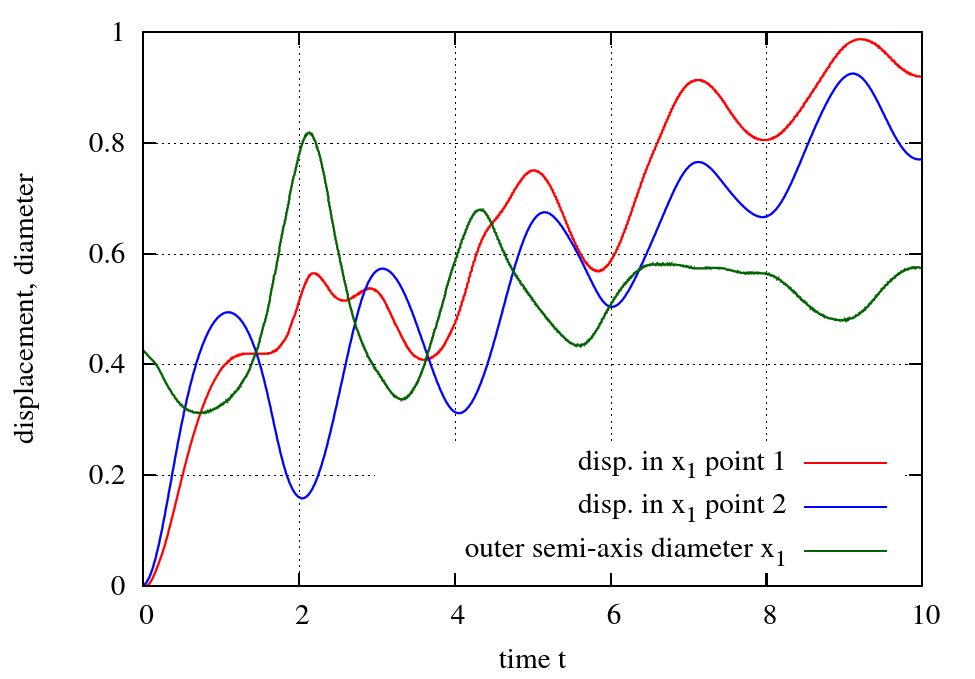}
\caption{Pressure solution in deformed domains at $T_7=3.0$, the boundaries between elements are visualized by black lines in the fluid domains and gray lines in the solid domains(left).
Displacement in $x_1$-direction of point 1 with $\bfX_1(t=0.0) = [-0.2375,0.0]$ and of point 2 with $\bfX_2(t=0.0) = [-0.6625,0.0]$ and evolution of the outer semi-axis diameter (right).}
\label{fig::chap_5_Filled_Ballon:Disp}
\end{figure}

\paragraph*{Results.}
Computed results for different selected time-steps are presented in figure \ref{fig:chap_5_Filled_Ballon:Pres}.
At $T_1=0.03$ the deformations is still small and therefore roughly the initial geometry can be seen.
A high pressure increase due to the inflow at the outer flow and slightly reduced for the enclosed fluid due to the container inertia can be observed.
The inflow velocity with its peak value at $T_2=0.5$, leads to large deformations of the tube and the fluid container.
Areas of low pressure can be identified close to the smallest restrictions at $T_3=0.7$.
At $T_4=2.0$ the resulting large stretch of the horizontal semi-axis due to the reversed flow direction at the inflow boundary can be seen.
The second inflow cycle, exemplarily shown at $T_5 = 2.9$, leads to a totally different deformation and pressure solution compared to the first cycle.
Due to the prevailing inflow direction, the fluid container passes threw the whole flexible pipe and finally reaches the outlet ($T_6 = 8.3$).\\
Figure \ref{fig::chap_5_Filled_Ballon:Disp} (left) shows the deformed computational mesh for $T_7=3.0$.
Although the deformation for this point in time is large, due to the beneficial separation of fitted and unfitted discretizations the distortion of the fluid 
elements is still moderate. Matching discretizations for all fluid-structure-interfaces would lead to a totally distorted computational mesh in domain \mbox{$\Dom^{\fd_1}$}.
Finally, figure~\ref{fig::chap_5_Filled_Ballon:Disp} (right) shows the evolution of the displacement in $x_1$-direction of both limiting points of the outer horizontal semi-axis in time.
The resulting diameter of this semi-axis quantifies the large deformation of the fluid container.

\subsection{Approach~\ref{approach:4_FXFSI}: Vibrating of a Flexible Structure}

The test example of a tail-shaped flexible structure,
which is vibrating due to vortex-shedding induced by a surrounding higher-Reynolds-number flow,
is perfectly suited to demonstrate the high capabilities of the composed unfitted discretization concept for \name{FSI} comprised in Approach~\ref{approach:4_FXFSI}.
This test case has been originally introduced and extensively studied in \cite{Wall1998}.
\begin{figure}
  \begin{center}
  \includegraphics[trim=0 5 0 0, clip, width=0.6\textwidth]{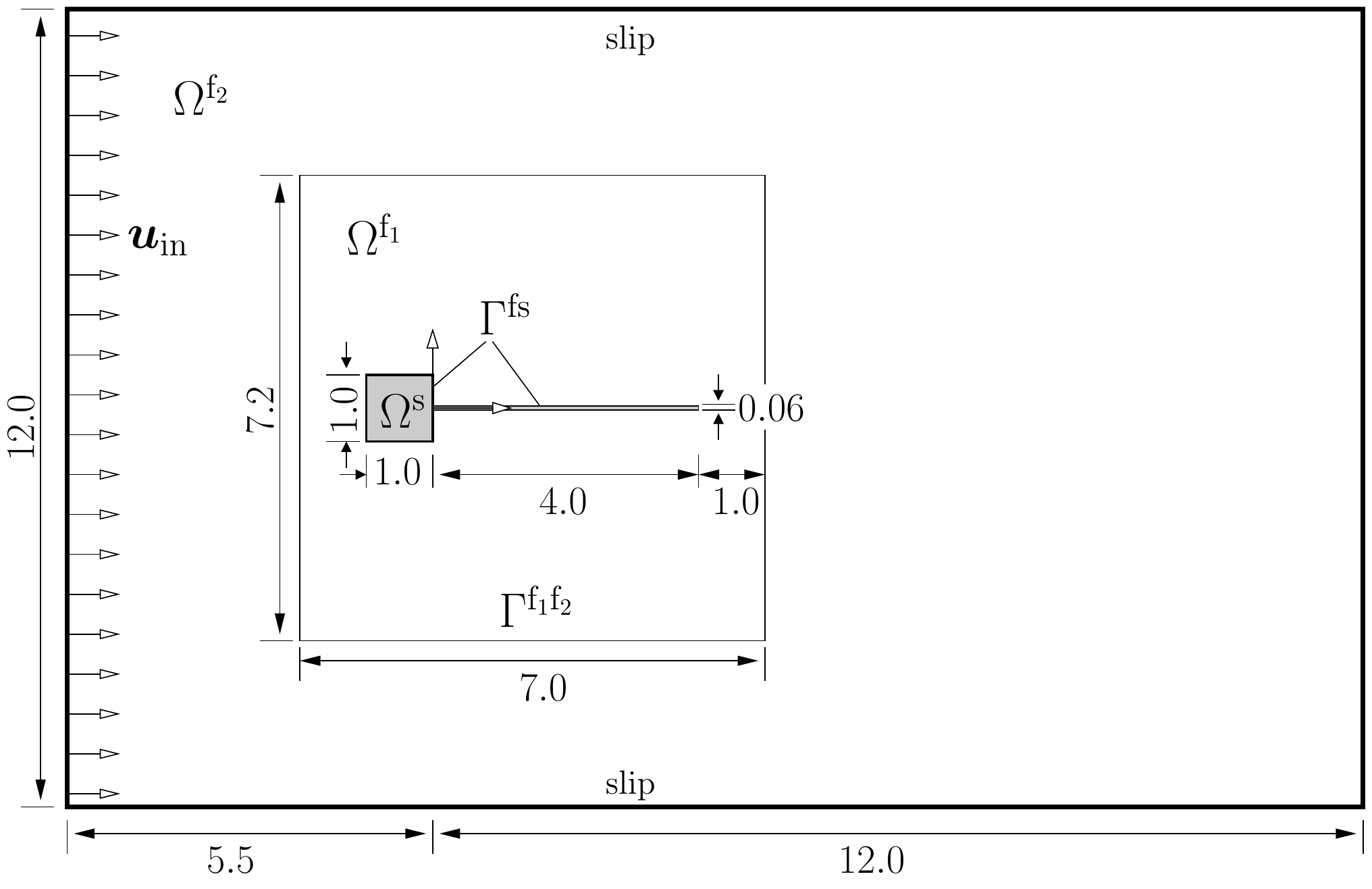}~
  \includegraphics[trim=0 0 0 0, clip, width=0.4\textwidth]{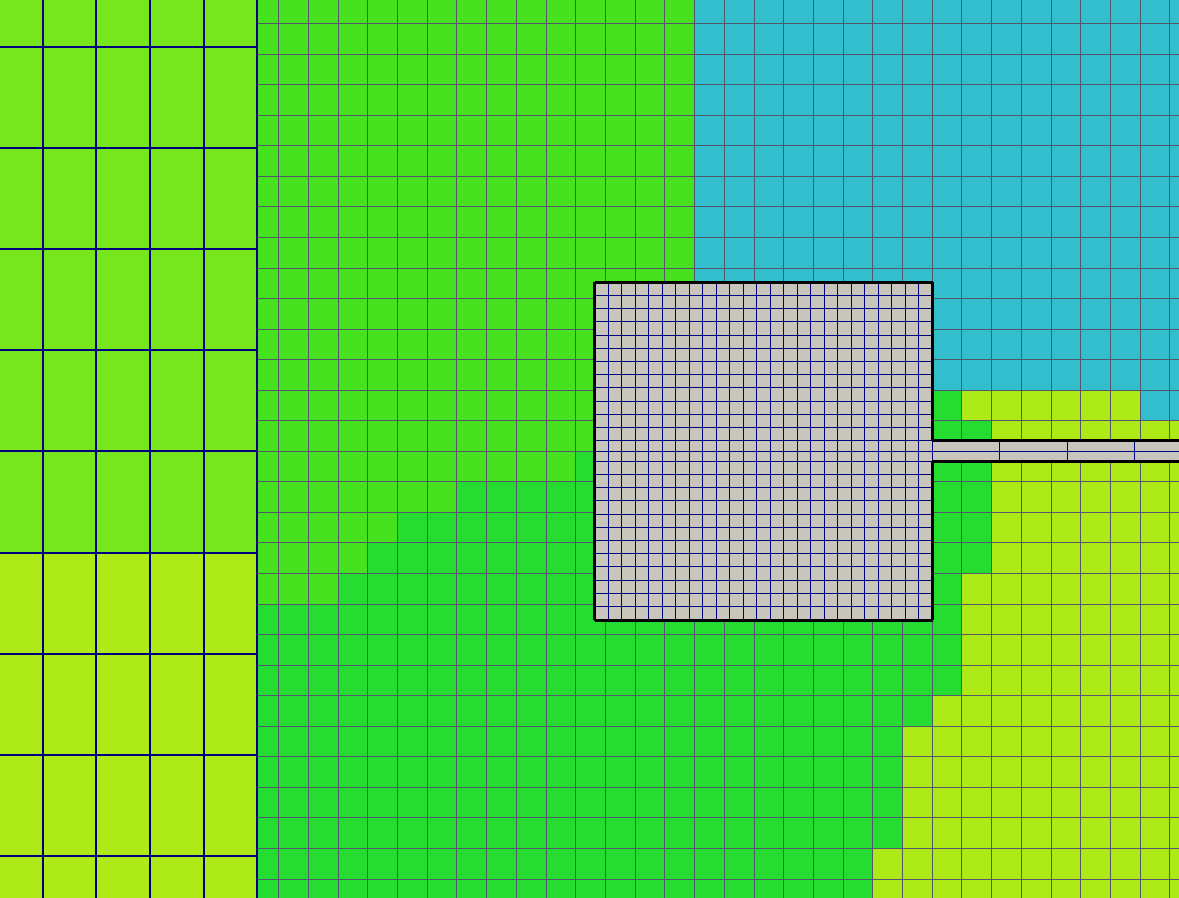} 
  \end{center}
  \vspace{-12pt}
  \caption{Vibrating of a flexible structure: (Left)~geometric setup for unfitted fluid-fluid-solid interaction.
A flexible tail is clamped by a fixed head and embedded into a surrounding flow fluid. Fluid domain is artificially decomposed
into an inner domain $\Dom^{\fd_1}$ and and outer template-shaped domain $\Dom^{\fd_2}$.
(Right)~Close-up view of overlapping meshes
in the vicinity of the structural head and the fluid-fluid interface.}
  \label{fig:chap_5_Vibrating_Flexible_Structure:Setup}
\end{figure}

\paragraph*{Problem Setup.}
The problem setting is taken unchanged from \cite{Wall1998} as drawn in \Figref{fig:chap_5_Vibrating_Flexible_Structure:Setup}.
A two-dimensional flexible structure of length \mbox{$4.0$} and height~\mbox{$0.06$} is clamped at its front end by 
a square-shaped obstacle of edge-length~$1.0$, which is assumed fixed.
The rest of the tail can arbitrarily move in a surrounding fluid
within a domain~$\Dom$ of length~$19.5$ and height~$12.0$.
The entire structural surface defines the fluid-structure interface~$\Int^{\fs}$.
Along the fixed head the fluid-solid interaction simplifies to a no-slip boundary condition for the fluid to be enforced weakly,
as mentioned in Remark~\ref{rem:Nitsche_weak_DBC} and presented in detail in the work \cite{MassingSchottWall2016_CMAME_Arxiv_submit}.
The origin of the setting is set to the front end of the flexible structure.
Constant inlet at \mbox{$x_1=-5.5$} with a fluid velocity of \mbox{$u^{\mathrm{max}}=51.3$} drives the flow
around the fixed structural head.
In its backflow vortices detach and strongly interact with the flexible structural tail.
These excite the structure to vibrate which in return cause further creation and detachment of vortices from the structure.
The flow velocity entering the domain is initially ramped up smoothly by a time curve factor
\mbox{$g(t)=\frac{1}{2}(1-\cos(\pi t/0.1))$} within \mbox{$t\in [0,0.1]$} and kept constant afterwards.
At the boarders perpendicular to the inlet, \ie~\mbox{$x_2=\pm 6.0$}, slip-conditions prevent the flow to escape and a zero-traction Neumann boundary condition \mbox{$\bfhN=\bfzero$}
is enforced at \mbox{$x_1=12.0$}.
The materials are chosen as follows: for the fluid, it is set \mbox{$\mu^\fd=1.82\cdot10^{-4}$} and
\mbox{$\rho^\fd=1.18\cdot 10^{-3}$} resulting in an approximate Reynold's number of $\RE\approx 333$ based on the structural head dimension.
For the structural tail, Neo-Hookean material is considered with \mbox{$\nu^\sd=0.35$} and two different characteristic material sets:
(A)~\mbox{$E= 2.5 \cdot 10^{-6}$}, \mbox{$\rho^\sd=0.1$} and
(B)~\mbox{$E= 2.0 \cdot 10^{-6}$}, \mbox{$\rho^\sd=2.0$}.

\paragraph*{Computational Approach.}

The structural tail is approximated by \mbox{$20\times 2$} two-dimensional $8$-node \mbox{$\mathbb{Q}^2$}-elements.
The fluid domain is decomposed at an artificial fluid-fluid interface \mbox{$\Int^{\fd_1 \fd_2}$}
given by the boundary of the box \mbox{$[-2.0,5.0]\times [-3.6,3.6]$} which defines \mbox{$\Domh^{\fd_1}\cup\Domh^{\sd}$}.
The inner fluid domain \mbox{$\Domh^{\fd_1}$} is approximated with a \mbox{$80\times 80$} bilinearly-interpolated fluid mesh
covering the fictitious domain \mbox{$\Domh^{\fd_1 \ast}=[-2.05,5.05]\times [-3.65,3.65]$}.
This mesh is unfitted to both interfaces, the fluid-solid interface \mbox{$\Int^{\fs}$} and the fluid-fluid interface \mbox{$\Int^{\fd_1 \fd_2}$}.
The outer template-shaped fluid subdomain \mbox{$\Domh^{\fd_2} = \Dom\backslash(\Domh^{\fd_1}\cup\Domh^{\sd})$}
is approximated by a mesh \mbox{$\mcT_h^{\fd_2}$} fitting to its outer and inner subdomain boundary \mbox{$\partial \Domh^{\fd_2}$}.
The mesh $\mcT_h^{\fd_2}$ exhibits a mesh size of \mbox{$h_1=0.1625$}, \mbox{$h_2=0.3$} in $x_1$- and $x_2$-directions, respectively.
This fluid mesh and the structural mesh overlap in an unfitted way with the inner fluid grid $\mcT_h^{\fd_1}$,
as visualized in \Figref{fig:chap_5_Vibrating_Flexible_Structure:Setup}.
As all fluid meshes exhibit perfect symmetry with respect to the $x_1$-axis, the entire structure needs to be slightly displaced
with a perturbation of \mbox{$\epsilon=0.001$} in positive $x_2$-direction
to introduce a small imperfection to later cause vortices to detach.
To reduce damping of the vortex shedding, it is chosen \mbox{$\theta=0.55$}
for the temporal approximation of the fluid with a time-step length of \mbox{$\Delta t = 0.001$}.
The composite flow field is approximated with a classical \name{RBVM} formulation
for $\mcT_h^{\fd_2}$ and an \name{RBVM/GP} method (see \Secref{ssec:stabilized_fitted_unfitted_FEM_fluids} and Remark~\ref{rem:notes_on_unfitted_approximations})
for the intersected mesh $\mcT_h^{\fd_1}$.
For preconditioning the linearized \name{FSI} system \Eqref{eq:chap_5_fsi_linearized_system_multifluid}
a \mbox{$2\times 2$} block-Gauss--Seidel preconditioner for the solid and the composed fluid block are utilized.

\paragraph*{Results.}

\begin{figure}
  \begin{center}
  \includegraphics[trim=0 0 0 0, clip, width=0.49\textwidth]{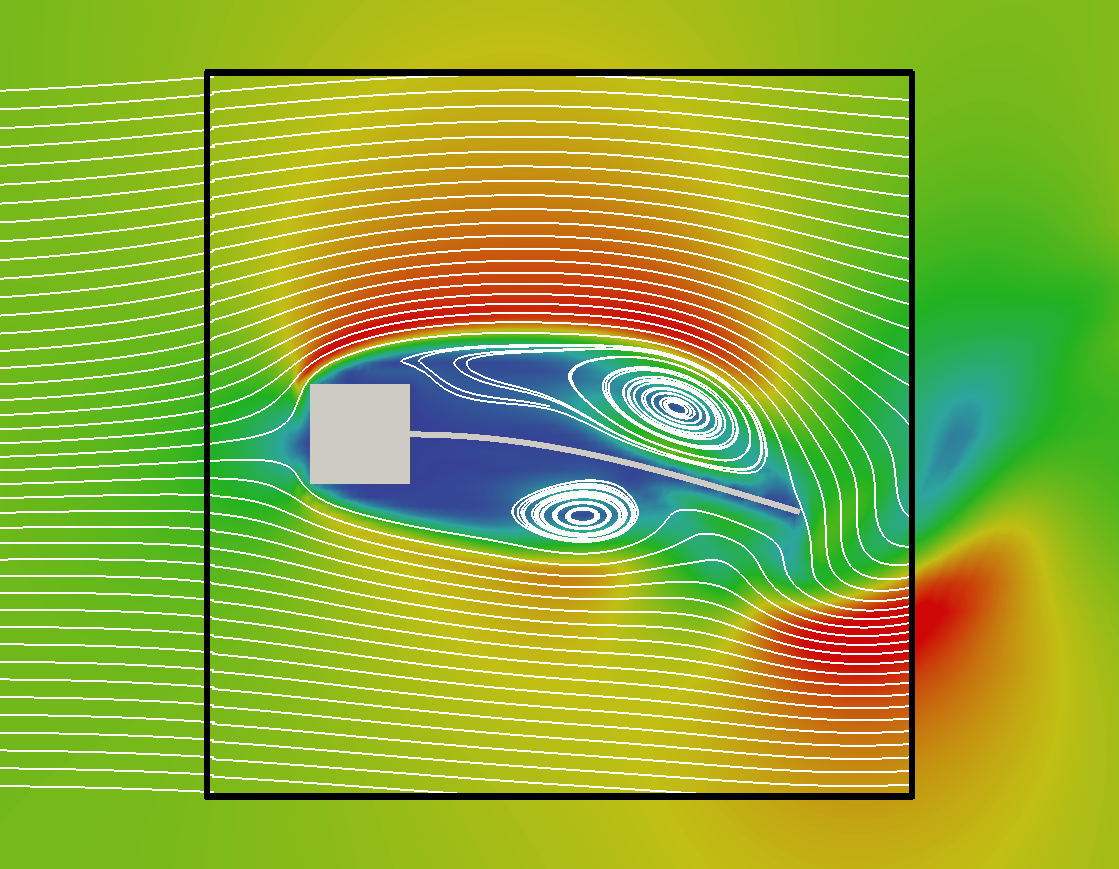} 
  \includegraphics[trim=0 0 0 0, clip, width=0.49\textwidth]{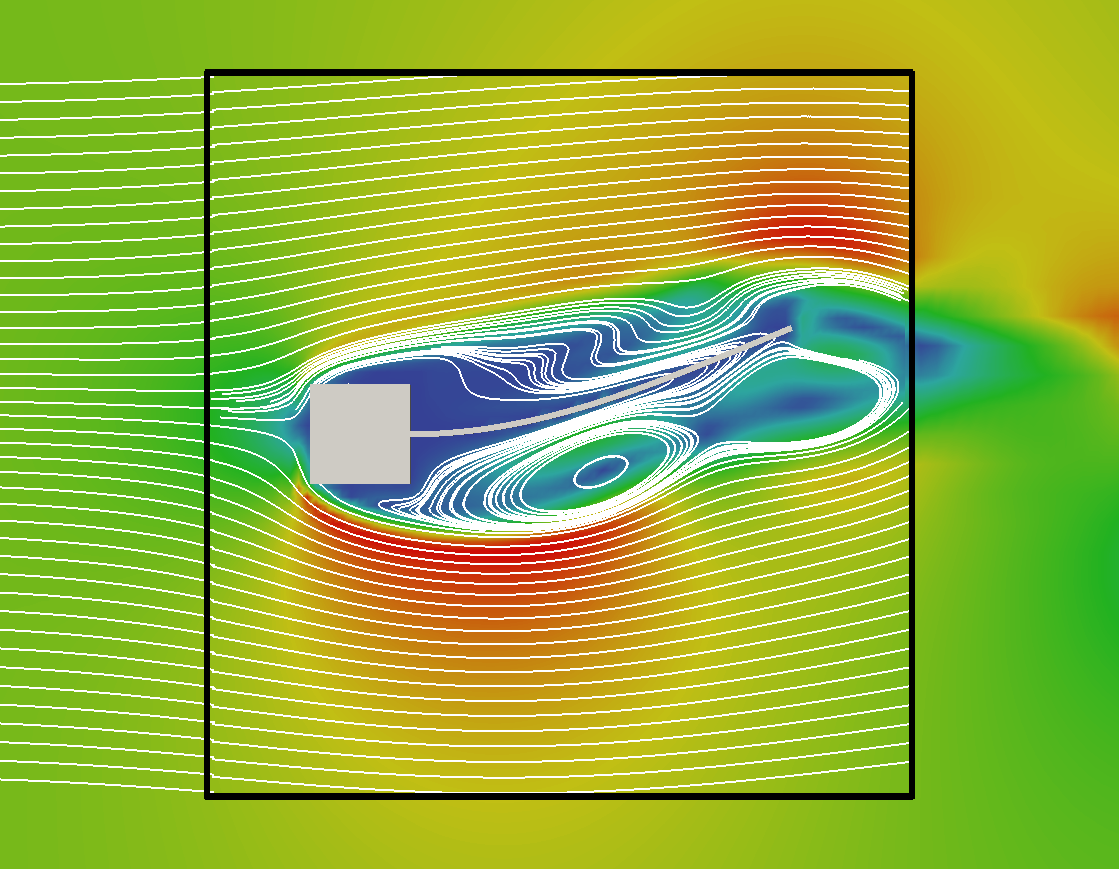} \\
  \vspace{0.1cm}
  \includegraphics[trim=0 0 0 0, clip, width=0.49\textwidth]{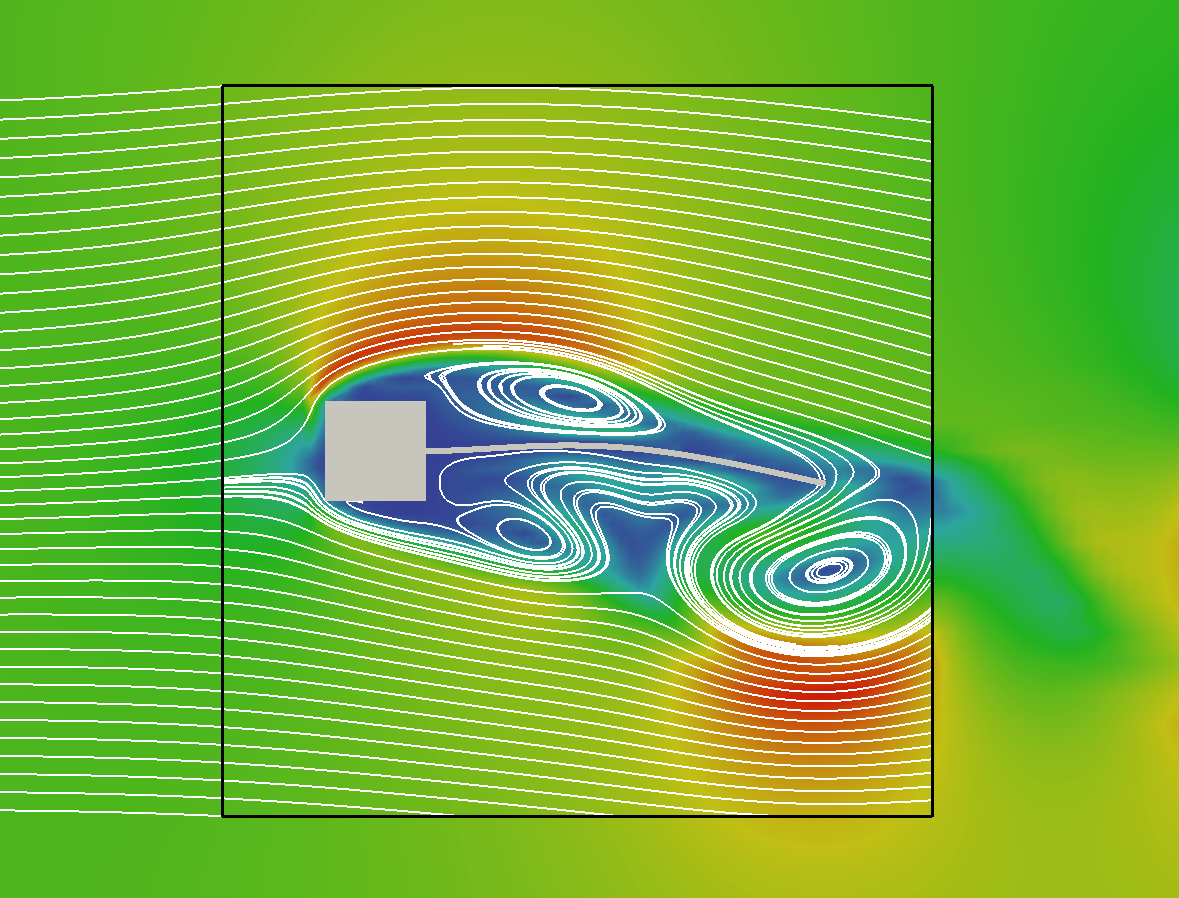} 
  \includegraphics[trim=0 0 0 0, clip, width=0.49\textwidth]{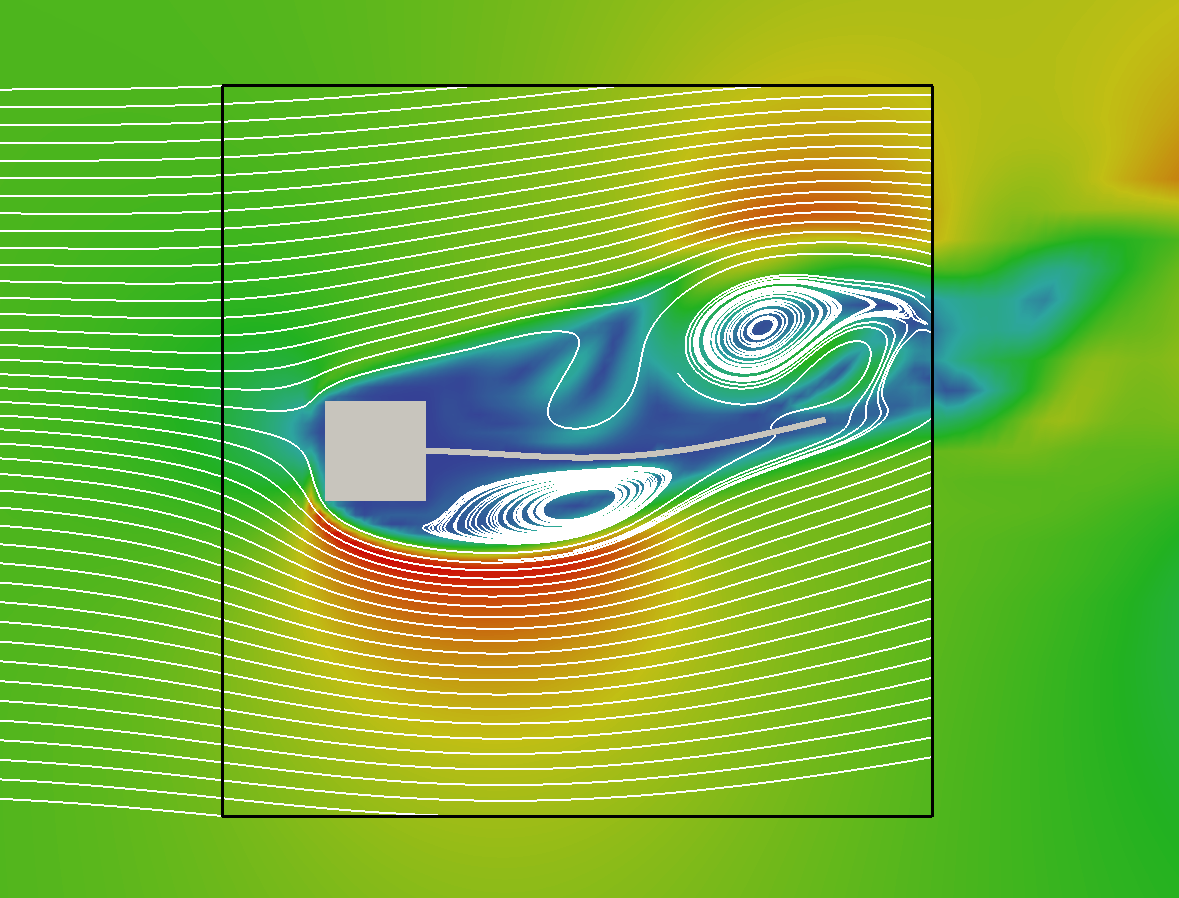} 
  \end{center}
  \vspace{-12pt}
  \caption{Vibrating of a flexible structure: visualization of the surrounding flow and deformation of the vibrating tail
at different times \mbox{$T_1=2.415$}, \mbox{$T_2=2.560$} for setting (A)~(top) and at \mbox{$T_3=3.520$} and \mbox{$T_4=3.880$}
for setting (B)~(bottom).
Streamlines indicate the flow field in the vicinity of the vibrating structure.
Fluid is convected from inflow across the fluid-fluid interface (colored as black lines) and streams around the structural head.
Vortices detach in its backflow and excite the tail to vibrate.
Created swirls are again convected across the fluid-fluid interface.
In contrast to setting (A), setting (B) exhibits vibration with even higher eigenfrequencies.
}
  \label{fig:chap_5_Vibrating_Flexible_Structure:time_evolution_struct}
\end{figure}

Simulation results for the two different material settings (A) and (B)
are visualized for different times in \Figref{fig:chap_5_Vibrating_Flexible_Structure:time_evolution_struct}.
For both settings, the flow is driven by the inlet and fluid is convected from the inflow across the fluid-fluid interface.
Accuracy of the fluid-fluid coupling is indicated by continuous streamlines crossing $\Int^{\fd_1 \fd_2}$.
Flow streams around the structural head at which no-slip boundary conditions are enforced weakly
using the Nitsche-type technique.
Due to the introduced small imperfection, flow develops slightly non-symmetric on the two sides of the flexible tail.
Vortices develop behind the structural step and are transported 
along the tail towards its end where they finally detach at slightly different times.
As a result, the tail is excited to deform which further induces strong detaching vortices.
The oscillation amplitude of the flexible tail increases and it starts to highly dynamically vibrate.
For setting~(A) vibration is expected to be dominated by the first structural eigenfrequency,
whereas for setting~(B) even higher modes are present.
This can be clearly seen from the simulation results for the two different material settings
in \Figref{fig:chap_5_Vibrating_Flexible_Structure:time_evolution_struct}.

The great advantage of this \name{CutFEM}-based discretization concept is, on the one hand, to allow for large structural deformations
and motions independent of the structural positioning within its surrounding fluid mesh,
similar to Approach~\ref{approach:2_XFSI}.
At this point it needs be pointed out that many people reported severe problems with fitted deforming mesh based approaches as soon as deformations get larger,
as especially the small fluid elements next to the tip of the tail tend to fail.
On the other hand, decomposing the fluid domain enables to utilize highly refined meshes in specific regions of interest,
while computational costs then can be kept at a minimum.
Meshing the entire \name{FSI} settings is as simple as possible due to the fixed-grid character of the approximation
and due to the possibility of unfittedness of the meshes with respect to boundaries and interfaces.
In particular when motion of fluid meshes is taken into account, which can be easily realized by utilizing \name{ALE} techniques,
this \name{CutFEM} approximation concept allows for a multitude of further developments and adoptions to different coupled problem settings.
Such schemes will be addressed in more detail in a future publication.

\section{Summary}
\label{sec:summary}

Different cut finite element based approaches for fluid-structure interaction (\name{FSI}) are juxtaposed and novel methods
fitting into a unified monolithic \name{FSI} framework are presented in this work.
Capabilities and limitations of classical interface-fitted \name{ALE}-based moving mesh techniques are discussed,
where most severe restrictions arise for large deformation and/or motion \name{FSI}.
Taking these as driving motivation, novel geometrically unfitted discretization concepts based on the powerful Cut Finite Element Method (\name{CutFEM}) are presented and compared.
First, an unfitted pure-fixed approach for the fluid is considered. It allows for large structural motions in a surrounding fluid, without the need for remeshing or mesh update techniques. Moreover, it is perfectly suited to simulate future fluid-structure-contact interaction problem settings. As a second advancement, classical \name{ALE}-based techniques for small deforming interfaces are combined with the unfitted approach to deal with largely-moving interfaces at the same time. Then, the unfitted fluid mesh does not remain fixed over time anymore. This introduces further algorithmic issues, however, provides a highly flexible approach to challenging \name{FSI}.
As a third development, to increase mesh resolution around largely-deforming bodies embedded into a fixed grid at reasonable computational costs, a CutFEM based fluid domain decomposition technique is incorporated. This is of particular interest, when the main region of interest is known a priori and most of the advantages of unfitted methods shall be preserved.
The latter provides vast ideas for future discretization concepts that are not limited to \name{FSI} even more.

In the present work, all interface couplings---whether with fitting or non-fitting meshes---are enforced weakly using Nitsche's methods, supported by ghost-penalties on cut fluid meshes. Stabilized spatial discretizations are provided for flow governed by the transient incompressible Navier-Stokes equations
coupled to the non-linear structural elastodynamics equations.
All proposed \name{FSI} systems are solved as a monolithic scheme to gain from the temporal stability properties of full-implicitly coupled systems.
Important algorithmic steps arising from cut finite element approximations are discussed and novel aspects regarding changes of fluid function
spaces within a monolithic Newton solve are proposed.
All methods are tested for several two- and three-dimensional \name{FSI} examples which demonstrate the potential of these novel \name{FSI} approaches.

\bibliography{bibliography}
\bibliographystyle{wileyj}
\end{document}